\documentstyle[12pt]{article}

\newcommand{\mysection}{\setcounter{equation}{0}\section}

\begin{document}
\vskip 0.2cm
\hfill{YITP-SB-01-73}
\vskip 0.2cm
\hfill{INLO-PUB-12/01}\\[0.5cm]
\vskip 0.2cm
\centerline{\large\bf {Next-to-leading order QCD corrections to differential}}
\centerline{\large\bf {distributions of Higgs boson production in hadron-hadron
collisions}}
\vskip 0.4cm
\centerline {\sc V. Ravindran}
\centerline{\it Harish-Chandra Research Institute,}
\centerline{\it Chhatnag Road, Jhunsi,}
\centerline{\it Allahabad, 211019, India.}
\vskip 0.2cm
\centerline {\sc J. Smith 
\footnote{partially supported
by the National Science Foundation grant PHY-0098527.}
}
\centerline{\it C.N. Yang Institute for Theoretical Physics,}
\centerline{\it State University of New York at Stony Brook,
New York 11794-3840, USA.}
\vskip 0.2cm
\centerline {\sc W.L. van Neerven 
\footnote{Work supported
by the EC network `QCD and Particle Structure' under contract 
\hspace*{5mm} No.~FMRX--CT98--0194.}}
\centerline{\it Instituut-Lorentz}
\centerline{\it University of Leiden,}
\centerline{\it PO Box 9506, 2300 RA Leiden,}
\centerline{\it The Netherlands.}
\vskip 0.2cm
\centerline{January 2002}
\vskip 0.2cm
\centerline{\bf Abstract}
\vskip 0.3cm
We present the full next-to-leading order corrected differential
distributions $d^2\sigma/dp_T/dy$, $d\sigma/dp_T$ and
$d\sigma/dy$ for the semi-inclusive
process $p + p\rightarrow H + 'X'$. Here $X$ denotes the inclusive hadronic 
state and $p_T$ and $y$ are the transverse momentum and rapidity of the
Higgs-boson $H$ respectively. 
All QCD partonic subprocesses have been included. 
The computation is carried out in the limit that the top-quark
mass $m_t \rightarrow \infty$
which is a very good approximation as long as $m_H,p_T < 200$ GeV. Our
calculations reveal that the dominant subprocess is given by 
$g + g \rightarrow H + 'X'$ but the reaction $g + q(\bar q) \rightarrow 
H + 'X'$ is not negligible. 
Another feature is that the $K$-factor representing the ratio between
the next-to-leading order and leading order differential distributions
is large. It varies from 1.4 to 1.7 depending on the kinematic region
and choice of parton densities. We show that a reliable determination of the 
differential cross sections requires good knowledge of the 
gluon density in the region where $x < 10^{-3}$.
Further we study whether the differential distributions are dominated at 
large transverse momentum by soft-plus-virtual gluon contributions. 
This is of interest for the resummation of large
corrections which occur near the boundary of phase space. We also compare
our results with those previously reported in the literature.

\vfill

\mysection{Introduction}
\newcommand{\be}{\begin{eqnarray}}
\newcommand{\ee}{\end{eqnarray}}
The Higgs boson, which is the corner stone of the standard model, is the 
only particle which has not been discovered yet. Its discovery or its
absence will shed light on the mechanism how particles acquire mass
as well as answer questions about supersymmetric extensions of the standard
model or about compositeness of the existing particles and the Higgs boson. 
The LEP experiments \cite{lep} give a lower mass limit of about 
$m_H\sim 114~{\rm GeV/c^2}$ and fits to the data using precision calculations
in the electro-weak sector of the standard model indicate an upper limit 
$m_H<200~{\rm GeV/c^2}$ with $95~\%$ confidence level. 
After the end of the LEP program the search for the Higgs will be
continued at hadron colliders in particular at the TEVATRON and the LHC.
If the Higgs mass is in the above range the 
principal production mechanisms at hadron
colliders are gluon-gluon fusion $g + g \rightarrow H + 'X'$
or $W^+W^-$-fusion appearing in the reaction 
$q + \bar q\rightarrow q + \bar q + H + 'X'$. In these processes $'X'$ denotes
an inclusive hadronic state. For $m_H<135~{\rm GeV/c^2}$ the process
$q + \bar q\rightarrow V \rightarrow H + V + 'X'$ with $V=\gamma,Z,W$ has
also to be taken into consideration. The Higgs boson will be observed via
its decay products among which the channels $H\rightarrow b + \bar b$ and
$H\rightarrow \gamma + \gamma$ are the most prominant ones although
$H\rightarrow \tau^+ + \tau^-$ should also be considered. However 
the large backgrounds make the observation
of these decays very difficult and it will take a lot of 
experimental and theoretical effort to detect the Higgs boson provided
it is there. 

In this paper we concentrate on Higgs production
channels where the lowest order reaction proceeds via the gluon-gluon fusion
mechanism. In the standard model the Higgs boson couples to the gluons
via heavy quark loops among which the top-quark loop is the most prominent since
the coupling of the Higgs to a fermion loop is proportional to the mass of
the fermion (for a review see \cite{ghkd}). In lowest order (LO) the 
gluon-gluon fusion process $g + g \rightarrow H$, represented
by the top-quark triangle graph, was computed in \cite{wil}. 
The next-to-leading (NLO) processes given by
gluon bremsstrahlung $g + g \rightarrow g + H$ and $g + q(\bar q) \rightarrow 
q(\bar q) + H$ were presented in \cite{hino}, \cite{ehsb} and \cite{kauff} 
from which one can derive
the transverse momentum ($p_T$) and rapidity ($y$) distributions of the
Higgs boson. The total integrated
cross section, which also involves the computation of the QCD corrections 
to the top-quark loop, has been calculated in \cite{gsz}.
This calculation is rather cumbersome since it involves the computation
of two-loop triangular graphs with massive quarks. Recently also the 
next-to-next-to-leading (NNLO) processes involving all two-to-three parton
processes have been computed in \cite{dkosz} using the helicity method
which means that the matrix elements are presented in four dimensions. 
From the experience gained from 
the NLO corrections it is clear that it will be very difficult to obtain 
the NNLO one-particle inclusive distributions from these calculations 
let alone the total cross sections. 

Fortunately one can simplify the calculations if one takes the
large top-quark mass limit $m_t \rightarrow \infty$. 
In this case the Feynman graphs are obtained from 
an effective Lagrangian describing the direct coupling of the Higgs boson
to the gluons. The LO and NLO contributions to the total cross section
in this approximation were computed in \cite{dawson} and they found that
the error introduced by taking the $m_t\rightarrow \infty$ limit is less than
about $5\%$ provided $m_H\le 2~m_t$. A similar investigation was done for
the differential distributions of the processes $g + g \rightarrow g + H$ and 
$g + q(\bar q) \rightarrow q(\bar q) + H$ in \cite{bagl}. Here the 
approximation is valid as long as $m_H$ and $p_T$ are smaller than $m_t$.
This is corroborated by the recent calculations in \cite{dkosz} which
show that for $m_H=120~{\rm GeV/c^2}$ the approximation is valid for
jets with transverse momenta smaller than $200~{\rm GeV/c}$.
The NNLO matrix elements using the effective Lagrangian were computed in
\cite{daka}, \cite{kdr} albeit in four dimensions. The one-loop corrections to
the two-to-two parton subprocesses were presented in \cite{schmidt},
where the computation of the loop integrals was performed in $n$-dimensions but
the matrix element was still presented in four dimensions. The results in
\cite{daka}, \cite{kdr} and \cite{schmidt} were used to compute the transverse 
momentum and rapidity distributions of the Higgs boson up to NLO \cite{fgk}.
The effective Lagrangian method was also applied to obtain the NNLO total 
cross section by the calculation of the two-loop corrections to the 
Higgs-gluon-gluon vertex in \cite{harland1}, the soft-plus-virtual gluon 
corrections in \cite{harland2} and the computation of the two to three 
body processes in \cite{harland3}.

In this paper we present the full NLO
computation of the double differential distributions $d^2\sigma/dp_T/dy$
for Higgs boson production in hadron-hadron collisions using the gluon-gluon
fusion mechanism in the $m_t\rightarrow \infty$ approximation. Here we have
included all partonic subprocesses. A similar
calculation has been performed in \cite{fgk} but our approach 
differs from it in various aspects. First our calculation
is purely analytical and follows the calculation carried
out for the Drell-Yan process describing vector boson production in 
hadron-hadron collisions (see \cite{arre}). The approach in \cite{fgk}
was mainly numerical and based on the methods explained in \cite{fks}.
Moreover it used the two-to-three particle matrix elements in 
\cite{daka}, \cite{kdr} and the two-to-two particle matrix elements including
virtual corrections in \cite{schmidt} which were all presented in four 
dimensions. In our calculation the matrix elements as well as the loop
integrals and phase space integrals are computed in $n$ dimensions.
The advantage of the analytical approach is that one gets more insight
into the structure of the radiative corrections. This is particularly
important for the large corrections, due to soft gluon radiation
and collinear fermion pair production, which arise 
near the boundary of phase space, where the $p_T$ of the Higgs 
boson gets large. Resummation of
this type of corrections has been carried out for the total cross section
in \cite{kls}. Resummation of small $p_T$ contributions due
to the Sudakov effect has been done in \cite{yuan}. 
In view of the experimental problems to observe the Higgs boson,
a recalculation of all the NLO corrections is necessary to be 
sure that the theoretical predictions are correct. 
We find that the contribution due to the (anti-) quark
gluon subprocess is substantial in particular at large transverse momentum
and that the important region of $x$ in the gluon density is
$x < 10^{-3}$. 
We also investigate the region of applicability 
of the soft-plus-virtual (S+V) approximation for the calculation of the
differential cross sections. Finally we mention that another paper 
has just appeared on the NLO corrections to the $g + g ->H + g$ channel,
using the helicity framework \cite{glosser}.
 
Our paper will be organized as follows. In section 2 we give an outline of
the kinematics and present the Born contributions. In section 3 we present
the virtual contributions. In section 4 the gluon bremsstrahlung corrections
to the Born reactions are computed and the soft gluon cross sections
are explicitly shown. In section 5 we show how the mass factorization is
carried out. In section 6 we give differential cross sections 
for proton-proton collisions at the LHC 
and make comparisons with results obtained earlier in the literature.
Some particular expressions are given in appendices A and B. 
Other formulae, in particular those for the two-to-three body
cross sections, are too long to be published. They
are available upon request as files written in the
the algebraic manipulation program FORM \cite{form}.

\pagestyle{myheadings}  
\mysection{Lowest order contributions to Higgs production}
In the large top-quark mass limit the Feynman rules (see e.g. \cite{kdr})
can be derived from the following effective Lagrangian density
\begin{eqnarray}
\label{eqn2.1}
{\cal L}_{eff}=G\,\Phi(x)\,O(x) \quad \mbox{with} \quad
O(x)=-\frac{1}{4}\,G_{\mu\nu}^a(x)\,G^{a,\mu\nu}(x)\,
\end{eqnarray}
where $\Phi(x)$ represents the Higgs field and $G$ is an effective coupling
constant given by
\begin{eqnarray}
\label{eqn2.2}
G^2=4\,\sqrt 2\,\left (\frac{\alpha_s(\mu_r^2)}{4\pi}\right )^2\,G_F\,\tau^2\,
F^2(\tau)\,{\cal C}^2\left (\alpha_s(\mu_r^2),\frac{\mu_r^2}{m_t^2}\right )\,.
\end{eqnarray}
In the expression above $m$ and $m_t$ denote the masses of the Higgs boson
and the top quark respectively. The running coupling is given
by $\alpha_s(\mu_r^2)$ where $\mu_r$ denotes the renormalization scale and
$G_F$ is the Fermi constant.
Further ${\cal C}$ is the coefficient function which originates from the
QCD corrections to the top-quark triangle graph describing the process
$H\rightarrow g + g$ in the limit $m_t \rightarrow \infty$. On the Born
level the width of this decay process is given by
\begin{eqnarray}
\label{eqn2.3}
\Gamma(H\rightarrow g + g)=\frac{(N^2-1)\,m^3\,G_F}{128\,\pi^3\,\sqrt 2}\,
\alpha_s^2(\mu_r^2)\,\tau^2\,F^2(\tau)\,,
\end{eqnarray}
where $N$ denotes the number of colours. Further the function $F(\tau)$
occurring in Eqs. (\ref{eqn2.2}), (\ref{eqn2.3}) is defined by
\begin{eqnarray}
\label{eqn2.4}
&& F(\tau)=1+(1-\tau)\,f(\tau)\,, \qquad \tau=\frac{4\,m_t^2}{m^2}\,,
\nonumber\\[2ex]
&&f(\tau)=\arcsin^2 \frac{1}{\sqrt\tau}\,, \quad \mbox{for} \quad \tau \ge 1\,,
\nonumber\\[2ex]
&& f(\tau)=-\frac{1}{4}\left ( \ln \frac{1-\sqrt{1-\tau}}{1+\sqrt{1-\tau}}
+\pi\,i\right )^2 \quad \mbox{for} \quad \tau < 1\,.
\end{eqnarray}
In the large $m_t$-limit $F(\tau)$ behaves as
\begin{eqnarray}
\label{eqn2.5}
 \mathop{\mbox{lim}}\limits_{\vphantom{\frac{A}{A}} \tau \rightarrow \infty}
F(\tau)=\frac{2}{3\,\tau}\,.
\end{eqnarray}
The coupling $G$ in Eq. (\ref{eqn2.2}) is presented for
general $m_t$ on the Born level only whereas ${\cal C}$ is computed 
in higher order for $m_t \rightarrow \infty$. 
In order to keep some part of the
top quark mass dependence we take for $G$ the expression in Eq. (\ref{eqn2.2}).
This is an approximation because the gluons which couple to the 
Higgs boson via the top-quark loop in the partonic processes
describing Higgs-production are often virtual. 
The virtual-gluon momentum dependence 
is neither described by $F(\tau)$ in Eq. (\ref{eqn2.4})
nor by ${\cal C}$. For on-mass-shell gluons the latter quantity has been
computed in the large top-quark mass limit up to order $\alpha_s$ in 
\cite{gsz}, \cite{dawson}, \cite{daka} and up to $\alpha_s^2$ in 
\cite{kls},\cite{cks}.
In second order it reads
\begin{eqnarray}
\label{eqn2.6}
{\cal C}\left (\alpha_s(\mu_r^2),\frac{\mu_r^2}{m_t^2}\right )&=&
1+\frac{\alpha_s^{(5)}(\mu_r^2)}{4\pi}\,\Big (11\Big )+
\left (\frac{\alpha_s^{(5)}(\mu_r^2)}{4\pi}\right )^2\,\left [\frac{2777}{18}
+19\,\ln \frac{\mu_r^2}{m_t^2}\right.
\nonumber\\[2ex]
&&\left.+ n_f\,\left (-\frac{67}{6}+\frac{8}{3}\,\ln \frac{\mu_r^2}{m_t^2}
\right ) \right ]\,.
\end{eqnarray}
Here $n_f$ denotes the number of light flavours and
$\alpha_s^{(5)}$ is presented in a five flavour number scheme.

In this paper we study the semi-inclusive reaction with one Higgs-boson $H$
in the final state. It will be denoted by
\begin{eqnarray}
\label{eqn2.7}
H_1(P_1)+H_2(P_2)\rightarrow H(-p_5) + 'X'\,,
\end{eqnarray}
where $H_1$ and $H_2$ denote the incoming hadrons and $X$ represents an
inclusive hadronic final state. In lowest order the partonic reactions
contributing to Eq. (\ref{eqn2.7}) are denoted by
\begin{eqnarray}
\label{eqn2.8}
&& a(p_1)+b(p_2)\rightarrow c(-p_3)+H(-p_5)\,, \qquad p_1+p_2+p_3+p_5=0\,,
\nonumber\\[2ex]
&& a,b,c=q,\bar q,g\,,
\end{eqnarray}
and the partonic cross sections in $n$-dimensions are given by
\begin{eqnarray}
\label{eqn2.9}
\sigma^{(1)}_{ab\rightarrow c~H}&=&K_{ab}\,G^2\,g^2\,\mu^{4-n}\,
\frac{1}{2s}\,\int \frac{d^np_3} {(2\pi)^{n-1}}\int \frac{d^np_5}{(2\pi)^{n-1}}
\,\delta^+(p_3^2)\, \delta^+( p_5^2-m^2)
\nonumber\\[2ex]
&&\times |M^{(1)}_{ab\rightarrow c~H}(\theta_1)|^2\,.
\end{eqnarray}
Here $M^{(1)}$ denotes the amplitude of the process and the strong
coupling constant is given by $g$ with $g^2=4\pi\alpha_s$. 
The scale $\mu$ originates
from the fact that the coupling constant acquires a dimension
in $n$ dimensions.  The quantity $K_{ab}$ represents the spin and colour 
average over the initial states including the
statistical factor $1/m!$ if one integrates over $m$ identical particles
in the final state. Further 
we have assumed that the Higgs boson is mainly produced on-mass-shell.
However one can also use the narrow width approximation. This can be achieved
by replacing
\begin{eqnarray}
\label{eqn2.10}
\delta(p_5^2-m^2) \quad \rightarrow \quad \frac{1}{\pi}\,
\frac{m\,\Gamma}{(p_5^2-m^2)^2+m^2\,\Gamma^2}\,,
\end{eqnarray}
where $\Gamma$ is the total width of the Higgs boson which is dominated by
the decay $H\rightarrow g + g$.
Choosing the C.M. frame of the incoming partons we have the following
parametrization
\begin{eqnarray}
\label{eqn2.11}
&&p_1=\frac{1}{2}\sqrt P_{12}\,(1,0,\cdots,0,1)\,,
\nonumber\\[2ex]
&&p_2=\frac{1}{2}\sqrt P_{12}\,(1,0,\cdots,0,-1)\,,
\nonumber\\[2ex]
&&-p_3=\frac{P_{12}-m^2}{2\sqrt P_{12}}\,(1,\cdots,-\sin \theta_1,
-\cos \theta_1)\,,
\nonumber\\[2ex]
&&-p_5=\frac{1}{2\sqrt P_{12}}\,(P_{12}+m^2,\cdots,(P_{12}-m^2)\sin \theta_1,
(P_{12}-m^2)\cos \theta_1)\,,
\nonumber\\
\end{eqnarray}
with
\begin{eqnarray}
\label{eqn2.12}
&&P_{ij}=(p_i+p_j)^2\,, \quad P_{12}=s\,, \quad P_{15}=t\,,\quad P_{25}=u\,, 
\quad s+t+u=m^2\,,
\nonumber\\[2ex]
&&\cos \theta_1=\frac{t-u}{s-m^2}\,.
\end{eqnarray}
{}From Eq. (\ref{eqn2.9}) we infer that
\begin{eqnarray}
\label{eqn2.13}
s^2 \frac{d^2~\sigma^{(1)}_{ab\rightarrow c~H}}{d~t~d~u}&=&
K_{ab}\,G^2\,\frac{\pi\,S_{\varepsilon}}{\Gamma(1+\varepsilon/2)}\,
\frac{\alpha_s(\mu_r^2)}{4\pi}\left (\frac{t~u}{\mu^2~s}\right )^{\varepsilon/2}
\,\delta(s+t+u-m^2)
\nonumber\\[2ex]
&&\times |M^{(1)}_{ab\rightarrow c~H}|^2\,, 
\quad \mbox{with}\quad n=4+\varepsilon\,.
\end{eqnarray}
The spherical factor $S_{\varepsilon}$ is defined by
\begin{eqnarray}
\label{eqn2.14}
S_{\varepsilon}=\exp\left (\frac{\varepsilon}{2}\Big (\gamma_E
-\ln 4\pi\Big )\right )\,.
\end{eqnarray}
On the Born level we have the following subprocesses
\begin{eqnarray}
\label{eqn2.15}
&&g + g \rightarrow g + H\,,
\\[2ex]
\label{eqn2.16}
|M^{(1)}_{gg\rightarrow g~H}|^2&=&N(N^2-1)\,\frac{1}{stu}\,\Bigg [\left (1
+\frac{\varepsilon}{2}\right )\Bigg \{s^4+t^4+u^4+m^8\Bigg\}
\nonumber\\[2ex]
&& +2\,\varepsilon \,m^2\,s\,t\,u\Bigg ]\,,
\\[2ex]
\label{eqn2.17}
K_{gg}&=&\frac{1}{4(1+\varepsilon/2)^2}\frac{1}{(N^2-1)^2}\,,
\\[2ex]
\label{eqn2.18}
&&q + \bar q \rightarrow g + H\,,
\\[2ex]
\label{eqn2.19}
|M^{(1)}_{q\bar q\rightarrow g~H}|^2&=&C_A\,C_F\,\frac{1}{s}\,
\Bigg [\left (1 +\frac{\varepsilon}{2}\right )\Bigg \{t^2+u^2\Bigg\}
+\varepsilon\, t\,u\Bigg ]\,,
\\[2ex]
\label{eqn2.20}
K_{q\bar q}&=&\frac{1}{4}\frac{1}{N^2}\,,
\\[2ex]
\label{eqn2.21}
&&q(\bar q) + g \rightarrow q(\bar q) + H\,,
\\[2ex]
\label{eqn2.22}
 |M^{(1)}_{qg\rightarrow q~H}|^2&=&C_A\,C_F\,\,\frac{1}{u}\,
\Bigg [-\left (1 +\frac{\varepsilon}{2}\right )\Bigg \{s^2+t^2\Bigg\}
-\varepsilon\, s\,t\Bigg ]\,,
\\[2ex]
\label{eqn2.23}
K_{qg}&=&\frac{1}{4(1+\varepsilon/2)}\frac{1}{N\,(N^2-1)}\,,
\end{eqnarray}
with
\begin{eqnarray}
\label{eqn2.24}
C_A=N\,, \qquad C_F=\frac{N^2-1}{2N}\,.
\end{eqnarray}
The $n$-dimensional matrix element $|M^{(1)}_{gg\rightarrow g~H}|^2$ is
proportional to $n-2$ provided $m=0$ so that it vanishes in two dimensions.
The same feature also appears in the lowest order matrix element for the decay
$H\rightarrow g + g$ derived from the effective Lagrangian in Eq. 
(\ref{eqn2.1}). Notice
that the factor $n-2=2(1+\varepsilon/2)$ also shows up in the spin average 
quantities $K_{ab}$.\\[3mm]

\mysection{One-loop corrections to the $2\rightarrow 2$-body\\ reactions}
The one-loop corrections to the gluon-gluon subprocess in Eq. (\ref{eqn2.15}) 
entails the computation of forty-two graphs. 
Fourteen graphs lead to independent expressions
whereas the remaining ones can be obtained via crossing from the $s$- to 
$t$-channel or from the $s$- to $u$-channel. 
The Feynman integrals show ultraviolet, infrared and collinear singularities
which will be regularized using $n$-dimensional regularization. Hence
the matrix element and the Feynman integrals have to be computed in $n$ 
dimensions for which we used the algebraic manipulation program FORM 
(version 3.0) \cite{form}. Since the Feynman integrals also contain
the integration momentum in the numerator we have to apply tensorial reduction.
Here we followed the procedure in \cite{pave}
which was extended to $n$-dimensions in \cite{been}. The scalar integrals
for the two-, three- and four-point functions can be found in \cite{mmn}.
The one-loop correction to the gluon-gluon differential distribution
(\ref{eqn2.15}) can be written as
\begin{eqnarray}
\label{eqn3.1}
s^2 \frac{d^2~{\hat \sigma}^{\rm VIRT}_{gg\rightarrow g~H}}{d~t~d~u}&
=&\pi\,\delta(s+t+u-m^2)\,
S_{\varepsilon}^2\,G^2\,\left (\frac{\alpha_s(\mu^2)}{4\pi}\right )^2\,
\frac{1}{(N^2-1)^2}
\nonumber\\[2ex]
&& \times\Bigg [N\,\Bigg \{-\frac{6}{\varepsilon^2}+\Big (2\ln\frac{s}{\mu^2}
-4\ln\frac{-t}{\mu^2}-4\ln\frac{-u}{\mu^2} +6\Big )\frac{1}{\varepsilon}
\nonumber\\[2ex]
&&+{\rm Li}_2\left(\frac{t}{m^2}\right )
+{\rm Li}_2\left(\frac{u}{m^2}\right )+{\rm Li}_2\left (\frac{s-m^2}{s}\right )
\nonumber\\[2ex]
&&-3\ln\frac{-t}{\mu^2}\ln\frac{-u}{\mu^2}
+\ln\frac{-t}{\mu^2}\ln\frac{s}{\mu^2}+\ln\frac{-u}{\mu^2}\ln\frac{s}{\mu^2}
-\ln^2\frac{-t}{\mu^2}
\nonumber\\[2ex]
&&-\ln^2\frac{-u}{\mu^2}
-\frac{1}{2}\ln^2\left (\frac{t-m^2}{t}\right )
-\frac{1}{2}\ln^2\left (\frac{u-m^2}{u}\right )
\nonumber\\[2ex]
&&+\frac{1}{2}\ln^2\left( \frac{m^2-t}{m^2}\right )
+\frac{1}{2}\ln^2 \left (\frac{m^2-u}{m^2}\right )
-2\ln\frac{s}{\mu^2}
\nonumber\\[2ex]
&&+4\ln\frac{-t}{\mu^2}+4\ln\frac{-u}{\mu^2}
+\frac{11}{2}\zeta(2)-\frac{9}{2} \Bigg \}\, |M^{(1)}_{gg\rightarrow g~H}|^2
\nonumber\\[2ex]
&& + \Big (N-n_f\Big )\Big \{\frac{1}{4} \Big \}
|MB^{(1)}_{gg\rightarrow g~H}|^2\Bigg ]\,,
\end{eqnarray}
\begin{eqnarray}
\label{eqn3.2}
|MB^{(1)}_{gg\rightarrow g~H}|^2=\frac{2}{3}\,N\,(N^2-1)\,\frac{m^2}{s\,t\,u}\,
\left [s\,t\,u+m^2\, \left (s\,t+s\,u+t\,u\right )\right ]\,.
\end{eqnarray}
Here the pole terms $1/\varepsilon^k$, $k=1,2$ (see Eq. (\ref{eqn2.13})) 
represent the three types of divergences mentioned above and
${\rm Li}_2(x)$ denotes the dilogarithmic function defined in 
\cite{lewin}. The virtual corrections to the quark-anti-quark reaction in Eq.
(\ref{eqn2.18}) require the computation of fourteen graphs. 
The procedure is the same as outlined above Eq. (\ref{eqn3.1}) and yields 
\begin{eqnarray}
\label{eqn3.3}
s^2 \frac{d^2~{\hat \sigma}^{\rm VIRT}_{q\bar q\rightarrow g~H}}{d~t~d~u}&
=&\pi\,\delta(s+t+u-m^2)\,
S_{\varepsilon}^2\,G^2\,\left (\frac{\alpha_s(\mu^2)}{4\pi}\right )^2\,
\nonumber\\[2ex]
&&\times \frac{1}{N^2}\,\Bigg [\Bigg \{n_f\,\Bigg (\frac{2}{3\varepsilon}
+\frac{1}{3}\ln \frac{-t}{\mu^2}+\frac{1}{3}\ln \frac{-u}{\mu^2}
-\frac{5}{9}\Bigg )
\nonumber\\[2ex]
&&+C_A\,\Bigg (-\frac{2}{\varepsilon^2}+\Big (2 \ln \frac{s}{\mu^2}
-2 \ln\frac{-t}{\mu^2}-2 \ln\frac{-u}{\mu^2}
-\frac{11}{3}\Big )\frac{1}{\varepsilon}
\nonumber\\[2ex]
&&+{\rm Li}_2\left (\frac{s-m^2}{s} \right )
-\ln \frac{-t}{\mu^2}\ln \frac{-u}{\mu^2}
+\ln \frac{-t}{\mu^2}\ln \frac{s}{\mu^2}
\nonumber\\[2ex]
&&+\ln \frac{-u}{\mu^2}\ln \frac{s}{\mu^2}-\ln^2 \frac{-t}{\mu^2}
-\ln^2 \frac{-u}{\mu^2} -\frac{11}{6}\ln \frac{-t}{\mu^2}
\nonumber\\[2ex]
&&-\frac{11}{6}\ln \frac{-u}{\mu^2}-\frac{7}{2}\zeta(2)+\frac{38}{9}\Bigg )
\nonumber\\[2ex]
&&+C_F\,\Bigg (-\frac{4}{\varepsilon^2}+\Big (-2\ln \frac{-t}{\mu^2}
-2\ln \frac{-u}{\mu^2}+3\Big )\frac{1}{\varepsilon}+
{\rm Li}_2 \left (\frac{t}{m^2}\right )
\nonumber\\[2ex]
&&+{\rm Li}_2\left (\frac{u}{m^2}\right)
 -2\ln \frac{-t}{\mu^2}\ln \frac{-u}{\mu^2}
-\frac{1}{2}\ln^2\left (\frac{t-m^2}{t}\right )
\nonumber\\[2ex]
&&-\frac{1}{2}\ln^2\left (\frac{u-m^2}{u}\right )+\frac{1}{2}\ln^2\left(
\frac{m^2-t}{m^2}\right )
+\frac{1}{2}\ln^2 \left (\frac{m^2-u}{m^2}\right )
\nonumber\\[2ex]
&&+\frac{3}{2}\ln\frac{-t}{\mu^2}
+\frac{3}{2}\ln\frac{-u}{\mu^2}+9\,\zeta(2)-4\Bigg )\Bigg \}
|M^{(1)}_{q\bar q\rightarrow g~H}|^2
\nonumber\\[2ex]
&&+\Big (C_A-C_F\Big )\,\Big \{\frac{1}{2}
\Big \} |MB^{(1)}_{q\bar q\rightarrow g~H}|^2\Bigg ]\,,
\end{eqnarray}
\begin{eqnarray}
\label{eqn3.4}
|MB^{(1)}_{q\bar q\rightarrow g~H}|^2=C_A\,C_F\,(-t-u)\,.
\end{eqnarray}
The square of the matrix element for the virtual corrections 
to the quark-gluon subprocess in Eq. (\ref{eqn2.21}) 
can be obtained from the one calculated for the
quark-anti-quark subprocess in Eq. (\ref{eqn2.18}) via crossing from 
the $s$-channel to the $u$-channel. This yields
\begin{eqnarray}
\label{eqn3.5}
s^2 \frac{d^2~{\hat \sigma}^{\rm VIRT}_{qg\rightarrow q~H}}{d~t~d~u}&
=&\pi\,\delta(s+t+u-m^2)\,
S_{\varepsilon}^2\,G^2\,\left (\frac{\alpha_s(\mu^2)}{4\pi}\right )^2\,
\frac{1}{N(N^2-1)}
\nonumber\\[2ex]
&& \times\Bigg [\Bigg \{n_f\,\Bigg (\frac{2}{3\varepsilon}
+\frac{1}{3}\ln \frac{-t}{\mu^2}
+\frac{2}{3}\ln \frac{-u}{\mu^2}-\frac{1}{3}\ln \frac{s}{\mu^2}
-\frac{8}{9}\Bigg )
\nonumber\\[2ex]
&&+C_A\,\Bigg (-\frac{2}{\varepsilon^2}+\Big (
-2 \ln\frac{t}{\mu^2}-\frac{8}{3}\Big )\frac{1}{\varepsilon}
+{\rm Li}_2\left (\frac{u}{m^2} \right )
\nonumber\\[2ex]
&&-\ln \frac{-t}{\mu^2}\ln \frac{-u}{\mu^2}+\ln \frac{-t}{\mu^2}
\ln \frac{s}{\mu^2}
-\ln \frac{-u}{\mu^2}\ln \frac{s}{\mu^2}-\ln^2 \frac{-t}{\mu^2}
\nonumber\\[2ex]
&&+\ln^2 \frac{-u}{\mu^2}
-\frac{1}{2}\ln^2\left (\frac{u-m^2}{u}\right )
+\frac{1}{2}\ln^2 \left (\frac{m^2-u}{m^2}\right )
\nonumber\\[2ex]
&&-\frac{5}{6}\ln \frac{-t}{\mu^2}
-\frac{11}{3}\ln \frac{-u}{\mu^2}+\frac{11}{6}\ln \frac{s}{\mu^2}
+\frac{9}{2}\zeta(2)+\frac{50}{9}\Bigg )
\nonumber\\[2ex]
&&+C_F\,\Bigg (-\frac{4}{\varepsilon^2}+\Big (-2\ln \frac{-t}{\mu^2}
-4\ln \frac{-u}{\mu^2}+2\ln \frac{s}{\mu^2}+5\Big )\frac{1}{\varepsilon}
\nonumber\\[2ex]
&&+{\rm Li}_2\left (\frac{s-m^2}{s} \right)
+{\rm Li}_2 \left (\frac{t}{m^2}\right )
-2\ln \frac{-t}{\mu^2}\ln \frac{-u}{\mu^2}
\nonumber\\[2ex]
&&+2\ln \frac{-u}{\mu^2}\ln \frac{s}{\mu^2}-2\ln^2 \frac{-u}{\mu^2}
-\frac{1}{2}\ln^2\left (\frac{t-m^2}{t}\right )
\nonumber\\[2ex]
&&+\frac{1}{2}\ln^2\left( \frac{m^2-t}{m^2}\right )
+\frac{5}{2}\ln\frac{-t}{\mu^2}
+5\ln\frac{-u}{\mu^2} -\frac{5}{2}\ln\frac{s}{\mu^2}
\nonumber\\[2ex]
&&+\zeta(2) -\frac{13}{2}\Bigg )\Bigg \} |M^{(1)}_{qg\rightarrow q~H}|^2
\nonumber\\[2ex]
&&+\Big (C_A-C_F\Big )\,\Big \{\frac{1}{2}\Big \}
|MB^{(1)}_{qg\rightarrow q~H}|^2 \Bigg ]\,,
\end{eqnarray}
\begin{eqnarray}
\label{eqn3.6}
|MB^{(1)}_{qg\rightarrow q~H}|^2=C_A\,C_F\,(s+t)\,.
\end{eqnarray}
Notice that the $MB^{(1)}$ in Eqs. (\ref{eqn3.2}), (\ref{eqn3.4}) and 
(\ref{eqn3.6}) are presented in four
dimensions and higher order terms in $\varepsilon$ are ignored. This is 
sufficient because the factors with which they are multiplied are finite
in the limit $\varepsilon \rightarrow 0$. Our results agree with those
presented in \cite{schmidt}. Notice that in the latter paper the Born 
amplitudes $M^{(1)}_{ab\rightarrow c~H}$ are presented in four dimensions 
because they are constructed using the helicity method. We have computed them
in $n$ dimensions. Apparently this does not affect the factors that multiply
the Born matrix elements and which contain all the pole terms in
$1/\varepsilon^k$.

\mysection{Gluon bremsstrahlung and other\\ $2\rightarrow 3$-body processes}
In this section we give an outline of the computation of the two-to-three
parton processes which show up in NLO. The calculation proceeds in
an analogous way as in the case for heavy flavour production presented
in \cite{bkns}. The processes under consideration will be denoted by
\begin{eqnarray}
\label{eqn4.1}
&&a(p_1)+b(p_2)\rightarrow c(-p_3)+d(-p_4)+H(-p_5)\,, \qquad 
\sum_{i=1}^5p_i=0\,,
\nonumber\\[2ex]
&& a,b,c,d=q,\bar q,g\,, \qquad p_i^2=0\,, \quad i=1-4\,, \quad p_5^2=m^2\,.
\end{eqnarray}
The cross section corresponding to the reaction above can be expressed
in $n$ dimensions as follows
\begin{eqnarray}
\label{eqn4.2}
{\hat \sigma}^{(2)}_{ab\rightarrow cd~H}&=&K_{ab}\,\frac{1}{2s}\,G^2\,g^4\,
\int \frac{d^np_3}
{(2\pi)^{n-1}} \int\frac{d^np_4}{(2\pi)^{n-1}}\int \frac{d^np_5}{(2\pi)^{n-1}}
\nonumber\\[2ex]
&& \delta^+(p_3^2)\,\delta^+(p_4^2)\,\delta^+(p_5^2-m^2)
(2\pi)^n\,\delta^{(n)}(p_1+p_2+p_3+p_4+p_5)\,
\nonumber\\[2ex]
&&\times |M^{(2)}_{ab\rightarrow cd~H}(\theta_1,\theta_2)|^2\,,
\end{eqnarray}
where $K_{ab}$ is defined below Eq. (\ref{eqn2.9}).
Choosing the C.M. frame of the outgoing partons 3 and 4 we get the following
parametrization in $n$ dimensions
\begin{eqnarray}
\label{eqn4.3}
&&p_1=\omega_1\,(1,0,\cdots,0,0,1)\,,
\nonumber\\[2ex]
&& p_2=(\omega_2,0,\cdots,0,|\vec p_5|\,\sin \psi,|\vec p_5|\,\cos \psi
-\omega_1)\,,
\nonumber\\[2ex]
&&-p_3=\omega_4\,(1,0,\cdots,\sin \theta_1\sin \theta_2,\sin \theta_1
\cos \theta_2,\cos \theta_1)\,,
\nonumber\\[2ex]
&&-p_4=\omega_4\,(1,0,\cdots,-\sin \theta_1\sin \theta_2,-\sin \theta_1
\cos \theta_2,-\cos \theta_1)\,,
\nonumber\\[2ex]
&&-p_5=(\omega_5,0,\cdots,0,|\vec p_5|\,\sin \psi,|\vec p_5|\,\cos \psi)\,,
\end{eqnarray}
where the energies $\omega_i$ are given by
\begin{eqnarray}
\label{eqn4.4}
&&\omega_1=\frac{P_{12}+P_{15}-m^2}{2\sqrt P_{34}}\,, \quad
\omega_2=\frac{P_{12}+P_{25}-m^2}{2\sqrt P_{34}}\,, \quad
\omega_3=\omega_4=\frac{1}{2}\sqrt P_{34}\,, \quad
\nonumber\\[2ex]
&& \omega_5=-\frac{P_{15}+P_{25}}{2\sqrt P_{34}}\,,
\qquad \cos \psi=\frac{P_{15}-m^2+2\,\omega_1\,\omega_5}{2\,\omega_1\,
|\vec p_5|}\,.
\end{eqnarray}
The invariants are denoted by
\begin{eqnarray}
\label{eqn4.5}
&& P_{ij}=(p_i+p_j)^2\,, \quad P_{12}=s\,, \quad P_{15}=t\,, \quad P_{25}=u\,,
\nonumber\\[2ex]
&&P_{34}=s_4=s+t+u-m^2\,.
\end{eqnarray}
From $|\cos \psi|\le 1$ and $s_4=P_{34}\ge 0$ we derive the boundary conditions
\begin{eqnarray}
\label{eqn4.6}
\frac{s\,m^2}{m^2-t}\le m^2-u \le s + t\,, \qquad m^2\le m^2-t\le s\,.
\end{eqnarray}
In the subsequent part of this paper it is more convenient to choose $s_4$ as
integration variable instead of $u$. In this case the integration boundaries are
\begin{eqnarray}
\label{eqn4.7}
0\le s_4 \le \frac{t(m^2-s-t)}{m^2-t}\,, \qquad m^2\le m^2-t\le s\,.
\end{eqnarray}
{}From the above kinematics the differential cross section 
in Eq. (\ref{eqn4.2}) becomes equal to
\begin{eqnarray}
\label{eqn4.8}
s^2 \frac{d^2~{\hat \sigma}^{(2)}_{ab\rightarrow cd~H}}{d~t~d~u}&=&\frac{1}{2}
K_{ab}\,\frac{S_{\varepsilon}^2}
{\Gamma(1+\varepsilon)}\,G^2\,\left (\frac{\alpha_s(\mu^2)}{4\pi}\right )^2
\left (\frac{t~u-m^2~s_4}{\mu^2~s}\right )^{\varepsilon/2}\,\left (
\frac{s_4}{\mu^2}\right)^{\varepsilon/2}
\nonumber\\[2ex]
&&\times \overline{|M^{(2)}_{ab\rightarrow cd~H}|^2}\,,
\end{eqnarray}
where $\overline{|M^{(2)}_{ab\rightarrow cd~H}|^2}$ is the second order
matrix element integrated over the polar angle $\theta_1$ and the
azimuthal angle $\theta_2$. It is given by
\label{eqn4.9}
\begin{eqnarray}
\overline{|M^{(2)}_{ab\rightarrow cd~H}|^2}&=&\int_0^{\pi}
d\theta_1\, (\sin \theta_1)^{1+\varepsilon}\int_0^{\pi}d\theta_2\,
(\sin \theta_2)^{\varepsilon}\,|M^{(2)}_{ab\rightarrow cd~H}
(\theta_1,\theta_2)|^2
\nonumber\\[2ex]
&\equiv& \int d\Omega_{n-1}
|M^{(2)}_{ab\rightarrow cd~H}(\theta_1,\theta_2)|^2\,.
\end{eqnarray}
As in the calculation of the
lowest order matrix elements in Eqs. (\ref{eqn2.16}), (\ref{eqn2.19})
(\ref{eqn2.22}) the next-to-leading order expressions  
$|M^{(2)}(\theta_1,\theta_2)|^2$ have to be computed in $n$ dimensions 
because we use $n$-dimensional regularization for the collinear and infrared
singularities which arise from the integration over the momenta of the
final state partons. 
We use the algebraic manipulation program FORM (version 3.0)
as in the earlier computation of the virtual corrections. Further we choose 
the Feynman gauge for the gluon propagators and the sum over the
physical polarizations of the external gluons is given by
\begin{eqnarray}
\label{eqn4.10}
\sum_{\alpha=L,R} \epsilon^{\mu}(p,\alpha)\,\epsilon^{\nu}(p,\alpha)
=P^{\mu\nu}(l,p)\,, \quad \mbox{with} \quad l_{\mu}\,P^{\mu\nu}=
P^{\mu\nu}\,l_{\nu}=0\,, \quad l^2=0\,,
\nonumber\\
\end{eqnarray}
with
\begin{eqnarray}
\label{eqn4.11}
P^{\mu\nu}=-g^{\mu\nu}+\frac{l^{\mu}\,p^{\nu}+l^{\nu}\,p^{\mu}}{l\cdot p}\,,
\end{eqnarray}
where $l$ is an arbitrary lightlike vector. There are three types of matrix 
elements depending on the number of gluons and (anti-) quarks appearing in
the initial and final state. They will be denoted by $ggggH$, $ggqqH$ and
$qqqqH$. Notice that the quark $q$ can also represent an anti-quark $\bar q$
depending on the type of process. The Feyman graphs can be found in \cite{kdr}.
The matrix element for $ggggH$ involves the computation of 
twenty-six graphs whereas eight graphs contribute to $ggqqH$. 
In the case of $qqqqH$ one
has to distinguish between identical and non-identical quarks. For identical
quarks we have two Feynman diagrams whereas the non-identical quarks are
represented by one graph only. In the case of $ggggH$ the amplitude is denoted
by $M^{(2)\mu\nu\kappa\rho}$ and the matrix element squared is obtained
by contraction over the Lorentz indices as follows
\begin{eqnarray}
\label{eqn4.12}
|M^{(2)}(\theta_1,\theta_2)|^2=P^{\mu\alpha}\,P^{\nu\beta}\,P^{\kappa\lambda}
\,P^{\rho\sigma}\,M^{(2)}_{\mu\nu\kappa\rho}\,
M^{(2)}_{\alpha\beta\lambda\sigma} \,.
\end{eqnarray}
We checked that the dependence on $l\cdot p_i$ disappears for arbitrary $l$. 
However this was only possible by writing the expression 
in Eq. (\ref{eqn4.12}) over a common denominator and expressing the result into
five independent kinematical invariants. It turns out that Eq. (\ref{eqn4.12})
has the same property as the square of the Born matrix 
element in Eq. (\ref{eqn2.16})
namely that it is proportional to $n-2$ provided $m=0$.
In the case of $ggqqH$ the amplitude is given by $M^{(2)}_{\mu\nu}$. 
Here we checked that the dependence
on $l\cdot p_i$ of the matrix element squared given by
\begin{eqnarray}
\label{eqn4.13}
|M^{(2)}(\theta_1,\theta_2)|^2=P^{\mu\alpha}\,P^{\nu\beta}
\,M^{(2)}_{\mu\nu}\,M^{(2)}_{\alpha\beta}\,,
\end{eqnarray}
already disappears after a simple partial fractioning. 
We made a comparison with the results in \cite{kdr} which are presented in four 
dimensions only because they are computed using the helicity method. 
After corrections for some misprints \footnote{We found additional misprints
in \cite{kdr}. The $S_{34}$ in the denominators of the two terms in (A.12) 
should read $S_{24}$. In (A.16) one has instead of
$n_{13}=-S_{24}~n_{12}(2\leftrightarrow 4)$
and $n_{23}=-S_{23}~n_{13}(3\leftrightarrow 4)$ the expressions
$n_{13}=-S_{24}~n_{12}(1\leftrightarrow 4)$ 
and $n_{23}=n_{13}(3\leftrightarrow 4)$.} the expressions obtained from
Eqs. (\ref{eqn4.12}) and (\ref{eqn4.13}) are in agreement with   
those presented in Appendix A of \cite{kdr}.
The procedure to bring the matrix element over a common denominator,
as discussed below Eq. (\ref{eqn4.12}), is not suitable for partial 
fractioning (see below) because it leads to high powers in the
angular dependent kinematical variables, which show up in the numerator and 
denominator. To avoid this complication
we proceed in a different way. Since the $l.p_i$ terms are
linear independent for $i=1-4$ they have to cancel among themselves.
This we checked by deleting them and comparing the truncated matrix element
squared with the one obtained by bringing 
all terms over a common denominator, which
is manifestly independent on $l.p_i$. The difference between these two
expressions turned out to be zero. Moreover the total power of the 
angular dependent kinematical invariants appearing in the numerator and 
denominator of the truncated matrix element does not exceed four so that
it becomes amenable for partial fractioning. After this procedure
the integration over the angles is performed in the following way. The 
expression $|M^{(2)}(\theta_1,\theta_2)|^2$ depends on the variables $P_{ij}$ 
defined in Eq. (\ref{eqn4.5}) among which five are linearly 
independent. Because of the parametrization in Eq. (\ref{eqn4.3}) the variables
$P_{12},P_{15},P_{25},P_{34}$ are independent of the angles $\theta_1,\theta_2$
whereas the remaining ones $P_{13},P_{23},P_{14},P_{24},P_{35}$ and $P_{45}$
are angular dependent. One can distinguish two types of integrals
\begin{eqnarray}
\label{eqn4.14}
J(P_{ij}^{-k},P_{m5}^{-l})&=&\int d\Omega_{n-1}\,P_{ij}^{-k}\,P_{m5}^{-l}\,,
\quad i,j=1-4 \quad m=3,4 \,,
\nonumber\\[2ex]
&& \mbox{with}\quad -4\le k\le 2\quad 0\le l \le 2\,,
\\[2ex]
\label{eqn4.15}
J(P_{ij}^{-k},P_{mn}^{-l})&=&\int d\Omega_{n-1}\,P_{ij}^{-k}\,P_{mn}^{-l}\,,
\quad i,j,m,n\not =5 \,,
\nonumber\\[2ex]
&& \mbox{with}\quad -3 \le k \le 2 \quad -3 \le l \le 2\,.
\end{eqnarray}
The basic integrals of the first class are given by
\begin{eqnarray}
\label{eqn4.16}
J(P_{13}^{-k},P_{45}^{-l})&=&2^k\,(P_{25}-P_{34})^{-k}\,I_{k,l}(A,B,C)\,,
\nonumber\\[2ex]
J(P_{13}^{-k},P_{35}^{-l})&=&2^k\,(P_{25}-P_{34})^{-k}\,I_{k,l}(A,-B,-C)\,,
\end{eqnarray}
with
\begin{eqnarray}
\label{eqn4.17}
I_{k,l}(A,B,C)=\int d\Omega_{n-1}(1-\cos \theta_1)^{-k}\,(A+B\cos \theta_1
+C\sin \theta_1\cos \theta_2)^{-l}\,.
\nonumber\\
\end{eqnarray}
The basic integrals of the second class are represented by
\begin{eqnarray}
\label{eqn4.18}
J(P_{13}^{-k},P_{24}^{-l})&=&2^{k+l}\,(P_{25}-P_{34})^{-k}\,(P_{15}-P_{34})^{-l}
\,I_{k,l}(\chi)\,,
\nonumber\\[2ex]
J(P_{13}^{-k},P_{23}^{-l})&=&2^{k+l}\,(P_{25}-P_{34})^{-k}\,(P_{15}-P_{34})^{-l}
\,I_{k,l}(\pi-\chi)\,,
\end{eqnarray}
with
\begin{eqnarray}
\label{eqn4.19}
I_{k,l}(\chi)&=&\int d\Omega_{n-1}(1-\cos \theta_1)^{-k}\,(1-\cos \theta_1
\cos \chi-\sin \theta_1\cos \theta_2\sin \chi)^{-l}
\nonumber\\[2ex]
&&= \pi\,2^{1-k-l}\,
\frac{\Gamma(n/2-1-k)\,\Gamma(n/2-1-l)\,\Gamma(n-3)}{
\Gamma^2(n/2-1)\,\Gamma(n-2-k-l)}
\nonumber\\[2ex]
&&\times F_{1,2} \left (k,l,\frac{n}{2}-1, \cos^2\frac{\chi}{2}\right )\,,
\end{eqnarray}
where $F_{1,2}(a,b,c;x)$ denotes the hypergeometric function which can be found
in \cite{abst}.
The other integrals
can be derived by interchanging the inclusive momenta $p_3$ and $p_4$
or interchanging the parametrization for $p_1$ and $p_2$.
The angular integrals above are presented in Appendix C of \cite{bkns} for
the case $-2\le k \le 2$. Because the square of the matrix
element for Higgs production contains two extra powers of $P_{ij}$ in the
numerator the previous computation has to be extended to 
cover the cases $k=-3,-4$. These extra powers can be attributed 
to the momentum dependence in the effective Higgs-gluon-gluon coupling
described by the Lagrangian in Eq. (\ref{eqn2.1}).
The colour decompositions of the integrated expressions 
read as follows
\begin{eqnarray}
\label{eqn4.20}
&&g + g \rightarrow g + g + H\,,
\\[2ex]
\label{eqn4.21}
\overline{|M^{(2)}_{gg\rightarrow gg~H}|^2}&=&N^2(N^2-1){\overline{
|M^{(2)}_{gg\rightarrow gg~H}|^2}}_N\,,
\\[2ex]
\label{eqn4.22}
&&g + g \rightarrow q_i + \bar q_i + H\,,
\\[2ex]
\label{eqn4.23}
\overline{|M^{(2)}_{gg\rightarrow q_i\bar q_i~H}|^2}
&=&n_f\,C_A\,C_F\,\left [C_A\,
{\overline{ |M^{(2)}_{gg \rightarrow q\bar q~H}|^2}}_A
+C_F\,{\overline{|M^{(2)}_{gg\rightarrow q\bar q~H}|^2}}_F\right ]\,,
\\[2ex]
\label{eqn4.24}
&&q + \bar q \rightarrow g + g + H\,,
\\[2ex]
\label{eqn4.25}
\overline{|M^{(2)}_{q\bar q\rightarrow gg~H}|^2}
&=&C_A\,C_F\,\left [C_A\,{\overline{
|M^{(2)}_{q\bar q\rightarrow gg~H}|^2}}_A
+C_F\,{\overline{|M^{(2)}_{q\bar q\rightarrow gg~H}|^2}}_F\right ]\,,
\\[2ex]
\label{eqn4.26}
&&q + g \rightarrow q + g + H\,,
\\[2ex]
\label{eqn4.27}
\overline{|M^{(2)}_{qg\rightarrow qg~H}|^2}
&=&C_A\,C_F\,\left [C_A\,{\overline{
|M^{(2)}_{qg\rightarrow qg~H}|^2}}_A
+C_F\,{\overline{|M^{(2)}_{qg\rightarrow qg~H}|^2}}_F\right ]\,.
\end{eqnarray}
The colour and spin average factors $K_{gg}$, $K_{q\bar q}$ and
$K_{qg}$ are given in Eqs. (\ref{eqn2.17}), (\ref{eqn2.20}) 
and (\ref{eqn2.23}) respectively. If two identical particles
in the processes above or below appear in an inclusive final state
a factor $1/2$ is implicitly understood. In NLO we encounter some new 
subprocesses. The first one is given
by the reaction in Eq. (\ref{eqn4.22}) where a sum over all light flavours
$q_i$ with $i=1\cdots n_f$ is understood. The next one is quark-quark 
scattering (non-identical and identical quarks) represented by
\begin{eqnarray}
\label{eqn4.28}
&&q_1 + q_2 \rightarrow q_1 + q_2 + H\,, \qquad q_1\not = q_2\,,
\\[2ex]
\label{eqn4.29}
\overline{|M^{(2)}_{q_1 q_2 \rightarrow q_1 q_2~H}|^2}&=&
C_A\,C_F\,{\overline{|M^{(2)}_{q_1q_2\rightarrow q_1q_2~H}|^2}}_A\,,
\\[2ex]
\label{eqn4.30}
&& q + q \rightarrow q + q + H\,,
\\[2ex]
\label{eqn4.31}
\overline{|M^{(2)}_{qq\rightarrow qq~H}|^2}&=&C_A\,C_F\,{\overline{
|M^{(2)}_{q_1q_2\rightarrow q_1q_2~H}|^2}}_A+C_F\,
{\overline{|M^{(2)}_{qq\rightarrow qq~H}|^2}}_F\,,
\end{eqnarray}
which have the colour and spin average factors
\begin{eqnarray}
\label{eqn4.32}
K_{q_1 q_2}=K_{qq}=\frac{1}{4}\frac{1}{N^2}\,.
\end{eqnarray}
The second subprocess is quark-anti-quark scattering
\begin{eqnarray}
\label{eqn4.33}
&&q_1 + \bar q_2 \rightarrow q_1 + \bar q_2 + H\,,\qquad q_1\not =q_2\,,
\\[2ex]
\label{eqn4.34}
\overline{|M^{(2)}_{q_1\bar q_2\rightarrow q_1\bar q_2~H}|^2}&=&C_A\,C_F\,
{\overline{ |M^{(2)}_{q_1q_2\rightarrow q_1q_2~H}|^2}}_A\,,
\\[2ex]
\label{eqn4.35}
&&q + \bar q \rightarrow q_i + \bar q_i + H\,,\qquad q_i\not =q \,,
\\[2ex]
\label{eqn4.36}
\overline{|M^{(2)}_{q\bar q\rightarrow q_i\bar q_i~H}|^2}&=&(n_f-1)\,C_A\,C_F\,
{\overline{ |M^{(2)}_{q\bar q\rightarrow q_a\bar q_a~H}|^2}}_A\,,\qquad
q_a\not =q\,,
\\[2ex]
\label{eqn4.37}
&&q + \bar q \rightarrow q + \bar q + H\,,
\\[2ex]
\label{eqn4.38}
\overline{|M^{(2)}_{q\bar q\rightarrow q\bar q~H}|^2}&=&C_A\,C_F\,\left (
{\overline{ |M^{(2)}_{q_1q_2\rightarrow q_1q_2~H}|^2}}_A
+{\overline{ |M^{(2)}_{q\bar q\rightarrow q_a\bar q_a~H}|^2}}_A\right )
\nonumber\\[2ex]
&& +C_F\,{\overline{|M^{(2)}_{q\bar q\rightarrow q\bar q~H}|^2}}_F\,,
\end{eqnarray}
and the colour and spin average factor reads
\begin{eqnarray}
\label{eqn4.39}
&& K_{q_1\bar q_2}=K_{q\bar q}=\frac{1}{4}\frac{1}{N^2}\,.
\end{eqnarray}
Further we have summed over all quark flavours $q_i$ in the final state 
provided they are not equal to the flavour of
the quark $q$ in the initial state (see Eq. (\ref{eqn4.35})). Notice that
the colour decomposition given above also holds before the squares of the 
matrix elements are integrated over the angles.
Some of the expressions in Eqs. (\ref{eqn4.20})-(\ref{eqn4.38})
are singular in the limit $s_4\rightarrow 0$, where $s_4$ is defined in
Eq. (\ref{eqn4.7}), which is due
to soft gluon radiation or soft (collinear) fermion pair production. 
This will lead to infrared singularities when the differential cross section
in Eq. (\ref{eqn4.8}) is convoluted with the parton densities even after
mass factorization is carried out. The convolution involves an integration
over $s_4$ so that one has to split up the partonic cross sections
into hard and soft gluonic parts as follows
\begin{eqnarray}
\label{eqn4.40}
s^2 \frac{d^2~{\hat \sigma}^{(2)}_{ab\rightarrow cd~H}}{d~t~d~u}
=s^2 \frac{d^2~{\hat \sigma}^{\rm HARD}_{ab\rightarrow cd~H}}{d~t~d~u}
(s_4>\Delta)+s^2 \frac{d^2~{\hat \sigma}^{\rm SOFT}_{ab\rightarrow cd~H}}
{d~t~d~u}(s_4\le\Delta)\,,
\nonumber\\
\end{eqnarray}
where $\Delta$ is chosen in such a way that it satisfies the inequalities
\begin{eqnarray}
\label{eqn4.41}
\Delta \ll s\,, \qquad \Delta \ll -t \,, \qquad \Delta \ll -u\,.
\end{eqnarray}
The soft gluon contribution to the differential cross section is defined by
\begin{eqnarray}
\label{eqn4.42}
s^2 \frac{d^2~{\hat \sigma}^{\rm SOFT}_{ab\rightarrow cd~H}}{d~t~d~u}&=&
\frac{1}{2} K_{ab}\,\frac{S_{\varepsilon}^2}
{\Gamma(1+\varepsilon)}\,G^2\,\left (\frac{\alpha_s(\mu^2)}{4\pi}\right )^2
\left (\frac{t~u}{\mu^2~s}\right )^{\varepsilon/2}
\nonumber\\[2ex]
&& \delta(s+t+u-m^2)\,\int_0^{\Delta}ds_4\,\left ( \frac{s_4}{\mu^2}\right)
^{\varepsilon/2} \overline{|M^{(2)}_{ab\rightarrow cd~H}|^2}\,.
\nonumber\\
\end{eqnarray}
Only the singular part of 
$\overline{|M^{(2)}_{ab\rightarrow cd~H}|^2}$, which behaves as $1/s_4$,
contributes to the above integral whereas the non-singular terms 
vanish in the limit $s_4\rightarrow 0$. This singular part is called
$\overline{|M^{\rm SOFT}_{ab\rightarrow cd~H}|^2}$ and behaves in two
different ways
\begin{itemize}
\item[I.]
\begin{eqnarray}
\label{eqn4.43}
\overline{|M^{\rm SOFT}_{ab\rightarrow cd~H}|^2}\sim \frac{1}{s_4}\,,
\end{eqnarray}
and
\item[II.]
\begin{eqnarray}
\label{eqn4.44}
\overline{|M^{\rm SOFT}_{ab\rightarrow cd~H}|^2}\sim \frac{1}{s_4}\,\left (
\frac{s~s_4}{t~u} \right )^{\varepsilon/2}\,.
\end{eqnarray}
\end{itemize}
The $d^2{\hat \sigma_{ab}}^{\rm SOFT}$ for the various subprocesses 
are presented in Appendix A. Adding the latter to the virtual 
contributions presented in section 3 leads to the soft-plus-virtual
differential cross sections defined by
\begin{eqnarray}
\label{eqn4.45}
s^2\,\frac{d^2{\hat \sigma_{ab}}^{\rm S+V}}{d~t~d~u}=
s^2\,\frac{d^2{\hat \sigma_{ab}}^{\rm SOFT}}{d~t~d~u}+
s^2\,\frac{d^2{\hat \sigma_{ab}}^{\rm VIRT}}{d~t~d~u}\,.
\end{eqnarray}
The infrared singularies cancel in this expression so that the double
pole terms $1/\varepsilon^2$, occuring in both the soft and the virtual
contributions, all vanish.

\mysection{Renormalization and mass factorization of the partonic 
differential cross sections}
The virtual cross sections in section 3 have two types of ultraviolet
divergences. One is removed by renormalization of the operator $O(x)$ 
occuring in Eq. (\ref{eqn2.1}) and the other by renormalization  
of the strong coupling constant. 
The renormalization constant corresponding to the former is
\begin{eqnarray}
\label{eqn5.1}
Z_O=\left (1+\frac{2}{\varepsilon}\beta(\alpha_s)\right )^{-1}
&=&1+\frac{\alpha_s}{4\pi}\,\frac{2}{\varepsilon}\,\beta_0+
\left (\frac{\alpha_s}{4\pi}\right )^2\left [\frac{4}{\varepsilon^2}\,\beta_0^2
+\frac{2}{\varepsilon}\,\beta_1\right ]+\cdots
\nonumber\\[2ex]
&=& 1+Z_O^{(1)}+Z_O^{(2)}+\cdots\,,
\end{eqnarray}
up to order $\alpha_s^2$ \cite{klzu},
where $\beta_0$ and $\beta_1$ are the lowest order coefficients of the 
beta-function given by
\begin{eqnarray}
\label{eqn5.2}
\beta(\alpha_s)&=&-\frac{\alpha_s}{4\pi}\,\beta_0
-\left (\frac{\alpha_s}{4\pi}\right )^2\,\beta_1+\cdots\,,
\nonumber\\[2ex]
\beta_0&=&\frac{11}{3}\,C_A-\frac{2}{3}\,n_f \,, \qquad 
\beta_1=\frac{34}{3}\,C_A^2- 2\,n_f\,C_F-\frac{10}{3}\,n_f\,C_A\,.
\end{eqnarray}
In lowest order $Z_O$ is the same as the renormalization constant for the
strong coupling constant which reads
\begin{eqnarray}
\label{eqn5.3}
Z_{\alpha_s}&=&1+\frac{\alpha_s}{4\pi}\,\frac{2}{\varepsilon}\,\beta_0+
\left (\frac{\alpha_s}{4\pi}\right )^2\left [\frac{4}{\varepsilon^2}\,\beta_0^2
+\frac{1}{\varepsilon}\,\beta_1\right ]+\cdots
\nonumber\\[2ex]
&=&1+Z_{\alpha_s}^{(1)}+Z_{\alpha_s}^{(2)}+\cdots\,.
\end{eqnarray}
Since the operator insertion always appears twice in the differential
cross sections the latter have to be multiplied by $Z_O^2$.

After renormalization one has still to perform mass factorization to
remove the remaining collinear divergences. This is achieved by the formula
\begin{eqnarray}
\label{eqn5.4}
s^2 \frac{d^2~{\hat \sigma}_{ab\rightarrow H + 'X'}}{d~t~d~u}(s,t,u,
\varepsilon)
&=&\int_0^1 \frac{dx_1}{x_1}\,\int_0^1 \frac{dx_2}{x_2}\,
\Gamma_{ea}(x_1,\varepsilon)\,\Gamma_{fb}(x_2,\varepsilon)
\nonumber\\[2ex]
&&\times {\hat s}^2\,\frac{d^2~\sigma_{ef\rightarrow H + 'X'}}
{d~\hat t~d~\hat u} (\hat s,\hat t,\hat u)
\nonumber\\[2ex]
&&\equiv \Gamma_{ea}\otimes\Gamma_{fb}\otimes {\hat s}^2\,
\frac{d^2~\sigma_{ef\rightarrow H + 'X'}}{d~\hat t~d~\hat u}\,,
\end{eqnarray}
where an implicit summation over $e,f$ is understood, 
with
\begin{eqnarray}
\label{eqn5.5}
\hat s = x_1\,x_2\,s\,, \qquad \hat t=x_1(t-m^2)+m^2\,,\qquad 
\hat u=x_2(u-m^2)+m^2\,.
\end{eqnarray}
In Eq. (\ref{eqn5.4}) $d^2\hat\sigma$ represents the singular cross section 
which still contains the collinear divergences indicated by $\varepsilon$.
These divergences are removed by the kernels $\Gamma_{ab}$ so that
the finite cross sections $d^2\sigma$ are left. 
Both therefore depend on the mass factorization scale $\mu$.
Since the mass factorization has to be carried out for the semi-inclusive
cross section up to NLO the kernels $\Gamma_{ab}$ have to be corrected
up to order $\alpha_s$ only. They are denoted by
\begin{eqnarray}
\label{eqn5.6}
\Gamma_{ab}(x,\varepsilon)=\delta_{ab}\delta(1-x)+\Gamma^{(1)}_{ab}
(x,\varepsilon)\,.
\end{eqnarray}
In section 4 the cross sections were split up into hard gluon ($H$) 
(Eq. (\ref{eqn4.40})) and soft-plus-virtual gluon parts ($S+V$) 
(Eq. (\ref{eqn4.45})). Therefore one proceeds with the kernels
in the same way so that the mass factorization can be carried out for
both parts separately
\begin{eqnarray}
\label{eqn5.7}
\Gamma^{(1)}_{ab}=\Gamma^{\rm HARD,(1)}_{ab}+\delta(1-x)\,
\Gamma^{\rm S+V,(1)}_{ab}\,.
\end{eqnarray}
The various kernels are expressed into the splitting functions $P_{ab}$, 
depending on the partons $a,b$, which represent the residues of the collinear
singularities
\begin{eqnarray}
\label{eqn5.8}
\Gamma^{(1)}_{qq}=\Gamma^{(1)}_{\bar q\bar q}=\frac{\alpha_s}{4\pi}\,
S_{\varepsilon}\,\frac{1}{\varepsilon}\,P^{(0)}_{qq}(x)\,,
\nonumber\\[2ex]
\Gamma^{(1)}_{gq}=\Gamma^{(1)}_{g\bar q}=\frac{\alpha_s}{4\pi}\,
S_{\varepsilon}\,\frac{1}{\varepsilon}\,P^{(0)}_{gq}(x)\,,
\nonumber\\[2ex]
\Gamma^{(1)}_{qg}=\Gamma^{(1)}_{\bar q g}=\frac{\alpha_s}{4\pi}\,
S_{\varepsilon}\,\frac{1}{2\varepsilon}\,P^{(0)}_{qg}(x)\,,
\nonumber\\[2ex]
\Gamma^{(1)}_{gg}=\frac{\alpha_s}{4\pi}\,S_{\varepsilon}\,
\frac{1}{\varepsilon}\,P^{(0)}_{gg}(x)\,.
\end{eqnarray}
The splitting functions are given by
\begin{eqnarray}
\label{eqn5.9}
P^{(0)}_{qq}(x)&=&4\,C_F\,\left [2\,\left (\frac{1}{1-x}\right)_+
-1-x+\frac{3}{2}\,\delta(1-x)\right ]\,,
\nonumber\\[2ex]
P^{(0)}_{gq}(x)&=&4\,C_F\,\left [\frac{1+(1-x)^2}{x}\right ]\,,
\nonumber\\[2ex]
P^{(0)}_{qg}(x)&=&8\,T_f\,\left [x^2+(1-x)^2\right ]\,,
\nonumber\\[2ex]
P^{(0)}_{gg}(x)&=&8\,C_A\,\left [\left (\frac{1}{1-x}\right)_+ +\frac{1}{x}
-2+x-x^2+\frac{11}{12}\,\delta(1-x)\right ]
\nonumber\\[2ex]
&& -\frac{4}{3}\,n_f\,\delta(1-x)\,.
\end{eqnarray}
The collinearly finite cross sections can be derived 
from Eq. (\ref{eqn5.4}) by using the above kernels. 
In LO they are equal to the Born results 
derived in section 2. In NLO the finite order $\alpha_s^2$ contributions
can be written as
\begin{eqnarray}
\label{eqn5.10}
&& g + g \rightarrow g + g + H\,,
\nonumber\\[2ex]
 s^2\,\frac{d^2~{\sigma}^{(2)}_{gg\rightarrow gg~H}}{d~t~d~u}&=&
s^2\,\frac{d^2~{\hat \sigma}^{(2)}_{gg\rightarrow gg~H}}{d~td~u}
\nonumber\\[2ex]
&&-\Gamma^{(1)}_{gg}\otimes {\hat s}^2\,
\frac{d^2~\sigma^{(1)}_{gg\rightarrow g~H}}{d~\hat t~d~\hat u}
-\Gamma^{(1)}_{gg}\otimes {\hat s}^2\,
\frac{d^2~\sigma^{(1)}_{gg\rightarrow g~H}}{d~\hat t~d~\hat u}\,,
\nonumber\\
\\[2ex]
\label{eqn5.11}
&& g + g \rightarrow q + \bar q + H\,,
\nonumber\\[2ex]
s^2\, \frac{d^2~{\sigma}^{(2)}_{gg\rightarrow q\bar q~H}}{d~t~d~u}&=&
s^2\, \frac{d^2~{\hat \sigma}^{(2)}_{gg\rightarrow q\bar q~H}}{d~t~d~u}
\nonumber\\[2ex]
&&-2\, \Gamma^{(1)}_{qg}\otimes {\hat s}^2\,
\frac{d^2~\sigma^{(1)}_{qg\rightarrow q~H}}{d~\hat t~d~\hat u}
-2\,\Gamma^{(1)}_{qg}\otimes {\hat s}^2\,
\frac{d^2~\sigma^{(1)}_{\bar qg\rightarrow \bar q~H}}{d~\hat t~d~\hat u}\,,
\nonumber\\
\\[2ex]
\label{eqn5.12}
&& q + \bar q \rightarrow g + g + H\,,
\nonumber\\[2ex]
s^2\, \frac{d^2~{\sigma}^{(2)}_{q\bar q\rightarrow gg~H}}{d~t~d~u}&=&
s^2\, \frac{d^2~{\hat \sigma}^{(2)}_{q\bar q\rightarrow gg~H}}{d~t~d~u}
\nonumber\\[2ex]
&&- \Gamma^{(1)}_{qq}\otimes {\hat s}^2\,
\frac{d^2~\sigma^{(1)}_{q\bar q\rightarrow g~H}}{d~\hat t~d~\hat u}
-\Gamma^{(1)}_{qq}\otimes {\hat s}^2\,
\frac{d^2~\sigma^{(1)}_{\bar q q\rightarrow g~H}}{d~\hat t~d~\hat u}\,,
\nonumber\\
\\[2ex]
\label{eqn5.13}
&& q + g \rightarrow q + g + H\,,
\nonumber\\[2ex]
s^2 \frac{d^2~{\sigma}^{(2)}_{qg\rightarrow qg~H}}{d~t~d~u}&=&
s^2 \frac{d^2~{\hat \sigma}^{(2)}_{qg\rightarrow qg~H}}{d~t~d~u}
\nonumber\\[2ex]
&&- \Gamma^{(1)}_{gg}\otimes {\hat s}^2\,
\frac{d^2~\sigma^{(1)}_{qg\rightarrow q~H}}{d~\hat t~d~\hat u}
-\Gamma^{(1)}_{gq}\otimes {\hat s}^2\,
\frac{d^2~\sigma^{(1)}_{gg \rightarrow g~H}}{d~\hat t~d~\hat u}
\nonumber\\[2ex]
&& - \Gamma^{(1)}_{qq}\otimes {\hat s}^2\,
\frac{d^2~\sigma^{(1)}_{qg\rightarrow q~H}}{d~\hat t~d~\hat u}
-\Gamma^{(1)}_{qg}\otimes {\hat s}^2\,
\frac{d^2~\sigma^{(1)}_{q \bar q\rightarrow g~H}}{d~\hat t~d~\hat u}\,,
\nonumber\\
\\[2ex]
\label{eqn5.14}
&&q_1 + q_2 \rightarrow q_1 + q_2 + H\,, \qquad q + q \rightarrow q + q + H\,,
\nonumber\\[2ex]
 s^2 \frac{d^2~{\sigma}^{(2)}_{qq \rightarrow qq~H}}{d~t~d~u}&=&
 s^2 \frac{d^2~{\hat \sigma}^{(2)}_{qq \rightarrow qq~H}}{d~t~d~u}
\nonumber\\[2ex]
&&- \Gamma^{(1)}_{gq}\otimes {\hat s}^2\,
\frac{d^2~\sigma^{(1)}_{qg\rightarrow q~H}}{d~\hat t~d~\hat u}
-\Gamma^{(1)}_{gq}\otimes {\hat s}^2\,
\frac{d^2~\sigma^{(1)}_{qg \rightarrow q~H}}{d~\hat t~d~\hat u}\,,
\nonumber\\
\\[2ex]
\label{eqn5.15}
&&q_1 + \bar q_2 \rightarrow q_1 + \bar q_2 + H\,,
\qquad q + \bar q \rightarrow q + \bar q + H\,,
\nonumber\\[2ex]
s^2 \frac{d^2~{\sigma}^{(2)}_{q \bar q \rightarrow q \bar q~H}}{d~t~d~u}&=&
s^2 \frac{d^2~{\hat \sigma}^{(2)}_{q \bar q \rightarrow q \bar q~H}}{d~t~d~u}
\nonumber\\[2ex]
&&- \Gamma^{(1)}_{gq}\otimes {\hat s}^2\,
\frac{d^2~\sigma^{(1)}_{qg\rightarrow q~H}}{d~\hat t~d~\hat u}
-\Gamma^{(1)}_{g\bar q}\otimes {\hat s}^2\,
\frac{d^2~\sigma^{(1)}_{\bar q g \rightarrow q~H}}{d~\hat t~d~\hat u}\,.
\nonumber\\
\end{eqnarray}
Notice that the finite parts of these cross sections 
can become negative due to the above subtractions.
The expressions for the finite hard gluon cross sections
$d^2~\sigma^{\rm HARD}_{ab\rightarrow cd~H}$ are too long to be published.
They exist as FORM \cite{form} files and they are available on request.
Here we can only show those parts which behave like $1/s_4$ since they are of
interest later on. In the soft gluon limit $s_4 \rightarrow 0$ they read
\begin{eqnarray}
\label{eqn5.16}
 \mathop{\mbox{lim}}\limits_{\vphantom{\frac{A}{A}} s_4 \rightarrow 0}
s^2 \frac{d^2~\sigma^{\rm HARD}_{gg\rightarrow gg~H}}{d~t~d~u}&=&\pi\,
G^2\,\left (\frac{\alpha_s(\mu^2)}{4\pi}\right )^2\,\frac{N}{(N^2-1)^2}\,
\frac{1}{s_4}\,\Bigg [3\ln \frac{s_4}{\mu^2}
\nonumber\\[2ex]
&&-\ln \frac{tu}{\mu^2 s}-\frac{11}{12}\Bigg ]\,|M^{(1)}_{gg\rightarrow g~H}|^2
\,,
\\[2ex]
\label{eqn5.17}
\mathop{\mbox{lim}}\limits_{\vphantom{\frac{A}{A}} s_4 \rightarrow 0}
s^2 \frac{d^2~\sigma^{\rm HARD}_{gg\rightarrow q \bar q~H}}{d~t~d~u}&=&
\pi\,G^2\,\left (\frac{\alpha_s(\mu^2)}{4\pi}\right )^2\,\frac{n_f}{(N^2-1)^2}\,
\frac{1}{s_4}\,\Bigg [\frac{1}{6} \Bigg ]
\nonumber\\[2ex]
&&\times |M^{(1)}_{gg\rightarrow g~H}|^2\,,
\\[2ex]
\label{eqn5.18}
 \mathop{\mbox{lim}}\limits_{\vphantom{\frac{A}{A}} s_4 \rightarrow 0}
s^2 \frac{d^2~\sigma^{\rm HARD}_{q \bar q\rightarrow gg~H}}{d~t~d~u}&=&
\pi\,G^2\,\left (\frac{\alpha_s(\mu^2)}{4\pi}\right )^2\,\frac{1}{N^2}\,
\frac{1}{s_4}\,\Bigg [C_A\,\Bigg \{-\ln \frac{s_4}{\mu^2}
\nonumber\\[2ex]
&&+\ln \frac{tu}{\mu^2 s}-\frac{11}{12}\Bigg \}+C_F\,\Bigg \{
4\ln \frac{s_4}{\mu^2}-2\ln \frac{tu}{\mu^2 s}\Bigg \} \Bigg ] 
\nonumber\\[2ex]
&&\times |M^{(1)}_{q\bar q\rightarrow g~H}|^2\,,
\\[2ex]
\label{eqn5.19}
 \mathop{\mbox{lim}}\limits_{\vphantom{\frac{A}{A}} s_4 \rightarrow 0}
s^2 \frac{d^2~\sigma^{\rm HARD}_{q g\rightarrow qg~H}}{d~t~d~u}&=&
\pi\,G^2\,\left (\frac{\alpha_s(\mu^2)}{4\pi}\right )^2\,\frac{1}{N(N^2-1)}\,
\frac{1}{s_4}\,\Bigg [C_A\,\Bigg \{2\ln \frac{s_4}{\mu^2}
\nonumber\\[2ex]
&&-\ln \frac{tu}{\mu^2 s}\Bigg \}+C_F\,\Bigg \{\ln \frac{s_4}{\mu^2}
-\frac{3}{4} \Bigg \} \Bigg ] \, |M^{(1)}_{qg\rightarrow q~H}|^2\,,
\\[2ex]
\label{eqn5.20}
 \mathop{\mbox{lim}}\limits_{\vphantom{\frac{A}{A}} s_4 \rightarrow 0}
s^2 \frac{d^2~\sigma^{\rm HARD}_{q \bar q\rightarrow q \bar q~H}}{d~t~d~u}&=&
\pi\,G^2\,\left (\frac{\alpha_s(\mu^2)}{4\pi}\right )^2\,\frac{n_f}{N^2}\,
\frac{1}{s_4}\,\Bigg [\frac{1}{6} \Bigg ]
\, |M^{(1)}_{q \bar q\rightarrow g~H}|^2\,.
\end{eqnarray}
Notice that the expressions above satisfy a supersymmetric relation.
It turns out that the expressions for $gg \rightarrow 'X' + H$ in
Eqs. (\ref{eqn5.16}) plus (\ref{eqn5.17}), $q\bar q \rightarrow 'X' + H$ in 
Eqs. (\ref{eqn5.18}) plus (\ref{eqn5.20}) and
$qg \rightarrow 'X' + H$ in Eq. (\ref{eqn5.19}) become equal for a
${\cal N}=1$ supersymmetry where $C_A=C_F=n_f=N$.

After renormalization of the the operator $O(x)$ and the 
coupling constant using $Z_O$ and $Z_{\alpha_s}$ in Eqs. (\ref{eqn5.1}) and 
(\ref{eqn5.3}) respectively the ultraviolet divergences vanish in the 
soft-plus-virtual cross section 
$d^2\,\sigma_{ab}^{\rm S+V}$ in Eq. (\ref{eqn4.45}). 
The remaining collinear divergences are
removed via mass factorization using $\Gamma_{ab}^{\rm S+V}$ in Eqs.
(\ref{eqn5.7})-(\ref{eqn5.9}). The combined effect of these three operations
leads to the following expressions 
\begin{eqnarray}
\label{eqn5.21}
s^2 \frac{d^2~\sigma^{\rm S+V}_{gg \rightarrow g~H}}{d~t~d~u}&=&
s^2 \frac{d^2~\hat{\sigma}^{\rm S+V}_{gg \rightarrow g~H}}{d~t~d~u}
\nonumber\\[2ex]
&& + \Bigg (Z_{\alpha_s}^{(1)}+2\,Z_{\rm O}^{(1)}-2\,\Gamma_{gg}^{\rm S+V,(1)}
\Bigg )\,
 s^2\, \frac{d^2~\sigma^{(1)}_{gg \rightarrow g~H}}{d~t~d~u}\,,
\\[2ex]
\label{eqn5.22}
s^2 \frac{d^2~\sigma^{\rm S+V}_{q\bar q\rightarrow g~H}}{d~t~d~u}&=&
s^2 \frac{d^2~\hat{\sigma}^{\rm S+V}_{q\bar q\rightarrow g~H}}{d~t~d~u}
\nonumber\\[2ex]
&&+ \Bigg (Z_{\alpha_s}^{(1)}+2\,Z_{\rm O}^{(1)}-2\,\Gamma_{qq}^{\rm S+V,(1)}
\Bigg )\, s^2\, \frac{d^2~\sigma^{(1)}_{q\bar q\rightarrow g~H}}{d~t~d~u}\,,
\\[2ex]
\label{eqn5.23}
s^2 \frac{d^2~\sigma^{\rm S+V}_{qg\rightarrow q~H}}{d~t~d~u}&=&
s^2 \frac{d^2~\hat{\sigma}^{\rm S+V}_{qg\rightarrow q~H}}{d~t~d~u}
\nonumber\\[2ex]
&&+ \Bigg (Z_{\alpha_s}^{(1)}+Z_{\rm O}^{(1)}\Bigg )\,
 s^2\, \frac{d^2~\sigma^{(1)}_{qg\rightarrow q~H}}{d~t~d~u}-\Gamma_{gg}^
{\rm S+V,(1)}\,s^2\,\frac{d^2~\sigma^{(1)}_{gg\rightarrow g~H}}{d~t~d~u}
\nonumber\\[2ex]
&&-\Gamma_{qq}^{\rm S+V,(1)}\,s^2\,\frac{d^2~\sigma^{(1)}_{q\bar q\rightarrow 
g~H}} {d~t~d~u}\,.
\end{eqnarray}
The results for the finite soft-plus-virtual differential cross sections read
\begin{eqnarray}
\label{eqn5.24}
s^2 \frac{d^2~\sigma^{\rm S+V}_{gg\rightarrow g~H}}{d~td~u}&=&
\pi\,\delta(s+t+u-m^2)\,G^2\,\left (\frac{\alpha_s(\mu^2)}{4\pi}\right )^2\,
\frac{1}{(N^2-1)^2}
\nonumber\\[2ex]
&&\times\Bigg [\Bigg \{n_f\,\Bigg (\frac{1}{6}\ln \frac{\Delta}{\mu^2}
-\frac{5}{18} \Bigg )
\nonumber\\[2ex]
&&+ N\,\Bigg (\frac{3}{2}\ln^2 \frac{\Delta}{\mu^2}+
\Big (-\ln\frac{tu}{\mu^2 s}-\frac{11}{12}\Big )\ln \frac{\Delta}{\mu^2}
+{\rm Li}_2\left(\frac{t}{m^2}\right )
\nonumber\\[2ex]
&&+{\rm Li}_2\left(\frac{u}{m^2}\right )
 +{\rm Li}_2\left (\frac{s-m^2}{s}\right )-\ln\frac{-t}{\mu^2}
\ln\frac{s}{\mu^2} -\ln\frac{-u}{\mu^2}\ln\frac{s}{\mu^2}
\nonumber\\[2ex]
&& +\frac{1}{2}\ln^2\frac{-t}{\mu^2}+\frac{1}{2}\ln^2\frac{-u}{\mu^2}
+\frac{1}{2}\ln^2\frac{s}{\mu^2}
-\frac{1}{2}\ln^2\left (\frac{t-m^2}{t}\right )
\nonumber\\[2ex]
&&-\frac{1}{2}\ln^2\left (\frac{u-m^2}{u}\right )+\frac{1}{2}\ln^2\left(
\frac{m^2-t}{m^2}\right )
+\frac{1}{2}\ln^2 \left (\frac{m^2-u}{m^2}\right )
\nonumber\\[2ex]
&&+3\zeta(2)+\frac{67}{36}\Bigg )\Bigg \}\, |M^{(1)}_{gg\rightarrow g~H}|^2
\nonumber\\[2ex]
&& + \Big (N-n_f \Big )\Big \{\frac{1}{4} \Big \}
|MB^{(1)}_{gg\rightarrow g~H}|^2\Bigg ]\,,
\end{eqnarray}
\begin{eqnarray}
\label{eqn5.25}
s^2 \frac{d^2~\sigma^{\rm S+V}_{q\bar q\rightarrow g~H}}{d~td~u}&=&
\pi\,\delta(s+t+u-m^2)\,G^2\,\left (\frac{\alpha_s(\mu^2)}{4\pi}\right )^2\,
\frac{1}{N^2}\,
\nonumber\\[2ex]
&&\times\Bigg [\Bigg \{n_f\,\Bigg (\frac{1}{6}\ln \frac{\Delta}{\mu^2}
+\frac{1}{3}\ln \frac{s}{\mu^2}-\frac{5}{6}\Bigg )
\nonumber\\[2ex]
&&+C_A\,\Bigg (-\frac{1}{2}\ln^2 \frac{\Delta}{\mu^2}
+\Big ( \ln \frac{tu}{\mu^2 s} -\frac{11}{12}\Big )\ln \frac{\Delta}{\mu^2}
+{\rm Li}_2\left (\frac{s-m^2}{s} \right )
\nonumber\\[2ex]
&&-\frac{1}{2}\ln^2 \frac{-t}{\mu^2}
-\frac{1}{2}\ln^2 \frac{-u}{\mu^2}
+\frac{1}{2}\ln^2 \frac{s}{\mu^2}-\frac{11}{6}\ln \frac{s}{\mu^2}
-5\zeta(2)+\frac{73}{12}\Bigg )
\nonumber\\[2ex]
&&+C_F\,\Bigg (2\ln^2 \frac{\Delta}{\mu^2}-2\ln \frac{tu}{\mu^2 s}
\ln \frac{\Delta}{\mu^2}+ {\rm Li}_2 \left (\frac{t}{m^2}\right )
+{\rm Li}_2\left (\frac{u}{m^2}\right)
\nonumber\\[2ex]
&&-\ln \frac{-t}{\mu^2}\ln \frac{s}{\mu^2}
-\ln \frac{-u}{\mu^2}\ln \frac{s}{\mu^2}+\ln^2 \frac{-t}{\mu^2}
+\ln^2 \frac{-u}{\mu^2}
\nonumber\\[2ex]
&&-\frac{1}{2}\ln^2\left (\frac{t-m^2}{t}\right )
-\frac{1}{2}\ln^2\left (\frac{u-m^2}{u}\right )
+\frac{1}{2}\ln^2\left( \frac{m^2-t}{m^2}\right )
\nonumber\\[2ex]
&&+\frac{1}{2}\ln^2 \left (\frac{m^2-u}{m^2}\right )
+\frac{3}{2}\ln \frac{s}{\mu^2}+8\,\zeta(2)-4\Bigg )\Bigg \}
|M^{(1)}_{q\bar q\rightarrow g~H}|^2
\nonumber\\[2ex]
&&+\Big (C_A-C_F \Big )\,\Big \{\frac{1}{2}
\Big \} |MB^{(1)}_{q\bar q\rightarrow g~H}|^2\Bigg ]\,,
\end{eqnarray}
\begin{eqnarray}
\label{eqn5.26}
s^2 \frac{d^2~\sigma^{\rm S+V}_{qg\rightarrow q~H}}{d~t~d~u}&=&
\pi\,\delta(s+t+u-m^2)\,G^2\,\left (\frac{\alpha_s(\mu^2)}{4\pi}\right )^2\,
\frac{1}{N(N^2-1)}
\nonumber\\[2ex]
&&\times \Bigg [\Bigg \{n_f\,\Bigg (\frac{1}{3}\ln \frac{-u}{\mu^2}
-\frac{5}{9}\Bigg )
\nonumber\\[2ex]
&&+C_A\,\Bigg (\ln^2\frac{\Delta}{\mu^2}-\ln \frac{tu}{\mu^2 s}
\ln \frac{\Delta}{\mu^2}
+{\rm Li}_2\left (\frac{u}{m^2} \right )
-\ln \frac{-u}{\mu^2}\ln \frac{s}{\mu^2} 
\nonumber\\[2ex]
&&+\ln^2 \frac{-u}{\mu^2}-\frac{1}{2}\ln^2\left (\frac{u-m^2}{u}\right )
+\frac{1}{2}\ln^2 \left (\frac{m^2-u}{m^2}\right )
-\frac{11}{6}\ln\frac{-u}{\mu^2} 
\nonumber\\[2ex]
&&+4\zeta(2)+\frac{38}{9}\Bigg )
\nonumber\\[2ex]
&&+C_F\,\Bigg (\frac{1}{2}\ln^2\frac{\Delta}{\mu^2}
-\frac{3}{4}\ln \frac{\Delta}{\mu^2}+{\rm Li}_2 \left (\frac{t}{m^2}\right )
+{\rm Li}_2\left (\frac{s-m^2}{s} \right)
\nonumber\\[2ex]
&&-\ln \frac{-t}{\mu^2}\ln \frac{s}{\mu^2}
+\frac{1}{2}\ln^2\frac{-t}{\mu^2}
-\frac{1}{2}\ln^2\frac{-u}{\mu^2}+\frac{1}{2}\ln^2\frac{s}{\mu^2}
\nonumber\\[2ex]
&&-\frac{1}{2}\ln^2\left (\frac{t-m^2}{t}\right )
+\frac{1}{2}\ln^2\left( \frac{m^2-t}{m^2}\right )
+\frac{3}{2}\ln\frac{-u}{\mu^2} 
\nonumber\\[2ex]
&&-\zeta(2) -\frac{9}{4}\Bigg )\Bigg \} |M^{(1)}_{qg\rightarrow q~H}|^2
\nonumber\\[2ex]
&&+\Big (C_A-C_F\Big )\,\Big \{\frac{1}{2}\Big \}
|MB^{(1)}_{qg\rightarrow q~H}|^2 \Bigg ]\,.
\end{eqnarray}
The expressions multiplying the Born matrix elements
$|M^{(1)}_{ab\rightarrow c~H}|^2$ satisfy a supersymmetry relation.
If we choose a ${\cal N}=1$ supersymmetry so that the quarks are put in the
adjoint representation i.e. $C_A=C_F=n_f=N$ then all expressions are equal
except for the rational constants which are equal to $19/12$ in 
Eq. (\ref{eqn5.24}), $15/12$ in Eq. (\ref{eqn5.25}) 
and $17/12$ in Eq. (\ref{eqn5.26}) respectively. 
These differences originate
from the fact that we use $n$-dimensional regularization rather than
$n$-dimensional reduction. The result for the latter case can be obtained
via a finite renormalization. It turns out that for $n$-dimensional reduction
the rational numbers become equal to $7/4$ for all three reactions which
provides us with a strong check on our calculations. 
Finally we want to comment on the scale $\mu$. 
In the computation of the radiative corrections we have
assumed that the renormalization scale $\mu_r$ is equal to the mass 
factorization scale $\mu$. If one wants to distinguish between both scales
one has to substitute 
\begin{eqnarray}
\label{eqn5.27}
\alpha_s(\mu^2)=\alpha_s(\mu_r^2)\left [1 + \frac{\alpha_s(\mu_r^2)}{4\pi}
\beta_0\,\ln \frac{\mu_r^2}{\mu^2}\right ]\,,
\end{eqnarray}
in all finite expressions. 

\mysection{Differential distributions for the process\\
 $p + p\rightarrow H +'X'$}
In this section we will present the differential cross sections for
Higgs-boson production in proton-proton collisions at the LHC. Here
we study the dependence of the cross sections on input
parameters like the QCD scale $\Lambda$, the renormalization/factorization
scale $\mu$ and the dependence on the chosen set of parton densities. 
We also study which region in the gluon density  
is the most important.  Furthermore
we make a comparison with similar results presented in previous papers.
Finally we also study at which values of the kinematical 
variables the soft-plus-virtual (S+V) gluons
start to dominate the cross sections. This will give an indication about
the validity of the S+V gluon approximation which is of
importance when one wants to improve the perturbation series via resummation
techniques leading to a better prediction for the differential cross sections.

The hadronic cross section $d\sigma$ is obtained from the partonic cross
section $d\sigma_{ab}$ as follows
\begin{eqnarray}
\label{eqn6.1}
S^2 \frac{d^2~\sigma^{{\rm H_1H_2}}}{d~T~d~U}(S,T,U,m^2)&=& \sum_{a,b=q,g}
\int_{x_{1,{\rm min}}}^1 \frac{dx_1}{x_1} \int_{x_{2,{\rm min}}}^1 
\frac{dx_2}{x_2}\,
f_a^{\rm H_1}(x_1,\mu^2)
\nonumber\\[2ex]
&&\times f_b^{\rm H_2}(x_2,\mu^2)\,s^2 
\frac{d^2~\sigma_{ab}}{d~t~d~u} (s,t,u,m^2,\mu^2)\,.
\end{eqnarray}
In analogy to Eq. (\ref{eqn2.12}) the hadronic kinematical variables are 
defined by
\begin{eqnarray}
\label{eqn6.2}
S=(P_1+P_2)^2 \,, \qquad T=(P_1+p_5)^2\,, \qquad U=(P_2+p_5)^2 \,,
\end{eqnarray}
where $P_1$ and $P_2$ denote the momenta of hadrons $H_1$ and $H_2$ 
respectively (see Eq. (\ref{eqn2.7})).
In the case parton $p_1$ emerges from hadron $H_1(P_1)$ and parton
$p_2$ emerges from hadron $H_2(P_2)$ we can establish the following relations
\begin{eqnarray}
\label{eqn6.3}
&& p_1=x_1\,P_1\,, \qquad p_2=x_2\,P_2 \,,
\nonumber\\[2ex]
&& s=x_1\,x_2\,S \,, \quad t=x_1(T-m^2)+m^2 \,, \quad u=x_2(U-m^2)+m^2\,,
\nonumber\\[2ex]
&& x_{1,{\rm min}}=\frac{-U}{S+T-m^2}\,, \qquad
x_{2,{\rm min}}=\frac{-x_1(T-m^2)-m^2}{x_1S+U-m^2}\,.
\end{eqnarray}
When the matrix element behaves as $1/s_4$, which occurs in the case
of gluon bremsstrahlung or collinear fermion pair emission, 
it is more convenient
to choose the integration variable $s_4$ instead of $x_2$ 
in order to get better numerical stability. The cross section becomes
\begin{eqnarray}
\label{eqn6.4}
S^2 \frac{d^2~\sigma^{{\rm H_1H_2}}}{d~T~d~U}(S,T,U,m^2)&=& \sum_{a,b=q,g}
\int_{x_{1,{\rm min}}}^1 \frac{dx_1}{x_1} \int_0^{s_{4,{\rm max}}}
\frac{ds_4}{s_4-x_1(T-m^2)-m^2}\,
\nonumber\\[2ex]
&&\times f_a^{\rm H_1}(x_1,\mu^2)\, f_b^{\rm H_2}(x_2^*(s_4),\mu^2)
\nonumber\\[2ex]
&&\times  s^2 \frac{d^2~\sigma_{ab}}{d~t~d~u} (s,t,s_4,m^2,\mu^2)\,.
\nonumber\\
\end{eqnarray}
The function $x_2^*(s_4)$ can be derived from
\begin{eqnarray}
\label{eqn6.5}
s_4=s+t+u-m^2=x_1\,x_2\,S+x_1(T-m^2)+x_2(U-m^2)+m^2\,,
\end{eqnarray}
and it reads
\begin{eqnarray}
\label{eqn6.6}
x_2^*(s_4)=\frac{s_4-x_1(T-m^2)-m^2}{x_1S+U-m^2} \,, \qquad
s_{4,{\rm max}}=x_1(S+T-m^2)+U\,.
\end{eqnarray}
When parton $p_1$ emerges from hadron $H_2(P_2)$ and parton $p_2$ emerges
from $H_1(P_1)$ one obtains the same expressions as in 
Eqs. (\ref{eqn6.3})-(\ref{eqn6.6}) except that $T$ and $U$ are interchanged.
This result has to be added to Eq. (\ref{eqn6.4}). When the partonic
cross section is symmetric under $t\leftrightarrow u$ one can also use the
representation in Eq. (\ref{eqn6.4}) without adding the result where
$T$ and $U$ are interchanged provided one makes the replacement
$f_a^{\rm H_1}f_b^{\rm H_2}\rightarrow f_a^{\rm H_1}f_b^{\rm H_2}+
f_a^{\rm H_2}f_b^{\rm H_1}$.

The expression in Eq. (\ref{eqn6.4}) simplifies when the partonic cross 
sections correspond to the Born reactions in Eq. (\ref{eqn2.13}) or 
the S+V contributions in Eqs. (\ref{eqn5.24})-(\ref{eqn5.26}). 
In this case $d^2\sigma \sim \delta(s_4)$ and the integral in 
Eq. (\ref{eqn6.4}) becomes one-dimensional. For two-to-three body 
processes with no soft gluons or collinear fermion pairs in the final
state the integral is two dimensional as given in Eq. (\ref{eqn6.4}). 
If the processes are of the gluon bremsstrahlung type, or contain 
collinear fermion pairs in the final state we have to split 
the two-to-three body partonic cross section 
in Eq. (\ref{eqn4.40}) 
into two parts i.e. a hard gluon part with 
$s_4>\Delta$ and a soft gluon part with $s_4\le \Delta$. The terms
containing the cut off parameter 
$\Delta$ cancel between the S+V cross sections originating
from Eqs. (\ref{eqn5.24})-(\ref{eqn5.26}) and the corresponding terms 
in the hard gluon hadronic cross section in Eq. (\ref{eqn6.4}), namely
\begin{eqnarray}
\label{eqn6.7}
&&\int_{x_{1,{\rm min}}}^1 \frac{dx_1}{x_1} \int_{\Delta}^{s_{4,{\rm max}}}
\frac{ds_4}{s_4-x_1(T-m^2)-m^2}\,
f_a^{\rm H_1}(x_1,\mu^2)\, f_b^{\rm H_2}(x_2^*(s_4),\mu^2)
\nonumber\\[2ex]
&&\times s^2 \frac{d^2~\sigma_{ab}^{\rm HARD}}{d~t~d~u} (s,t,s_4,m^2,\mu^2)\,.
\nonumber\\
\end{eqnarray}
In order to achieve the cancellation analytically we rewrite the expression
above as follows
\begin{eqnarray}
\label{eqn6.8}
&&\int_{x_{1,{\rm min}}}^1 \frac{dx_1}{x_1} 
\Bigg [\int_{\Delta}^{s_{4,{\rm max}}}
\frac{ds_4}{s_4-x_1(T-m^2)-m^2}\,
f_a^{\rm H_1}(x_1,\mu^2)\, f_b^{\rm H_2}(x_2^*(s_4),\mu^2)
\nonumber\\[2ex]
&&\times s^2 \frac{d^2~\sigma_{ab}^{\rm HARD}}{d~t~d~u} (s,t,s_4,m^2,\mu^2)
- \int_{\Delta}^{s_{4,{\rm max}}}\frac{ds_4}{-x_1(T-m^2)-m^2}\,
f_a^{\rm H_1}(x_1,\mu^2)
\nonumber\\[2ex]
&& \times f_b^{\rm H_2}(x_2^*(0),\mu^2)
\mathop{\mbox{lim}}\limits_{\vphantom{\frac{A}{A}} s_4 \rightarrow 0}
s^2 \frac{d^2~\sigma_{ab}^{\rm HARD}}{d~t~d~u} (s,t,s_4,m^2,\mu^2) \Bigg ]
\nonumber\\[2ex]
&&+\int_{x_{1,{\rm min}}}^1 \frac{dx_1}{x_1}
\int_{\Delta}^{s_{4,{\rm max}}}\frac{ds_4}{-x_1(T-m^2)-m^2}
 f_a^{\rm H_1}(x_1,\mu^2)\, f_b^{\rm H_2}(x_2^*(0),\mu^2)
\nonumber\\[2ex]
&&\times \mathop{\mbox{lim}}\limits_{\vphantom{\frac{A}{A}} s_4 \rightarrow 0}
s^2 \frac{d^2~\sigma_{ab}^{\rm HARD}}{d~t~d~u} (s,t,s_4,m^2,\mu^2)\,.
\end{eqnarray}
The expression between the square brackets is integrable in $s_4$ so we
can put $\Delta=0$ in it. One can perform the integration
over $s_4$ analytically in the second part by using the simple expressions
$d^2\sigma_{ab}^{\rm HARD}$ in the limit $s_4\rightarrow 0$. The 
$\ln^i\Delta/\mu^2$ terms, which arise from the integration
over $s_4$, then cancel those appearing in the S+V cross sections.

In practice one is not interested in the Higgs boson differential cross sections
depending on T and U but rather in the rapidity $y$ and the transverse
momentum $p_T$ distributions. Neglecting the masses of the incoming
hadrons we have the following relations
\begin{eqnarray}
\label{eqn6.9}
&&T=m^2-\sqrt S\,\sqrt{p_T^2+m^2}\,\cosh y+\sqrt S\,\sqrt{p_T^2+m^2}
\,\sinh y\,,
\nonumber\\[2ex]
&&U=m^2-\sqrt S\,\sqrt{p_T^2+m^2}\,\cosh y-\sqrt S\,\sqrt{p_T^2+m^2}
\,\sinh y\,,
\end{eqnarray}
so that the cross section becomes
\begin{eqnarray}
\label{eqn6.10}
S \frac{d^2~\sigma^{{\rm H_1H_2}}}{d~p_T^2~d~y}(S,p_T^2,y,m^2)=
S^2 \frac{d^2~\sigma^{{\rm H_1H_2}}}{d~T~d~U}(S,T,U,m^2)\,.
\end{eqnarray}
The kinematical boundaries are
\begin{eqnarray}
\label{eqn6.11}
m^2-S\le T \le 0 \,, \qquad -S-T+m^2\le U \le \frac{S~m^2}{T-m^2}+m^2\,,
\end{eqnarray}
from which one can derive
\begin{eqnarray}
\label{eqn6.12}
&& 0\le p_T^2 \le p^2_{T,{\rm max}}\,, \qquad
- \frac{1}{2}\ln \frac{S}{m^2}\le y \le \frac{1}{2}\ln \frac{S}{m^2}\,,
\nonumber\\[2ex]
&&\mbox{with} \quad p^2_{T,{\rm max}}=\frac{(S+m^2)^2}{4~S~\cosh^2 y}-m^2\,,
\end{eqnarray}
or
\begin{eqnarray}
\label{eqn6.13}
&&- y_{{\rm max}}\le y \le y_{{\rm max}}\,, 
\qquad 0\le p_T^2 \le \frac{(S-m^2)^2}{4~S}\equiv {\bar p}^2_{T,{\rm max}} \,,
\nonumber\\[2ex]
&&\mbox{with} \quad y_{{\rm max}}=
\frac{1}{2}\ln\frac{1+\sqrt{1-sq}}{1-\sqrt{1-sq}}\,,
\qquad sq=\frac{4~S~(p_T^2+m^2)}{(S+m^2)^2}\,.
\end{eqnarray}
Since the cross section diverges for $p_T\rightarrow 0$ we cannot perform
the integral over this kinematical variable down to zero. However we can
perform the integral over the rapidity and obtain the transverse momentum
distribution
\begin{eqnarray}
\label{eqn6.14}
\frac{d~\sigma^{{\rm H_1H_2}}}{d~p_T}(S,p_T^2,m^2)=
\int_{-y_{{\rm max}}}^{y_{{\rm max}}} dy
\,\frac{d^2~\sigma^{{\rm H_1H_2}}}{d~p_T~d~y}
(S,p_T^2,y,m^2)\,,
\end{eqnarray}
with $y_{{\rm max}}$ given in Eq. (\ref{eqn6.13}).
There is an alternative way to obtain the distribution above. This is shown  
in Appendix B. We checked that both procedures lead to the same 
numerical result.
In the case we plot the rapidity distribution we have to impose a
cut $p_{T,{\rm min}}$ on the transverse momentum integration i.e.
\begin{eqnarray}
\label{eqn6.15}
\frac{d~\sigma^{{\rm H_1H_2}}}{d~y}(S,y,m^2)=\int_{p_{T,{\rm min}}}
^{p_{T,{\rm max}}} dp_T\,\frac{d^2~\sigma^{{\rm H_1H_2}}}{d~p_T~d~y}
(S,p_T^2,y,m^2)\,,
\end{eqnarray}
with $p_{T,{\rm max}}$ given in Eq. (\ref{eqn6.12}).
Actually the differential cross section dies off so fast that
it is sufficient to put $p_{T,{\rm max}} = 8 \times p_{T,{\rm min}}$.
Finally we define what we mean by leading order (LO) and next-to-leading 
order (NLO). In LO the differential cross section is defined by
\begin{eqnarray}
\label{eqn6.16}
\frac{d^2~\sigma^{\rm LO}}{d~p_T~d~y}(S,p_T^2,y,m^2)=
\frac{d^2~\sigma^{(1)}}{d~p_T~d~y}(S,p_T^2,y,m^2)\,,
\end{eqnarray}
and the gluon-gluon-Higgs coupling is given by $G$ in Eq. (\ref{eqn2.2}) with
${\cal C}=1$ (see Eq. (\ref{eqn2.6})). We also adopt the leading logarithmic
representation for the running coupling and the parton densities. 
The NLO corrected differential cross section reads
\begin{eqnarray}
\label{eqn6.17}
\frac{d^2~\sigma^{\rm NLO}}{d~p_T~d~y}(S,p_T^2,y,m^2)&=&
\left [1+22\,\left (\frac{\alpha_s^{(5)}(\mu^2)}{4\pi}\right )\right ]\,
\frac{d^2~\sigma^{(1)}}{d~p_T~d~y}(S,p_T^2,y,m^2)
\nonumber\\[2ex]
&&+ \frac{d^2~\sigma^{(2)}}{d~p_T~d~y}(S,p_T^2,y,m^2)\,,
\end{eqnarray}
where the LO contribution in this formula is now multiplied with 
${\cal C}^2=1+22~\alpha_s/4\pi$. Furthermore 
the running coupling and parton densities are 
represented in next-to-leading order for which we have chosen
the $\overline{\rm MS}$-scheme.
In this way one obtains a result which is consistently corrected up to 
NLO.
\begin{table}
\begin{center}
\begin{tabular}{|c|c|c|}\hline
MRST98 (LO, lo05a.dat) & $\Lambda_5^{\rm LO}=130.5~{\rm MeV}$  & 
$\alpha_s^{\rm LO}(M_Z)=0.125$      \\
MRST98 (NLO, ft08a.dat)    & $\Lambda_5^{\rm NLO}=220~{\rm MeV}$ & 
$\alpha_s^{\rm NLO}(M_Z)=0.1175$       \\
CTEQ4 (LO, cteq4l.tbl)     & $\Lambda_5^{\rm LO}=181~{\rm MeV}$   & 
$\alpha_s^{\rm LO}(M_Z)=0.132$       \\
CTEQ4 (NLO, cteq4m.tbl)    & $\Lambda_5^{\rm NLO}=202~{\rm MeV}$  & 
$\alpha_s^{\rm NLO}(M_Z)=0.116$    \\
GRV98 (LO, grv98lo.grid) & $\Lambda_5^{\rm LO}=131~{\rm MeV}$   & 
$\alpha_s^{\rm LO}(M_Z)=0.125$   \\
GRV98 (NLO, grvnlm.grid) & $\Lambda_5^{\rm NLO}=173~{\rm MeV}$  & 
$\alpha_s^{\rm NLO}(M_Z)=0.114$       \\
MRST99 (NLO, cor01.dat) & $\Lambda_5^{\rm NLO}=220~{\rm MeV}$ & 
$\alpha_s^{\rm NLO}(M_Z)=0.1175$ \\
\hline
\end{tabular}
\end{center}
\caption{Various parton density sets with the values for the QCD scale
$\Lambda$ and the running coupling $\alpha_s$.}
\label{table1}
\end{table}
In our computations the number of light flavours is taken 
to be $n_f=5$ which holds for the running coupling,
the partonic cross sections and the number of quark flavour densities. 
Further we have chosen for our plots the parton densities obtained
from the sets MRST98 \cite{mrst98} CTEQ4 \cite{lai}, GRV98 \cite{grv}
and MRST99 \cite{mrst99} (see Table \ref{table1}). 
Notice that the GRV sets do not contain charm and bottom quark densities.
For simplicity the factorization scale $\mu$ is set equal to the 
renormalization scale $\mu_r$. For our plots we take 
$\mu^2=m^2+p_T^2$
unless mentioned otherwise. Here we want to emphasize that the 
magnitudes of the cross sections are extremely sensitive to the 
choice of the renormalization scale because the effective coupling 
constant $G\sim \alpha_s(\mu_r)$,
which implies that $d\sigma^{\rm LO}\sim \alpha_s^3$ and
$d\sigma^{\rm NLO}\sim \alpha_s^4$. However the slopes 
of the differential distributions are less sensitive to the 
scale choice if they are only plotted over a limited range.

For the computation of the Higgs-gluon-gluon effective coupling constant
$G$ given in Eq. (\ref{eqn2.2}) we choose the top-quark mass 
$m_t=173.4~{\rm GeV/c^2}$ and the Fermi constant 
$G_F=1.16639~{\rm GeV}^{-2}=4541.68~{\rm pb}$.
In this paper we will only study Higgs boson production in proton-proton 
collisions at the center of mass energy $\sqrt S=14~{\rm TeV}$ characteristic
of the LHC. Since the hadrons $H_1$ and $H_2$ are now identical
the $y$ differential cross sections are symmetric.

The effect of the NLO corrections to Higgs-boson production were already 
studied earlier by the authors in \cite{fgk}. 
In order to compare with their results we present LO and NLO differential
cross sections in $p_T$, integrated over $y$, for $m=120$ GeV$/c^2$ and
$\mu^2= m^2 + p_T^2$ in Figs. 1a and 1b respectively. The MRST98 parton 
densities \cite{mrst98} were used for these plots. We note that the NLO
results from the $q(\bar q) g$ and $qq$ channels are negative at small
$p_T$ so we have plotted their absolute values multiplied by 100. 
It is clear that the 
$gg$ subprocess dominates but the $q(\bar q)g$-subprocess is also important.
The corresponding results for the $y$ distributions integrated over
the $p_T$ region between $p_{T,{\rm min}}=$ 30 GeV$/c$ and 
$p_{T,{\rm max}}=240~{\rm GeV/c}$ are given in Figs. 2a and 2b
respectively. The latter value was chosen because the cross 
section above 240 ${\rm GeV/c}$ is extremely small and can be neglected. 
Here the scale is $\mu^2= m^2 + p_{T,{\rm min}}^2$. 
Again we see that the $gg$-subprocess dominates. 
Using the recent MRST99 set \cite{mrst99} we present 
the dependence of the NLO $p_T$- and $y$-distributions on the Higgs mass 
in Figs. 3 and 4 respectively.

Next we plot the scale dependence of the above differential cross sections
in Figs. 5 and 6. For the $p_T$-distributions we have chosen the scale factors 
$\mu=2\mu_0$, $\mu=\mu_0$ and $\mu=\mu_0/2$ with $\mu_0^2= m^2 + p_T^2$.
In the case of the $y$-distributions we adopted 
$\mu_0^2= m^2 + p_{T,{\rm min}}^2$ where 
$p_{T,{\rm min}}= 30~{\rm GeV/c}$
(see Eq. (\ref{eqn6.15})). Furthermore we have again adopted the MRST98 parton
density set in \cite{mrst98} since it contains both LO and NLO versions. 
In the case of the $p_T$-distribution in Fig. 5a one observes a small 
reduction in the scale dependence while going from LO to NLO. This
reduction becomes more visible when we plot the quantity
\begin{eqnarray}
\label{eqn6.18}
N\left (p_T,\frac{\mu}{\mu_0}\right )
=\frac{d\sigma(p_T,\mu)/dp_T}{d\sigma(p_T,\mu_0)/dp_T} 
\end{eqnarray}
in the range $0.1 < \mu/\mu_0 < 10$ at fixed values of $p_T=$ 30, 70 and 
~{\rm 100 GeV/c}.
The upper set of curves at small $\mu/\mu_0$ are for LO and the
lower set are for NLO. Notice that the NLO plots at 70 and 100 are 
extremely close to each other and it is hard to distinguish between them.
Further one sees that the slopes of the LO curves are
larger that the slopes of the NLO curves. This is an indication that
there is better stability in NLO, which was expected. 
However there is no sign of a flattening or an optimum in either of these
curves. This implies that one will have to calculate the
differential cross sections in NNLO to find a better stability under
scale variations.
In the case of the $y$-distributions we adopted the scale
$\mu_0^2= m^2 + p_{T,{\rm min}}^2$ where $p_{T,{\rm min}}= 30~{\rm GeV/c}$
(see Eq. (\ref{eqn6.15})). As shown in Fig.6 there is 
hardly any reduction in the scale dependence for the $y$-distributions 
between the LO and NLO curves.

Besides the dependence on the factorization and renormalization scales there
are two other uncertainties which affect the predictive power of the
theoretical cross sections. The first one concerns the rate of convergence
of the perturbation series which is indicated by the $K$-factor defined by
\begin{eqnarray}
\label{eqn6.19}
K=\frac{d~\sigma^{\rm NLO}}{d~\sigma^{\rm LO}} \,. 
\end{eqnarray}
Another uncertainty is the dependence of the cross section on the specific
choice of parton densities, which can be expressed by the factors
\begin{eqnarray}
\label{eqn6.20}
R^{\rm CTEQ}=\frac{d~\sigma^{\rm CTEQ}}{d~\sigma^{\rm MRST}}\,,
\quad \mbox \quad R^{\rm GRV}=\frac{d~\sigma^{\rm GRV}}{d~\sigma^{\rm MRST}}\,.
\end{eqnarray}
The quantities in Eqs. (\ref{eqn6.19}) and (\ref{eqn6.20}) are plotted for
the $p_T$ distributions in Figs. 7a and 7b respectively 
where we have chosen the same parameters and scales ($\mu=\mu_0$) as in 
Figs. 5a,b and 6. Depending on the parton density set the $K$-factors
shown in Fig. 7a are pretty large and vary from 1.4 at $p_T=30~{\rm GeV/c}$ 
to 1.7 at $p_T=150~{\rm GeV/c}$. Here the CTEQ4 parton densities
lead to the smallest $K$-factor whereas the MRST98 set provides us
with the largest one with the results from the GRV98 set
in between.
In Fig. 7b we show the dependences of the ratios defined in Eq. (6.20) 
as a function of $p_T$. From this figure we infer that both the GRV98 
and the CTEQ4 densities lead to larger cross sections than those 
computed from MRST98. 
The difference between the results obtained from the 
latter set with respect to the
other ones is smaller in NLO than in LO. Furthermore In LO there is 
a small decrease in the $R$ values as a function of $p_T$ which is in 
contrast to NLO where we observe an increase.
In the case of the $y$-distributions in Fig. 8a
again the CTEQ4 parton densities yield the smallest $K$-factor which
does not vary much as a function of $y$. The latter also holds for
the GRV98 set where however the $K$-factor is larger than the one
obtained from CTEQ4. The largest $K$-factor is obtained from the 
the MRST98 densities which show a much stronger dependence on the rapidity
$y$ than the ones obtained from the other sets. It becomes maximal
at $y=0$ and decreases at larger absolute values of the rapidity.
In Fig. 8b we show the ratio $R$ as a function of the rapidity.
Like in Fig. 7b the difference between the cross sections obtained
from the MRST set and the two other sets becomes smaller while
going from LO to NLO. At $y=0$ the discrepancy between the cross sections
attains a maximum in the case of LO whereas it reaches a minimum for NLO.
Notice that our results for the $K$-factors computed in NLO in 
Figs. 7a, 8a agree with those shown in the corresponding Figs. 2a, 2b in 
\cite{fgk}. The same also holds for the NLO $R$-factors in Figs. 7b, 8b. 
when compared with the same figures in \cite{fgk}.
The authors of \cite{fgk} informed us that they did not
adopt a fixed scale at 
$\mu_0^2= p_{T,{\rm min}}^2 + m^2$ but they used a variable scale 
$\mu_0^2= p_{T}^2 + m^2$  when integrating over $p_T$ to calculate the 
rapidity $y$-distributions. We have also run our programs with the 
latter scale, which yields different rapidity-distributions. The 
rapidity cross sections become smaller due to the smaller running coupling 
constant but the $R$-ratios hardly change w.r.t. the ones shown in Fig. 8b.
Therefore the agreement between our results and those obtained in 
\cite{fgk} are not spoiled. We also made a comparison between our results
obtained for the differential cross section $d~\sigma/d~p_T$
and those shown in Fig. 1a of \cite{fgk}. To compare we have read off 
the central values of the bins from their Fig. 1a and, after changing 
bin sizes, have replotted their values versus ours in Fig. 9 for the 
MRST98 set and the same input parameters. We only show the 
case where the scale is $\mu=\mu_0$. Fig. 9 shows a small difference
in LO and NLO between our cross sections (indicated in Fig. 9 by RSN(1)) 
and the those given by \cite{fgk} (indicated by FGK). However the slopes 
are exactly the same. The difference might be due to a different choice
of the effective Higgs-gluon-gluon coupling constant in Eq. (\ref{eqn2.2}).
We have chosen $m_t=173.4~{\rm GeV/c^2}$ whereas the authors in \cite{fgk} 
took the limit $m_t\rightarrow \infty$ (see Eq. (\ref{eqn2.5})).
If we take the value $m_t=10^4~{\rm GeV/c^2}$, which is rather close
to an infinite top quark mass, our curves (indicated in Fig. 9 by RSN(2)) 
approach the ones given in Fig. 1a of \cite{fgk}. For further checks
we also made a comparison with 
the LO $p_T$-distribution in Fig. 6 of \cite{ehsb} by reading
off their values. Choosing the same input parameters and parton density 
set we completely obtain the same results as in \cite{ehsb}.
Finally we checked several of the figures in \cite{glosser}. In the
latter reference only the NLO corrections to the $gg$-subprocess
have been completed. Using the same parton densities and parameters 
we found excellent agreement between our results for this
subprocess and those obtained in \cite{glosser}.

We now analyse why the MRST98, GRV98 and CTEQ4 parton densities yield 
different results for the differential cross sections.
This can be mainly attributed to the small $x$-behaviour of the
gluon density because gluon-gluon fusion is the dominant
production mechanism.
To investigate the small $x$-behaviour we consider the product of
the gluon flux with the corresponding partonic cross section
\begin{eqnarray}
\label{eqn6.21}
F_{gg}(xS,p_T^2,\mu^2)=\Phi_{gg}(x,\mu^2)\,
\frac{d~\sigma_{gg}}{d~p_T}(xS,p_T^2,\mu^2)
\end{eqnarray}
In this case we have removed the factor $G^2~\alpha_s/4\pi$ from the LO 
(Born) cross section and the factor $G^2~(\alpha_s/4\pi)^2$ from the NLO
contribution to the partonic cross section. This was done to suppress the
dependence on the strong coupling constant $\alpha_s$ so that the
differences between the several $F_{gg}$ can be only attributed to the
various parton density sets. In Fig. 10a we have plotted the LO partonic
cross section $d~\sigma_{gg}^{(1)}/d~p_T$ versus $\log_{10} x$ 
for different values for $p_T$. 
For $x<10^{-3}$ this cross section rises steeply whereas it flattens out 
for $x>10^{-3}$. The latter behaviour is changed when we look at the NLO
partonic cross section $d~\sigma_{gg}^{(2)}/d~p_T$ plotted 
versus $\log_{10} x$ in Fig. 10b. Here
the rise at small $x$ becomes even steeper whereas for $x>10^{-3}$ the 
flatness shown by the Born cross section is replaced by a steep decrease 
when $x\rightarrow 1$ except at high $p_T$. Notice that the NLO partonic 
cross section becomes negative (here for $x \ge 8.0 \times 10^{-2}$) due 
to mass factorization as explained below Eq. (5.15).

To show the effect of the gluon flux
we have plotted $F_{gg}(xS,p_t^2,m^2)$ as a function
of $\log_{10} x$ in Fig. 11 for $p_T=100~{\rm GeV/c}$
using three parton density sets in table \ref{table1}. From Fig. 11
we infer that the GRV98 set contains the steepest gluon density whereas
the gluon densities from the other sets are about equal. From this 
observation one can understand the relative ordering of the plots in 
Figs. 7a and 7b. Another feature is that
the hadronic cross section, which is represented by the integral 
of $F_{gg}$ over $x$, receives its main support from the small $x$-region.
Using the representation for the hadronic cross section $d~\sigma/d~p_T$
as given in Eq. (\ref{eqnB.7}) we have computed this quantity where now
all subprocesses are included. Further we have chosen two different
integration regions. The first one is the full range 
$x_{\rm min} \le x \le x_{\rm max}$ 
with $x_{\rm max}=1$ (see Eq. (\ref{eqnB.8}) and the second range is given by 
$x_{\rm max}=5 \times x_{\rm min}$. 
In Fig. 12 we have shown the $p_T$-distributions due to the two different
integration regions. They do not differ by more than 10\% which shows 
that the whole differential cross section is dominated by the small 
$x$-region $x<5 \times x_{\rm min}$. Hence apart from the different 
coupling constants the difference between the cross sections is due to 
the different gluon densities. Future HERA data will have to provide us 
with unique gluon
densities before we can make more accurate predictions for the Higgs
differential distributions. Note that we have chosen the axes in Fig. 12
to coincide with those in Fig. 1b so that one can see the $p_T$ dependence
of the $K$-factor for these two densities by overlaying the plots.

After having studied the small $x$-region we now investigate the
large $x$-region of the partonic cross sections where
S+V gluons and collinear quark anti-quark pairs
dominate the radiative corrections. The S+V gluon part
of the partonic cross section is obtained by omitting the 
hard contributions which are regular at $s_4=0$ so that we only 
keep the singular parts presented in Eqs. (\ref{eqn5.16})-(\ref{eqn5.20}).
Furthermore we include the S+V partonic cross sections in Eqs.
(\ref{eqn5.24})-(\ref{eqn5.26}). These two contributions constitute the
S+V gluon approximation. To study its validity we compute in
NLO the ratio
\begin{eqnarray}
\label{eqn6.22}
R^{\rm S+V}=\frac{d~\sigma^{\rm S+V}}{d~\sigma^{\rm EXACT}}\,,
\end{eqnarray}
for the $p_T$ and $y$ distributions which are given in Figs. 13 and 14 
respectively. Here we use the MRST99 parton density set. 
One expects that the approximation becomes better 
at larger transverse momenta where 
$p_T={\bar p}_{T,{\rm max}}$ in Eq. (\ref{eqn6.13}) 
at the boundary of phase space, which leads to $x=1$ using 
Eq. (\ref{eqnB.8}). However in Fig. 13 the highest value of
$p_T$, given by $p_T=150~{\rm GeV/c}$, is still very small with respect to 
$p_{T,{\rm max}}\sim \sqrt{S}/2=7\times 10^{3}~{\rm GeV/c}$. Therefore it
is rather fortuitous that the approximation works so well 
for $p_T>100~{\rm GeV/c}$ where one obtains $R^{\rm S+V}<1.2$.
The boundary of phase space is also approached
when the Higgs mass increases. This can also be inferred from 
Eq. (\ref{eqnB.8}) where at fixed $p_T$, $x_{\rm min}\rightarrow 1$ 
when $m^2 \rightarrow S-2\sqrt{S}\,p_T$.
Therefore we have plotted $R^{\rm S+V}$ for various Higgs boson masses
in Fig. 13. This figure reveals that the largest Higgs mass leads to the 
worst approximation contrary to our expectations which means that the 
kinematics is not in the large $x$-region.
In Fig. 14 we plot the ratio $R^{\rm S+V}$ for the $y$-distributions for 
the same mass range as in Fig. 13 
where we have chosen the cut $p_{\rm T,min}=30$ ${\rm GeV/c}$
in Eq. (\ref{eqn6.15}). 
Here the approximation is less good than for the $p_T$ distributions in 
Fig. 13 and like in the latter figure it becomes better when
the mass of the Higgs gets smaller. At the lowest value chosen for
the mass i.e. $m=120~{\rm GeV/c^2}$ we observe that $R^{\rm S+V}\sim 1.3$.
Furthermore we see no variation in $R^{\rm S+V}$ with respect 
to the values of $y$. The approximation becomes better, see Fig. 15, if 
$p_{T,{\rm min}}$ is chosen to be larger than $30$ ${\rm GeV/c}$ so that 
it becomes closer to the values
given in Fig. 13. Finally the figures discussed above show
that the S+V gluon approximation overestimates
the exact NLO result. However this overestimate becomes smaller when 
the transverse momentum gets larger. In particular for $p_T>200~{\rm GeV/c}$
the S+V approximation is good enough so that resummation 
techniques can be used to give a better estimate of Higgs boson production
corrected up to all orders in perturbation theory. This statement will 
also hold when Higgs boson production is described
according to the standard model approach where the boson is coupled via 
top-quark loops to the gluons without taking $m_t \rightarrow \infty$ as 
we did above. Although the $p_T$ distributions will change for the exact
and S+V gluon approximation at large $p_T$,  
the ratio $R^{\rm S+V}$ in Fig. 13 will be less affected by the large
top quark mass approach because of cancellations between the numerator
and denominator in Eq. (\ref{eqn6.22}). Notice that the S+V 
gluon approach in the case of the exact cross section can be obtained from
Eqs. (\ref{eqn5.16})-(\ref{eqn5.20}) and (\ref{eqn5.24})-(\ref{eqn5.26}) 
by replacing the approximate Born matrix elements 
$M^{(1)}_{ij \rightarrow k~H}$, obtained from the effective Lagrangian 
in Eq. (\ref{eqn2.1}), by the exact expressions presented in 
\cite{hino}, \cite{ehsb}, \cite{kauff}.

Summarizing the above we have calculated the NLO corrections
to the differential cross section for Higgs boson
production in the large top quark mass approach which can be obtained
from an effective Lagrangian. The calculation was carried out in standard
$n$-dimensional regularization where the $\overline{\rm MS}$
scheme was chosen for renormalization and mass factorization. 
We have presented some of the NLO results in the region where the 
effective Lagrangian should be trustworthy. It turns out that the 
gluon-gluon subprocess dominates the hadronic cross section but
the (anti-)quark-gluon subprocess is certainly not negligible. For 
the transverse momentum distributions there is a small reduction
in scale dependence while going from the LO to the NLO cross section
which is not observed for the rapidity distributions. Further
there is still a large uncertainty in our predictions because the
$K$-factor varies from approximately 1.4 to 1.7 depending on the parton
density set. Also the dependence on the running coupling and
the parton density set is appreciable. The latter is mainly due to the
small $x$ behaviour of the various gluon densities since 
both the partonic cross sections and the gluon densities increase very 
steeply at decreasing $x$. Finally we have shown that the S+V
approximation is quite reasonable provided 
$p_{T,{\rm min}}>100~{\rm GeV/c}$ in spite of the fact that $x$ is still
too small to belong to the large $x$-region. This means that this 
approximation can be used to resum the large corrections due to
S+V gluons in order to obtain a better extimate of the
all order corrected cross section. 

Acknowledgement: We thank S. Dawson, B. Field and R. Kauffman for 
their assistance in tracking down the misprints in \cite{kdr}.
We also thank D. de Florian, M. Grazzini and Z. Kunszt for their
comments on the first version of this manuscript.


\appendix
\mysection*{Appendix A}
\setcounter{section}{1}
Here we list the soft contributions to the partonic cross sections as
defined in Eq. (\ref{eqn4.42}) which were not explicitly given in section 4.
For the various processes we obtain
\begin{eqnarray}
\label{eqnA.1}
s^2 \frac{d^2~{\hat \sigma}^{\rm SOFT}_{gg\rightarrow gg~H}}{d~t~d~u}&
=&\pi\,\delta(s+t+u-m^2)\,
S_{\varepsilon}^2\,G^2\,\left (\frac{\alpha_s(\mu^2)}{4\pi}\right )^2\,
\frac{N}{(N^2-1)^2}
\nonumber\\[2ex]
&&\times \Bigg [\frac{6}{\varepsilon^2}
+\Big (4 \ln\frac{\Delta}{\mu^2}+2\ln\frac{tu}{\mu^2 s}
 -\frac{47}{6}\Big )\frac{1}{\varepsilon}+\ln\frac{\Delta}{\mu^2}
\ln \frac{tu}{\mu^2 s}
\nonumber\\[2ex]
&&+\frac{3}{2}\ln^2\frac{\Delta}{\mu^2}
+\frac{1}{2}\ln^2\frac{tu}{\mu^2 s}-\frac{59}{12}\ln\frac{\Delta}{\mu^2}
-\frac{35}{12}\ln \frac{tu}{\mu^2 s}
\nonumber\\[2ex]
&&-\frac{5}{2}\zeta(2)+\frac{295}{36}\Bigg ]\,
|M^{(1)}_{gg\rightarrow g~H}|^2\,,
\end{eqnarray}
\begin{eqnarray}
\label{eqnA.2}
s^2 \frac{d^2~{\hat \sigma}^{\rm SOFT}_{gg\rightarrow q\bar q~H}}{d~t~d~u}&=&
\pi\,\delta(s+t+u-m^2)\,
S_{\varepsilon}^2\,G^2\,\left (\frac{\alpha_s(\mu^2)}{4\pi}\right )^2\,
n_f\,\frac{1}{(N^2-1)^2}
\nonumber\\[2ex]
&&\times \Bigg [ \frac{1}{3\varepsilon} +\frac{1}{6}\ln\frac{\Delta}{\mu^2}
+\frac{1}{6}\ln \frac{tu}{\mu^2 s}-\frac{11}{18}\Bigg ]\,
|M^{(1)}_{gg\rightarrow g~H}|^2\,,
\end{eqnarray}
\begin{eqnarray}
\label{eqnA.3}
s^2 \frac{d^2~{\hat \sigma}^{\rm SOFT}_{q\bar q\rightarrow gg~H}}{d~t~d~u}&=&
\pi\,\delta(s+t+u-m^2)\,
S_{\varepsilon}^2\,G^2\,\left (\frac{\alpha_s(\mu^2)}{4\pi}\right )^2\,
\frac{1}{N^2}
\nonumber\\[2ex]
&&\times \Bigg [C_A\,\Bigg \{\frac{2}{\varepsilon^2}
+\Big (2 \ln\frac{tu}{\mu^2 s}-\frac{11}{6}\Big )\frac{1}{\varepsilon}
+\ln\frac{\Delta}{\mu^2}\ln \frac{tu}{\mu^2 s}
-\frac{1}{2}\ln^2\frac{\Delta}{\mu^2}
\nonumber\\[2ex]
&&+\frac{1}{2}\ln^2\frac{tu}{\mu^2 s}
-\frac{11}{12}\ln\frac{\Delta}{\mu^2}
-\frac{11}{12}\ln \frac{tu}{\mu^2 s}
-\frac{3}{2}\zeta(2)+\frac{67}{36}\Bigg \}
\nonumber\\[2ex]
&&+C_F\,\Bigg \{\frac{4}{\varepsilon^2}+\left (4 \ln\frac{\Delta}{\mu^2}
\right )\frac{1}{\varepsilon}+2\ln^2\frac{\Delta}{\mu^2}-\zeta(2)
\Bigg \}\Bigg ]
\nonumber\\[2ex]
&&\times |M^{(1)}_{q\bar q\rightarrow g~H}|^2\,,
\end{eqnarray}
\begin{eqnarray}
\label{eqnA.4}
s^2 \frac{d^2~{\hat \sigma}^{\rm SOFT}_{qg\rightarrow qg~H}}{d~t~d~u}&=&
\pi\,\delta(s+t+u-m^2)\,
S_{\varepsilon}^2\,G^2\,\left (\frac{\alpha_s(\mu^2)}{4\pi}\right )^2\,
\frac{1}{N(N^2-1)}
\nonumber\\[2ex]
&&\times \Bigg [C_A\,\Bigg \{\frac{2}{\varepsilon^2}
+\Big (2 \ln\frac{\Delta}{\mu^2}
-1 \Big )\frac{1}{\varepsilon}+\ln^2\frac{\Delta}{\mu^2}
-\ln\frac{\Delta}{\mu^2}
\nonumber\\[2ex]
&&-\frac{1}{2}\zeta(2)+\frac{1}{2}\Bigg \}
\nonumber\\[2ex]
&&+C_F\,\Bigg \{\frac{4}{\varepsilon^2}+\Big (2\ln\frac{\Delta}{\mu^2}
+2\ln \frac{tu}{\mu^2 s}-\frac{7}{2}\Big )\frac{1}{\varepsilon}
+\ln\frac{\Delta}{\mu^2}\ln \frac{tu}{\mu^2 s}
\nonumber\\[2ex]
&&+\frac{1}{2}\ln^2\frac{\Delta}{\mu^2}
+\frac{1}{2}\ln^2\frac{tu}{\mu^2 s}-\frac{7}{4}\ln\frac{\Delta}{\mu^2}
-\frac{7}{4}\ln \frac{tu}{\mu^2 s}
\nonumber\\[2ex]
&&-2\zeta(2)+\frac{7}{2}\Bigg \}\Bigg ]\,
|M^{(1)}_{qg\rightarrow q~H}|^2\,,
\end{eqnarray}
\begin{eqnarray}
\label{eqnA.5}
s^2 \frac{d^2~{\hat \sigma}^{\rm SOFT}_{q\bar q\rightarrow q\bar q~H}}{d~t~d~u}
&=&\pi\, \delta(s+t+u-m^2)\,S_{\varepsilon}^2\,G^2\,
\left (\frac{\alpha_s(\mu^2)}{4\pi}\right )^2\,n_f \,\frac{1}{N^2}
\nonumber\\[2ex]
&&\Bigg [ \frac{1}{3\varepsilon}+\frac{1}{6}\ln\frac{\Delta}{\mu^2}
+\frac{1}{6}\ln \frac{tu}{\mu^2 s}
-\frac{5}{18}\Bigg ]\, |M^{(1)}_{q\bar q\rightarrow g~H}|^2\,.
\end{eqnarray}

\mysection*{Appendix B}
\setcounter{section}{2}
There is an alternative way to obtain the transverse momentum distribution 
as presented in Eq. (\ref{eqn6.14}). First one computes the partonic cross 
section
\begin{eqnarray}
\label{eqnB.1}
\frac{d~\sigma_{ab}}{d~p_T^2}(s,p_T^2,m^2,\mu^2)=
\int_{0}^{s_{4,{\rm max}}} ds_4\frac{d^2~\sigma_{ab}}{d~p_T^2~d~s_4}
(s,p_T^2,s_4,m^2,\mu^2)\,,
\end{eqnarray}
with
\begin{eqnarray}
\label{eqnB.2}
&& \frac{d^2~\sigma_{ab}}{d~p_T^2~d~s_4}(s,p_T^2,s_4,m^2,\mu^2)=
\frac{s}{\sqrt{(s+m^2-s_4)^2-4~s~(p_T^2+m^2)}}
\nonumber\\[2ex]
&&\times \left [
\frac{d^2~\sigma_{ab}}{d~t~d~u}(s,t,u,m^2,\mu^2)
+\frac{d^2~\sigma_{ab}}{d~t~d~u}(s,u,t,m^2,\mu^2) \right ]\,,
\end{eqnarray}
and
\begin{eqnarray}
\label{eqnB.3}
s_{4,{\rm max}}=s+m^2-2\,\sqrt{s(p_T^2+m^2)}\,,
\end{eqnarray}
The reason is that if one changes the variables $t$ and $u$ into $s_4$ and 
$p_T^2$ one has two possibilities
\begin{eqnarray}
\label{eqnB.4}
t&=&\frac{1}{2}\,\Big [s_4+m^2-s+\sqrt{(s+m^2-s_4)^2-4~s~(p_T^2+m^2)}\Big ]
\equiv t_1\,,
\nonumber\\[2ex]
u&=&\frac{1}{2}\,\Big [s_4+m^2-s-\sqrt{(s+m^2-s_4)^2-4~s~(p_T^2+m^2)}\Big ]
\equiv u_1\,,
\end{eqnarray}
or
\begin{eqnarray}
\label{eqnB.5}
t&=&\frac{1}{2}\,\Big [s_4+m^2-s-\sqrt{(s+m^2-s_4)^2-4~s~(p_T^2+m^2)}\Big ]
\equiv t_2=u_1\,,
\nonumber\\[2ex]
u&=&\frac{1}{2}\,\Big [s_4+m^2-s+\sqrt{(s+m^2-s_4)^2-4~s~(p_T^2+m^2)}\Big ]
\equiv u_2=t_1\,,
\nonumber\\
\end{eqnarray}
which follows from the substitution
\begin{eqnarray}
\label{eqnB.6}
\sinh y=\pm\frac{\sqrt{(s+m^2-s_4)^2-4~s~(p_T^2+m^2)}}
{2\sqrt s \sqrt{p_T^2+m^2}}\,,
\qquad
\cosh y=\frac{s+m^2-s_4}{2\sqrt s \sqrt{p_T^2+m^2}}\,,
\nonumber\\
\end{eqnarray}
in Eq. (\ref{eqnB.2}).
Hence one has to compute the sum $d\sigma(t_1,u_1)+d\sigma(t_2,u_2)$
which is equal to $d\sigma(t_1,u_1)+d\sigma(u_1,t_1)$. Notice that when the 
partonic cross section is symmetric in $t$ and $u$ one can replace the sums 
above by $2~d\sigma(t_1,u_1)$. This does not apply to the quark-gluon 
subprocess because here the cross section is asymmetric in $t$ and $u$.
The hadronic cross section is now obtained from
\begin{eqnarray}
\label{eqnB.7}
\frac{d~\sigma^{{\rm H_1H_2}}}{d~p_T}(S,p_T^2,m^2)=\sum_{a,b=q,g}
\int_{x_{\rm min}}^{\rm x_{max}} dx\, \Phi_{ab}^{\rm H_1H_2}(x,\mu^2)\,
\frac{d~\sigma_{ab}}{d~p_T}(x~S,p_T^2,m^2,\mu^2)\,,
\nonumber\\
\end{eqnarray}
with
\begin{eqnarray}
\label{eqnB.8}
x_{\rm min}=\frac{m^2+2p_T^2+2\,\sqrt{p_T^2(p_T^2+m^2)}}{S}\,,\qquad
x_{\rm max}=1\,,
\end{eqnarray}
and $\Phi_{ab}$ denotes the partonic flux defined by
\begin{eqnarray}
\label{eqnB.9}
\Phi_{ab}^{\rm H_1H_2}(x,\mu^2)=\int_0^1 dx_1\int_0^1dx_2\,\delta(x-x_1\,x_2)\,
f_a^{\rm H_1}(x_1,\mu^2)\,f_b^{\rm H_2}(x_2,\mu^2)\,.
\end{eqnarray}
The lower boundary $x_{\rm min}$ follows from the inequality in Eq. (6.13)
where $s=x~S$. We have checked that expressions  Eqns. (6.14) and 
(B.7) lead to the same result. 

%

\centerline{\bf \large{Figure Captions}}
\begin{description}
\item[Fig. 1a.]
The differential cross section $d~\sigma/dp_T$
integrated over the whole rapidity range (see Eq. (\ref{eqn6.14}))
with $m=120~{\rm GeV/c^2}$ and $\mu^2=m^2+p_T^2$.
The LO plots are presented for the subprocesses $gg$ (long-dashed line),
$q(\bar q)g$ (dot-dashed line) 
and  $100 \times (q\bar q)$ (dotted line)
using the parton density set MRST98(lo05a.dat). 
\item[Fig. 1b.]
Same as Fig. 1a in NLO except for $100*{\rm abs}(q\bar q)$ 
(dotted line) and
the additional subprocess $100 \times {\rm abs}(qq)$ (short-dashed line) 
using the parton density set MRST98(ft08a.dat).
\item[Fig. 2a.]
The differential cross section $d~\sigma/dy$ for 
$p_{T,{\rm min}}=30~{\rm GeV/c}$ and $p_{T,{\rm max}}=240~{\rm GeV/c}$  
(see Eq. (\ref{eqn6.15}))
with $m=120~{\rm GeV/c^2}$ and $\mu^2=m^2+p_{T,{\rm min}}^2$.
The LO plots are presented for the subprocesses
$gg$ (long-dashed line),
$q(\bar q)g$ (dot-dashed line)
$100 \times q(\bar q)$ (dotted line)
using the parton density set MRST98(lo05a.dat). 
\item[Fig. 2b.]
Same as Fig. 2a in NLO with the additional subprocess
$100 \times qq$ (short-dashed line) 
using the parton density set MRST98(ft08a.dat).
\item[Fig. 3.]
The mass dependence of $d~\sigma^{\rm NLO}/dp_T$
(Eq. (\ref{eqn6.14})) using the set MRST99 with $\mu^2=m^2+p_T^2$
for Higgs masses
$m=120~{\rm GeV/c^2}$ (solid line),
$m=160~{\rm GeV/c^2}$ (dashed line) and
$m=200~{\rm GeV/c^2}$ (dot-dashed line).
\item[Fig. 4.]
The mass dependence of $d~\sigma^{\rm NLO}/dy$
for $p_{T,{\rm min}}=30~{\rm GeV/c}$ (Eq. (\ref{eqn6.15}))
using the set MRST99 with $\mu^2=m^2+p_{T,{\rm min}}^2$.
The notation is the same as in Fig. 3.
\item[Fig. 5a.]
The scale dependence of $d~\sigma^{\rm LO}/dp_T$ 
integrated over the whole rapidity range (see Eq. (\ref{eqn6.14}))
with $m=120~{\rm GeV/c^2}$, $\mu_0^2=m^2+p_T^2$ 
using the MRST98 parton density sets.
The results are shown for  
$\mu=\mu_0/2$ (dashed line), $\mu=\mu_0$ (solid line)
and $\mu=2\times \mu_0$ (dot-dashed line).
The upper three curves are the NLO results.
\item[Fig. 5b.]
The quantity $N(p_T,\mu/\mu_0)$ (see Eq. (\ref{eqn6.18})) plotted in the range
$0.1<\mu/\mu_0<10$ at fixed values of $p_T$
with $m=120~{\rm GeV/c^2}$ and $\mu_0^2=m^2+p_T^2$
using the MRST98 parton density sets.
The results are shown for
$p_T=30~{\rm GeV/c}$ (solid line), $p_T=70~{\rm GeV/c}$ (dashed line),
$p_T=100~{\rm GeV/c}$ (dot-dashed line).
The upper three curves on the left hand side are the LO results whereas
the lower three curves refer to NLO.
\item[Fig. 6.]
The scale dependences of $d~\sigma^{\rm LO}/dy$ 
and $d~\sigma^{\rm NLO}/dy$ integrated over the region 
$p_{T,{\rm min}}=30~{\rm GeV/c}$ and $p_{T,{\rm max}}=240~{\rm GeV/c}$ 
(see Eq. (\ref{eqn6.15})). The notation and parameters are as in Fig.5.
\item[Fig. 7a.]
The $K$-factors in Eq. (\ref{eqn6.19}) for $d~\sigma/dp_T$
integrated over the whole rapidity range (see Eq. (\ref{eqn6.14})) 
with $m=120~{\rm GeV/c^2}$ and $\mu^2=m^2+p_T^2$ for 
the MRST98 sets (solid line), the 
GRV98 sets (dot-dashed line),
and the CTEQ98 sets, (dashed line). 
\item[Fig. 7b.]
The ratios $R$
in Eq. (\ref{eqn6.20}) for the same differential cross 
sections as in Fig. 7a,
$R^{\rm GRV}$ in LO (solid line), 
$R^{\rm GRV}$ in NLO (dot-dashed line), 
$R^{\rm CTEQ}$ in LO (dashed line) and 
$R^{\rm CTEQ}$ in NLO (dotted line). 
\item[Fig. 8a]
The $K$-factors in Eq. (\ref{eqn6.19}) for $d~\sigma/dy$
integrated over the region 
$p_{T,{\rm min}}=30~{\rm GeV/c}$ and $p_{T,{\rm max}}=240~{\rm GeV/c}$ 
(see Eq. (\ref{eqn6.15}))
with $m=120~{\rm GeV/c^2}$ and $\mu^2=m^2+p_{T,{\rm min}}^2$. 
The parton density sets and notation are as in Fig. 7a.
\item[Fig. 8b.]
The ratios $R$
in Eq. (\ref{eqn6.20}) for the same differential cross sections as in Fig. 8a.
The notation is the same as in Fig. 7b.
\item[Fig. 9.]
Comparisons of $d~\sigma/dp_T$ in LO and NLO obtained from our calculation 
(RSN) versus those in Fig. 1a in \cite{fgk} (FGK), for 
$m_t=173.4~{\rm GeV/c^2}$ RSN(1) (solid line),
$m_t=10^4~{\rm GeV/c^2}$ RSN(2) (long dashed line) and
$m_t=\infty$ FGK (dot-dashed line).
The upper curves are for NLO and the lower ones for LO.
\item[Fig. 10a.]
The partonic differential cross section $d~\sigma^{(1)}_{gg}/dp_T$ (Born)
with $m=120~{\rm GeV/c^2}$ and $\mu^2=m^2+p_T^2$ for
$p_T=200~{\rm GeV/c}$ (solid line), 
$p_T=150~{\rm GeV/c}$ (long-dashed line), 
$p_T=100~{\rm GeV/c}$ (dot-dashed line), 
$p_T=75~{\rm GeV/c}$ (short-dashed line) and
$p_T=50~{\rm GeV/c}$ (dotted line). 
\item[Fig. 10b.]
Same as in Fig. 10a but now for the NLO 
differential cross section $d~\sigma^{(2)}_{gg}/dp_T$.
\item[Fig. 11.]
The flux multiplied by the partonic cross section
represented by the quantity $F_{gg}(xS,p_T^2,m^2)$ in Eq. (\ref{eqn6.21})
with $p_T=100~{\rm GeV/c}$, $m=120~{\rm GeV/c^2}$ and $\mu^2=m^2+p_T^2$
plotted as a function of $\log_{10}x$ for
MRST98 (solid lines), GRV98 (dot-dashed lines) and CTEQ4 (dashed lines).
The upper curves are for NLO and the lower ones for LO.
\item[Fig. 12.]
The NLO hadronic cross section $d~\sigma/dp_T$ obtained from Eq. (\ref{eqnB.7})
(all subprocesses included) with $m=120~{\rm GeV/c^2}$ and $\mu^2=m^2+p_T^2$.
The parton density set is GRV98 with $x_{\rm max}=1$ 
(solid line) and with $x_{\rm max}=5~x_{\rm min}$ (dashed line).
\item[Fig. 13.]
The ratio $R^{\rm S+V}$ in Eq. (\ref{eqn6.22}) for the $p_T$
distributions using the set MRST99 with $\mu^2=m^2+p_{T,{\rm min}}^2$
and various Higgs masses given by
$m=120~{\rm GeV/c^2}$ (solid line),
$m=160~{\rm GeV/c^2}$ (dashed line) and 
$m=200~{\rm GeV/c^2}$ (dot-dashed line).
\item[Fig. 14.]
The ratio $R^{\rm S+V}$ in Eq. (\ref{eqn6.22}) for the $y$ distributions 
using the MRST99 set with $\mu^2=m^2+p_{T,{\rm min}}^2$
for $p_{T,{\rm min}}=30$ ${\rm GeV/c}$ and various Higgs masses given by
$m=120~{\rm GeV/c^2}$ (solid line),
$m=160~{\rm GeV/c^2}$ (dashed line) and
$m=200~{\rm GeV/c^2}$ (dot-dashed line).
\item[Fig. 15.]
The ratio $R^{\rm S+V}$ in Eq. (\ref{eqn6.22}) for the $y$ distributions
using the MRST99 set with $\mu^2=m^2+p_{T,{\rm min}}^2$ 
and $m=120~{\rm GeV/c^2}$,
for various values of $p_{T,{\rm min}}$, namely 
$p_{T,{\rm min}} = 100$ ${\rm GeV/c}$ (solid line),
$p_{T,{\rm min}} = 200$ ${\rm GeV/c}$ (dashed line),
$p_{T,{\rm min}} = 250$ ${\rm GeV/c}$ (dot-dashed line) and
$p_{T,{\rm min}} = 300$ ${\rm GeV/c}$ (dotted line).
\end{description}

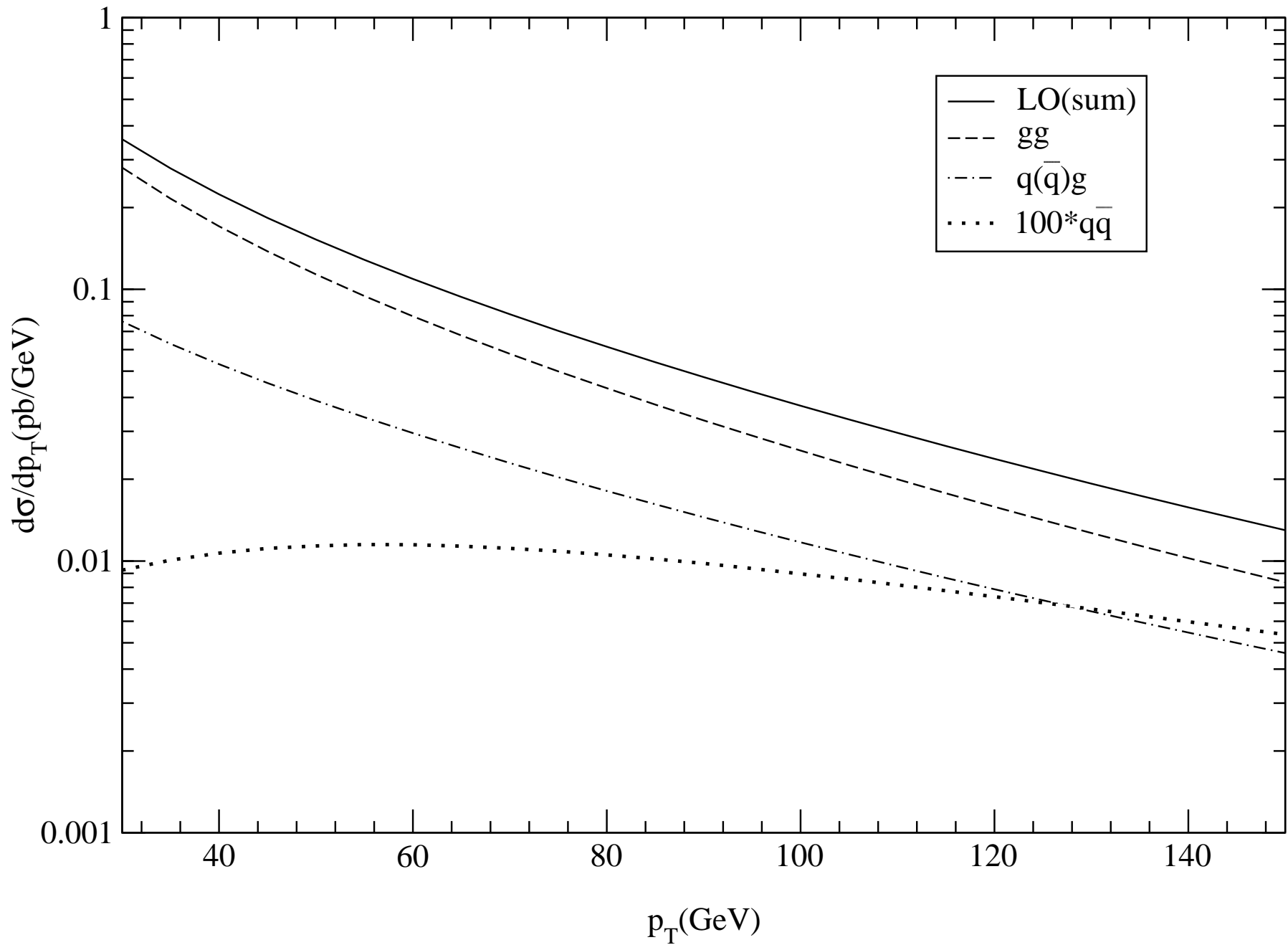
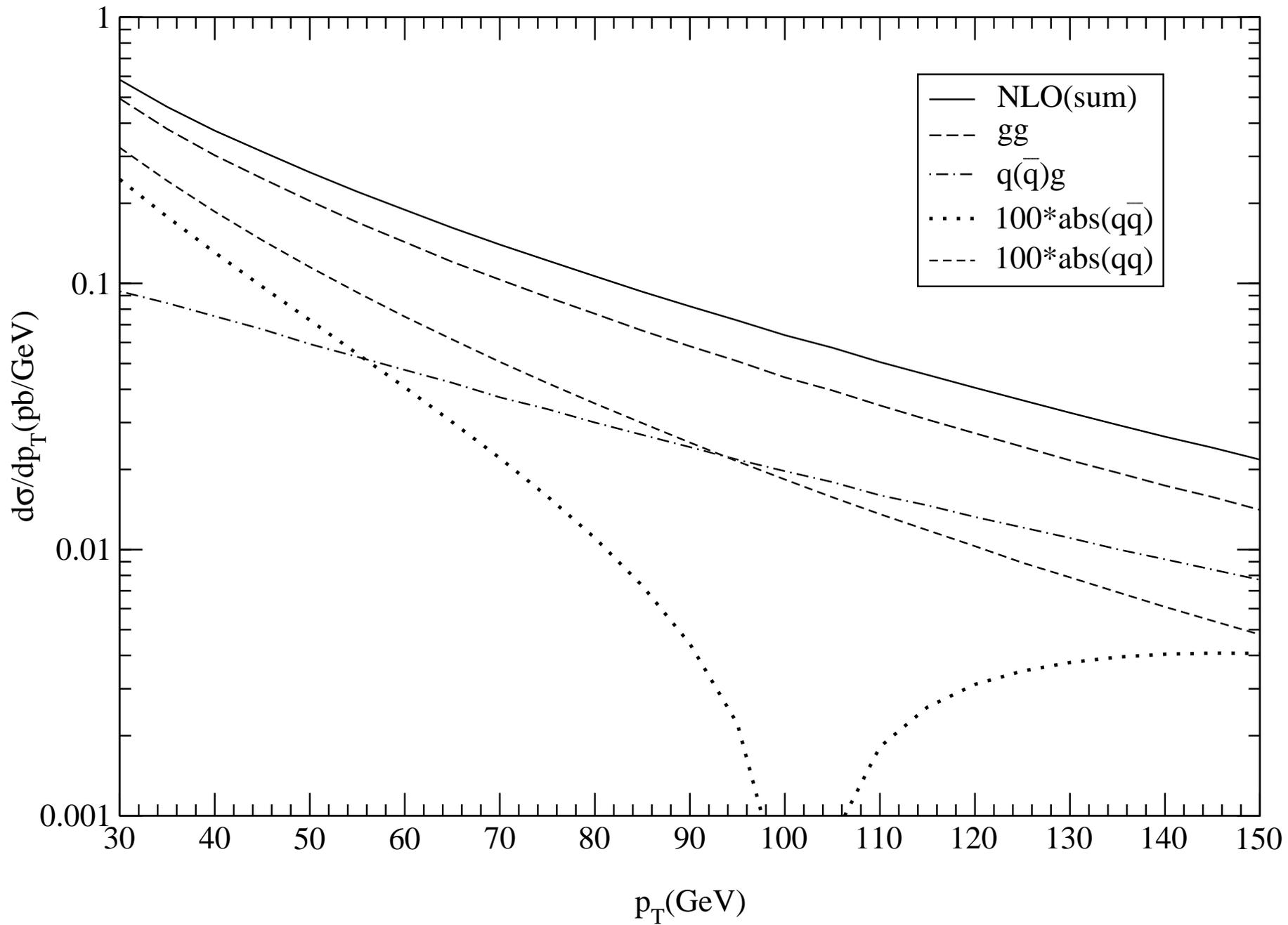
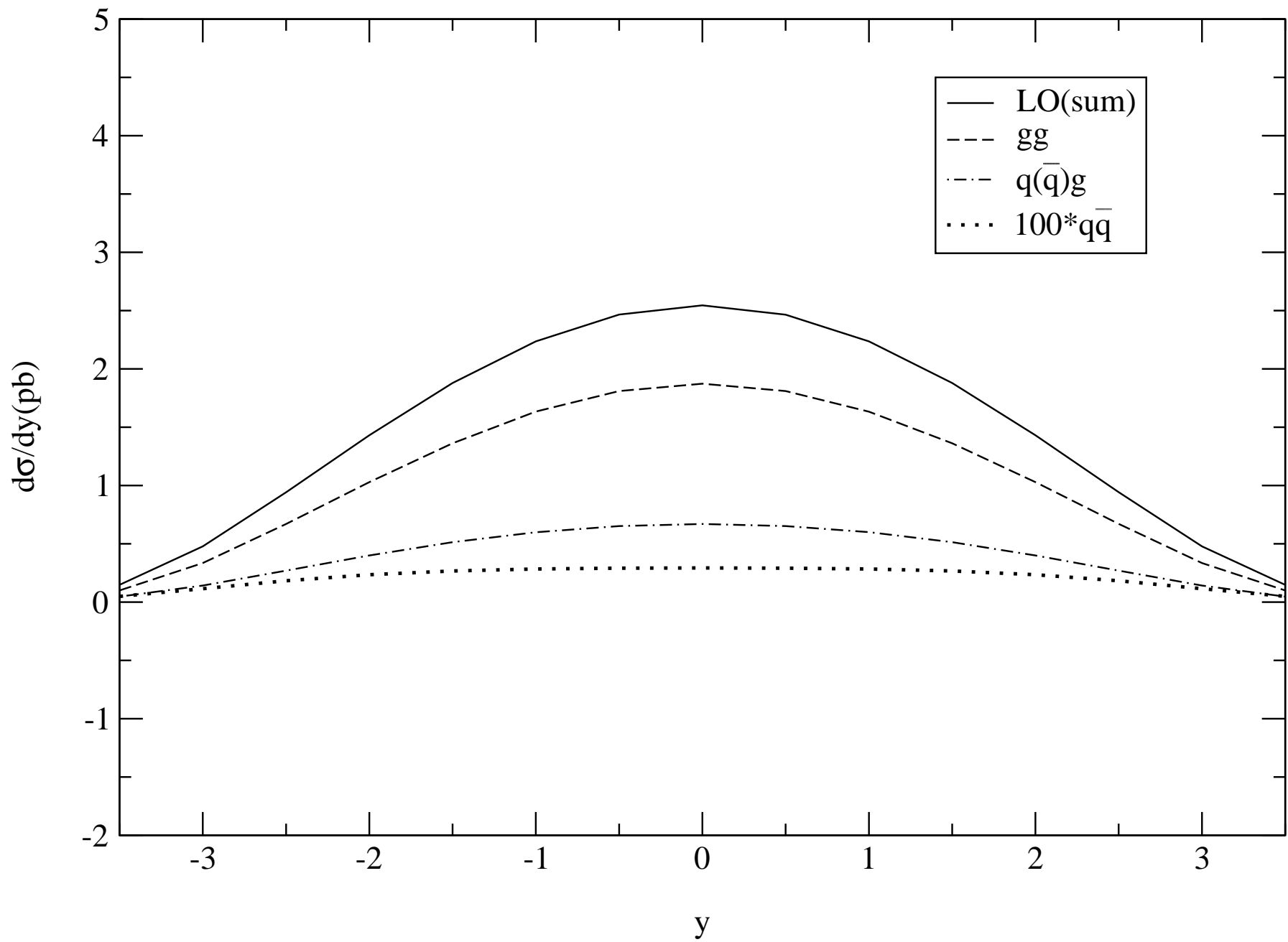
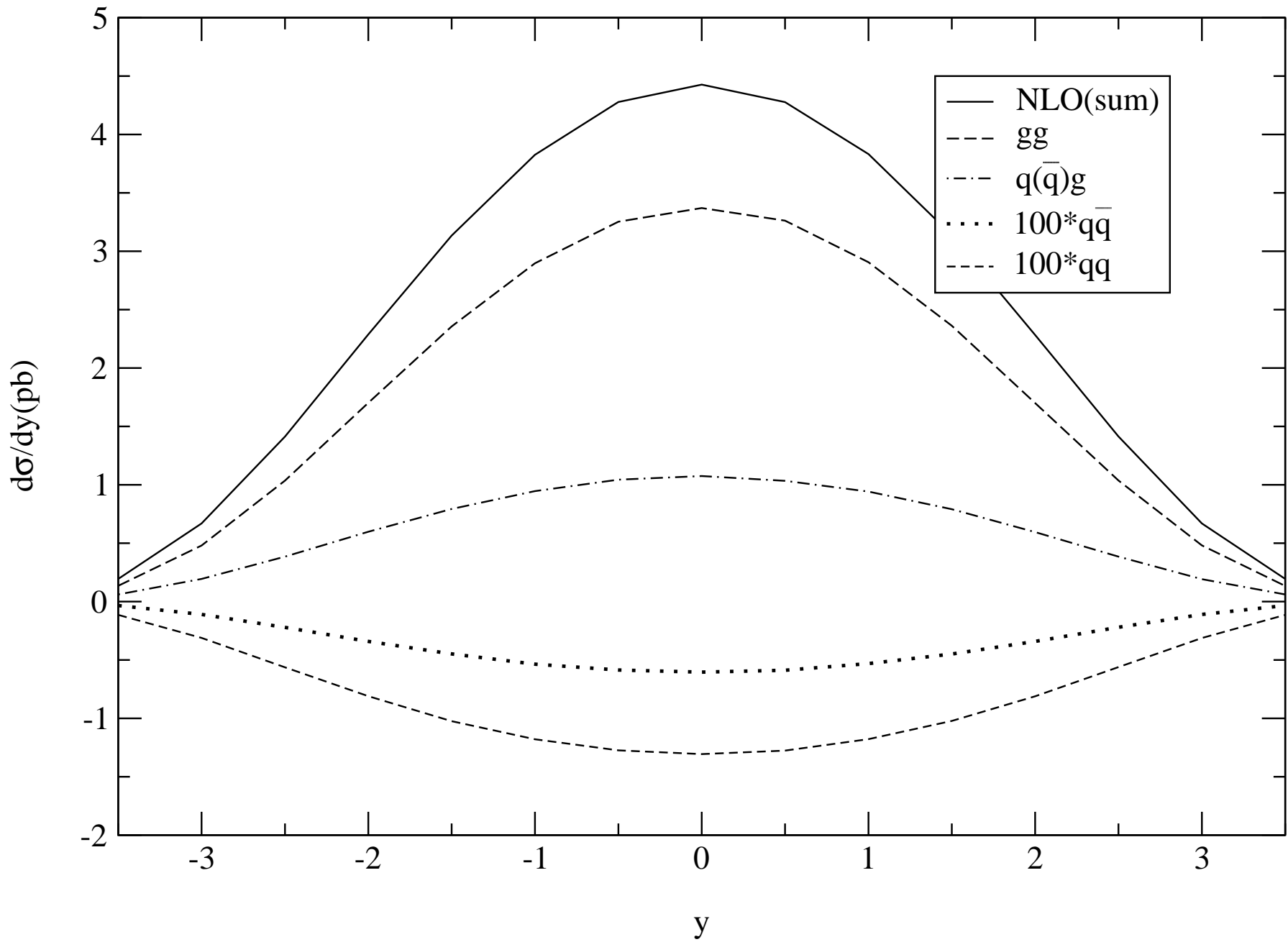
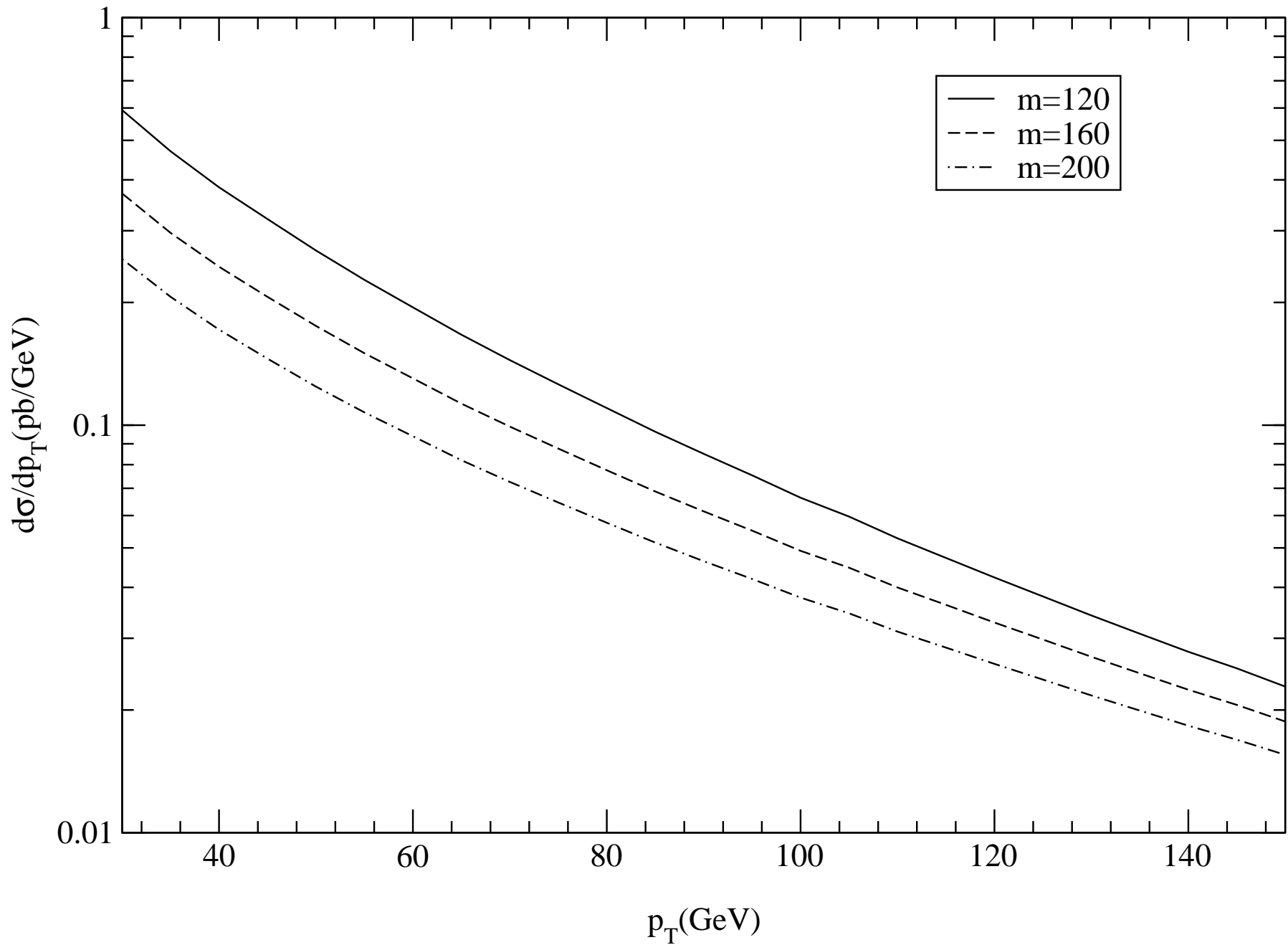
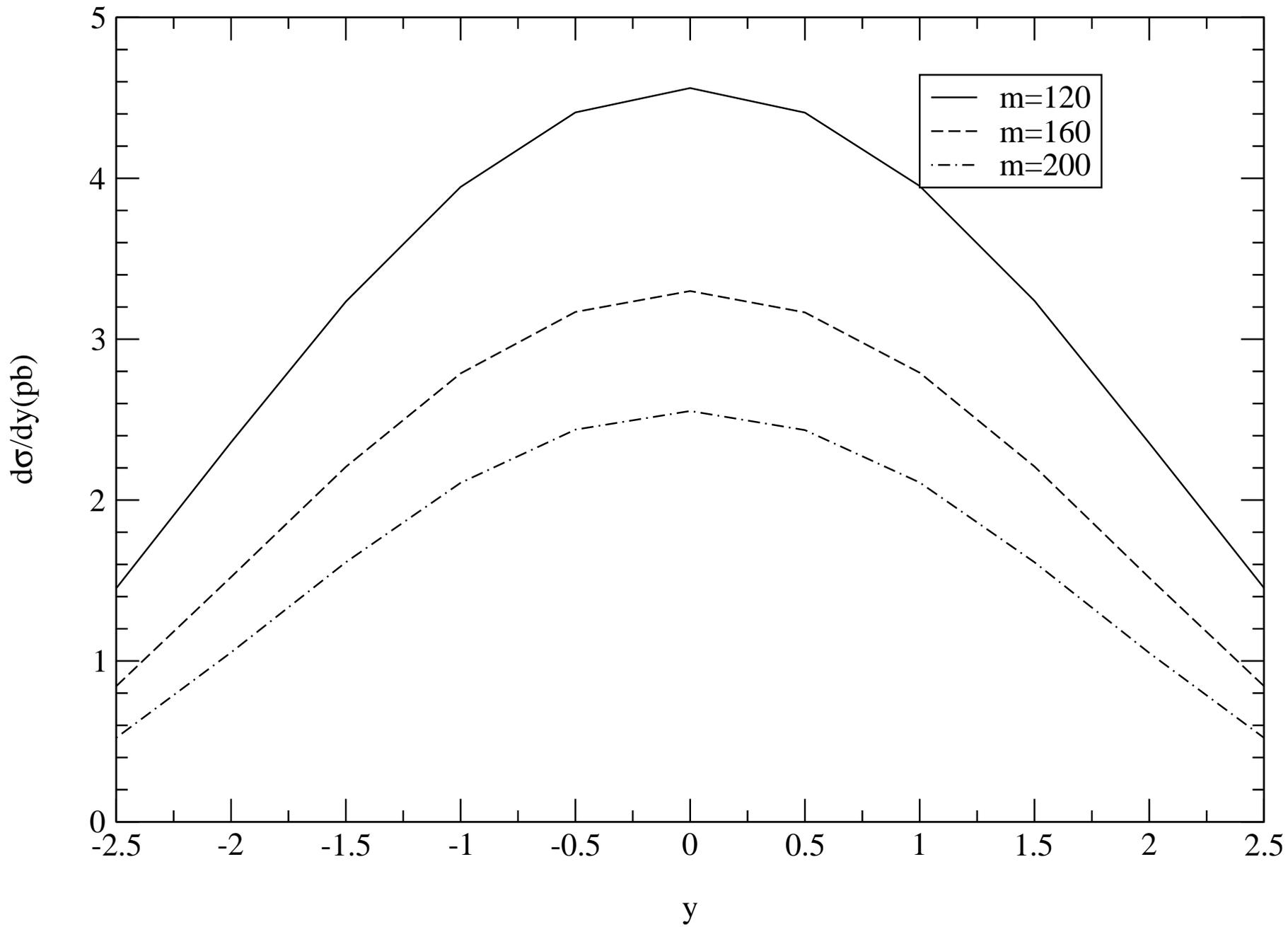
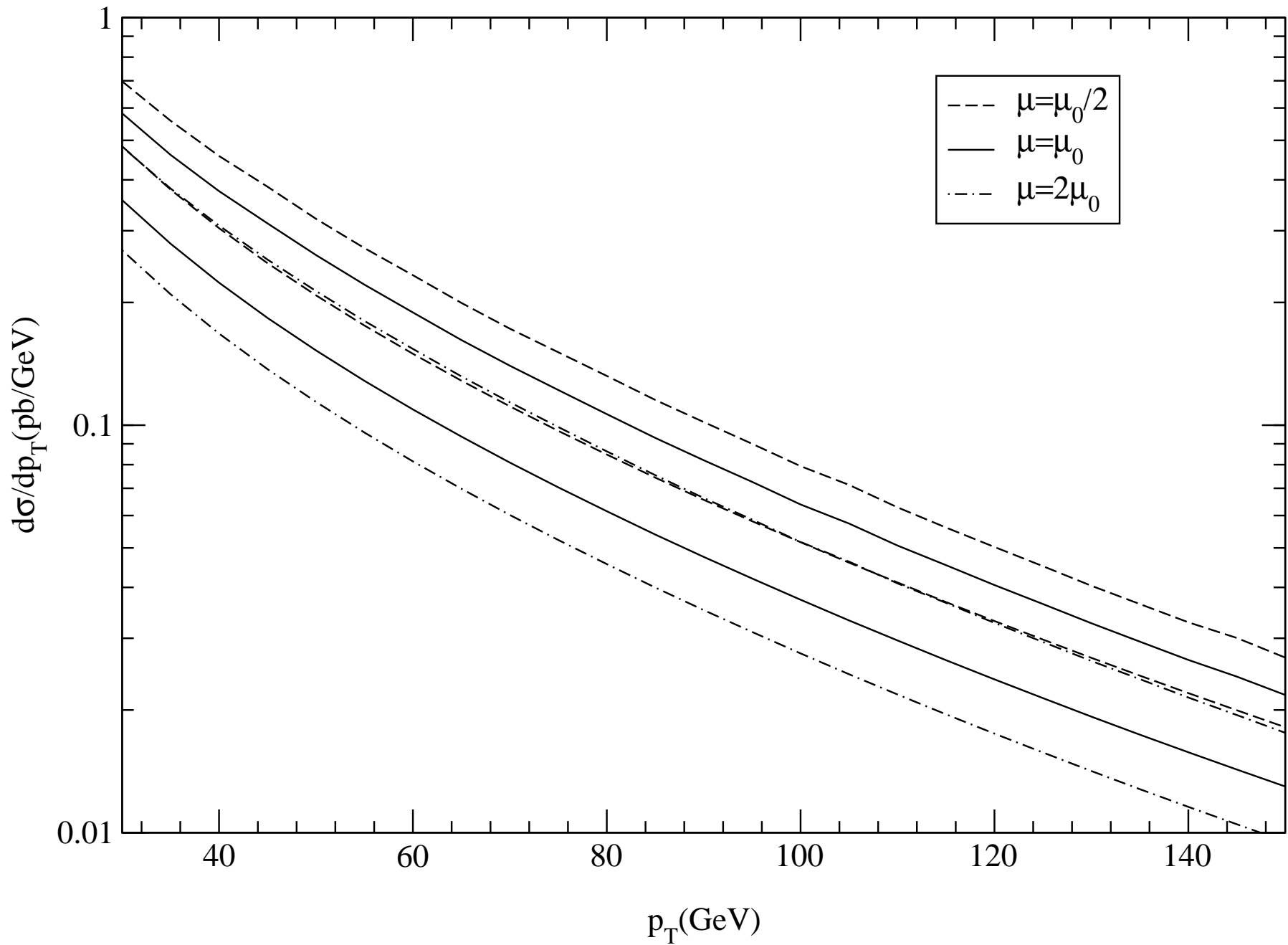
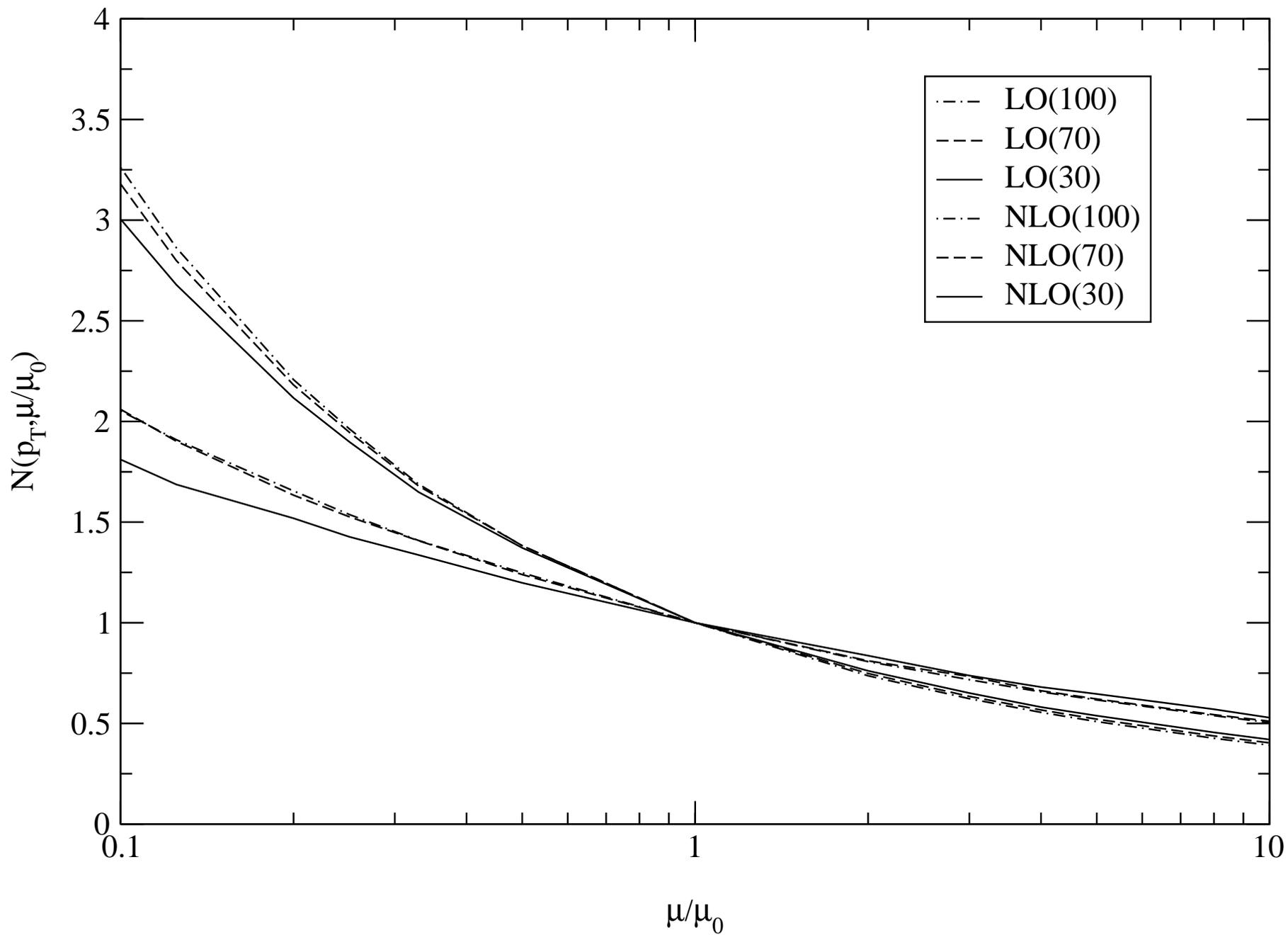
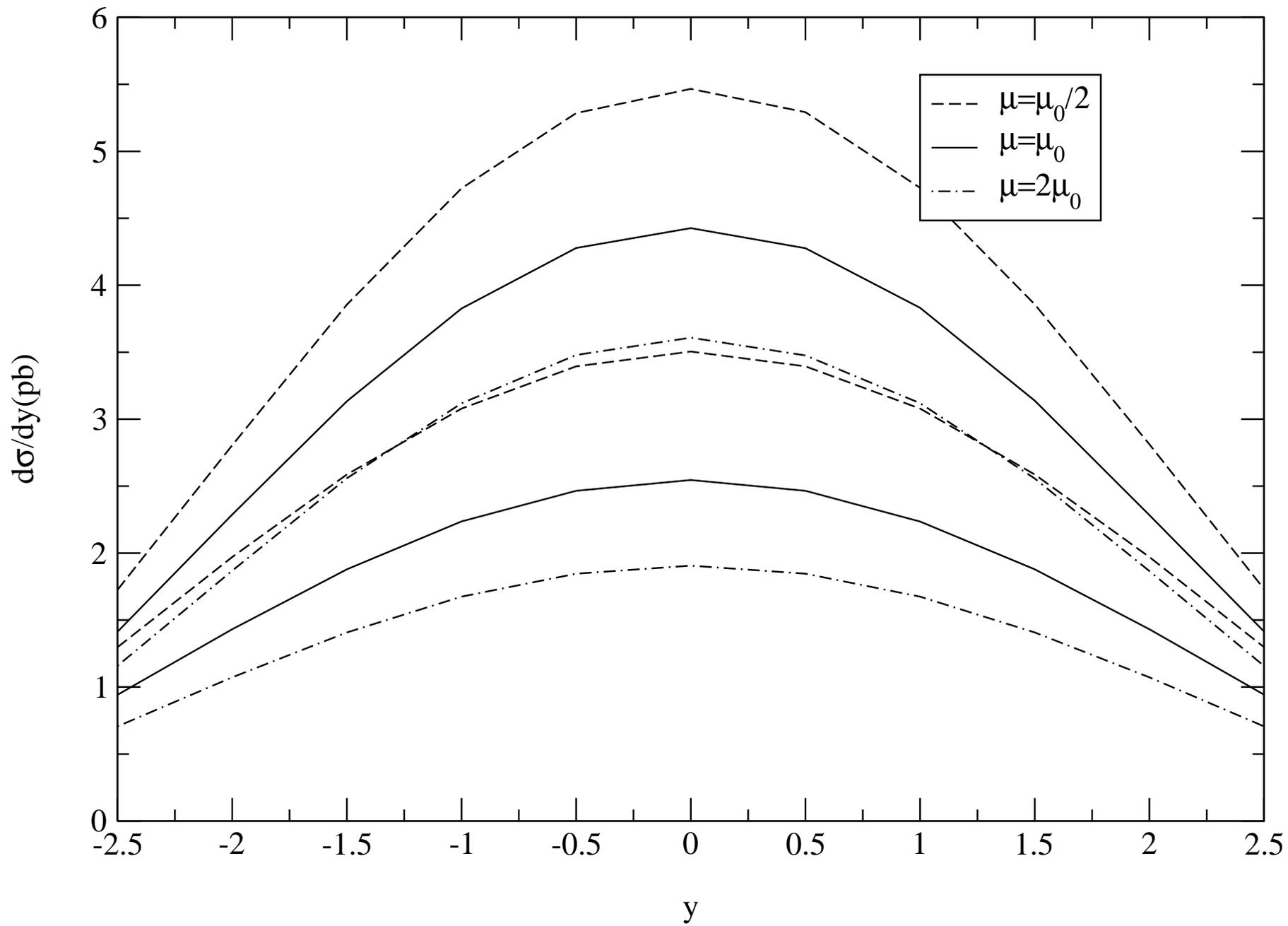
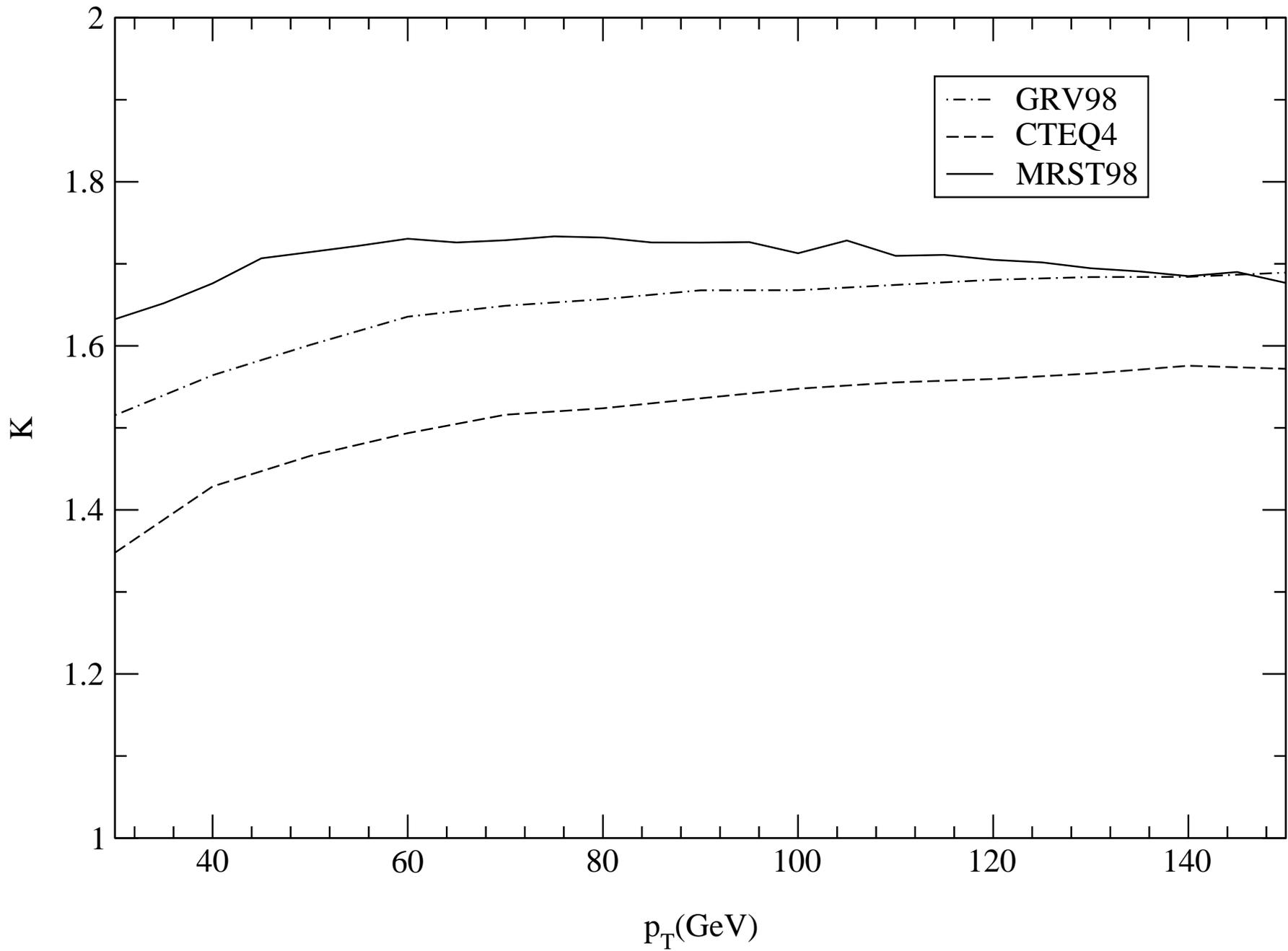
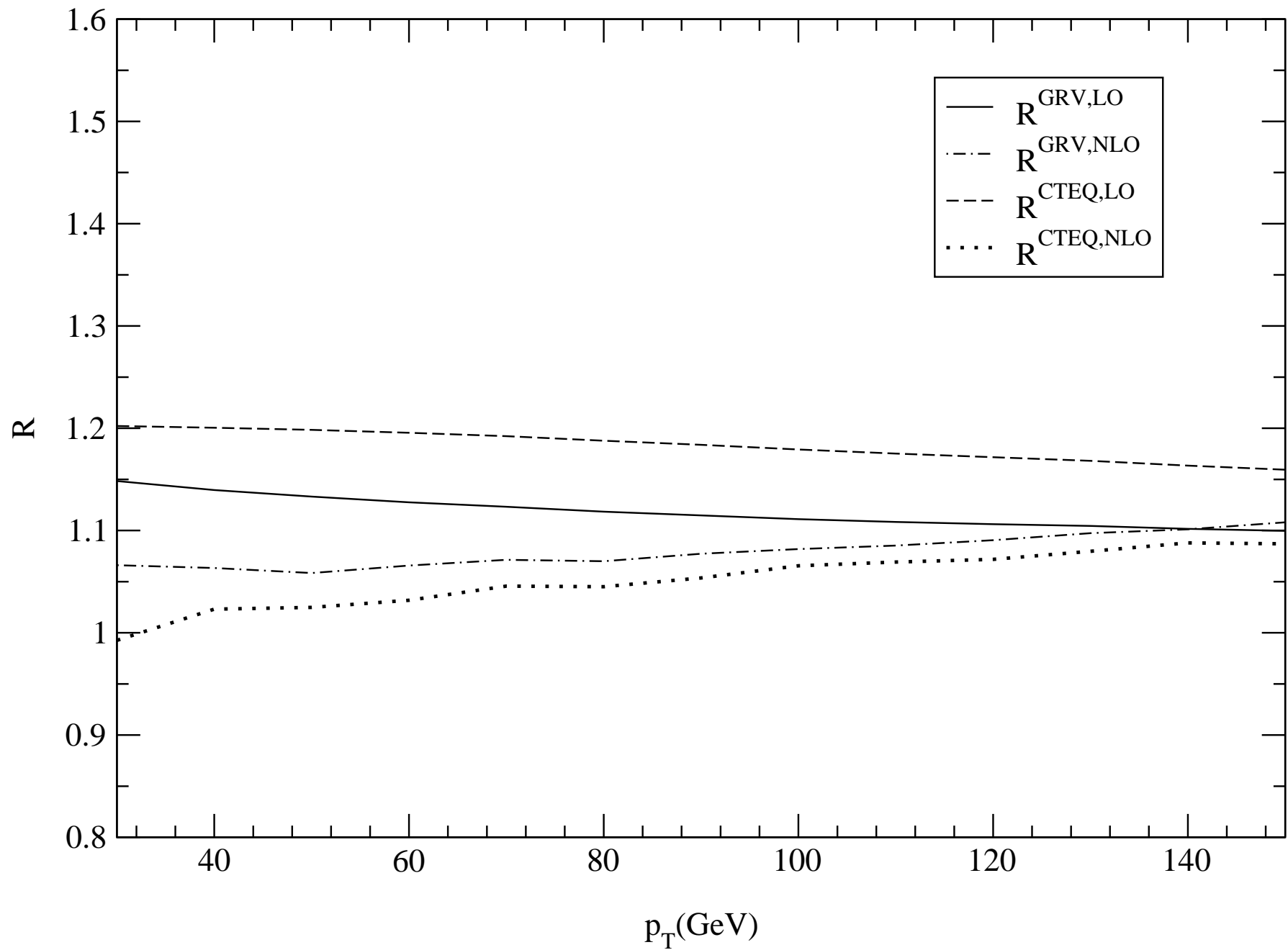
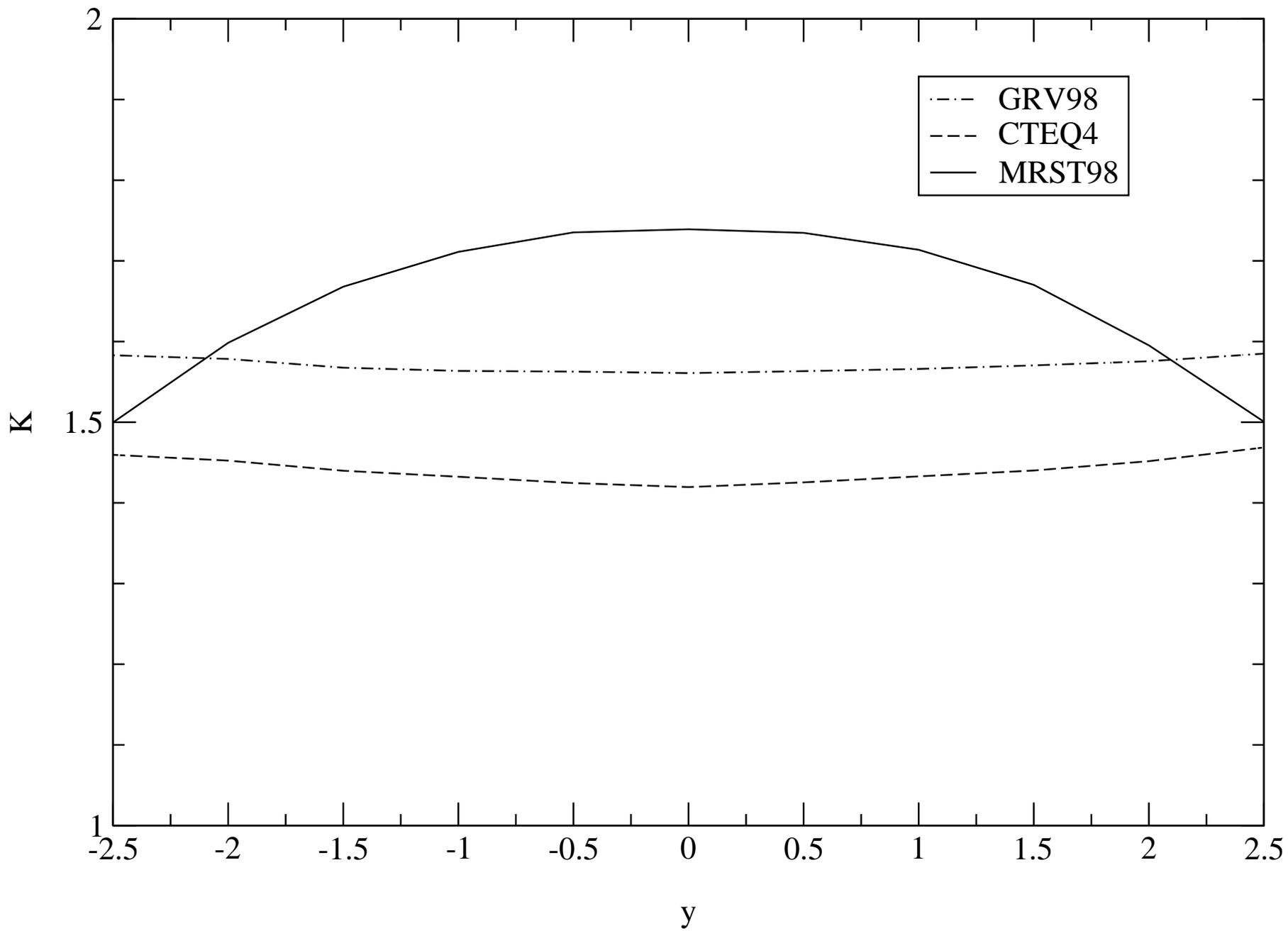
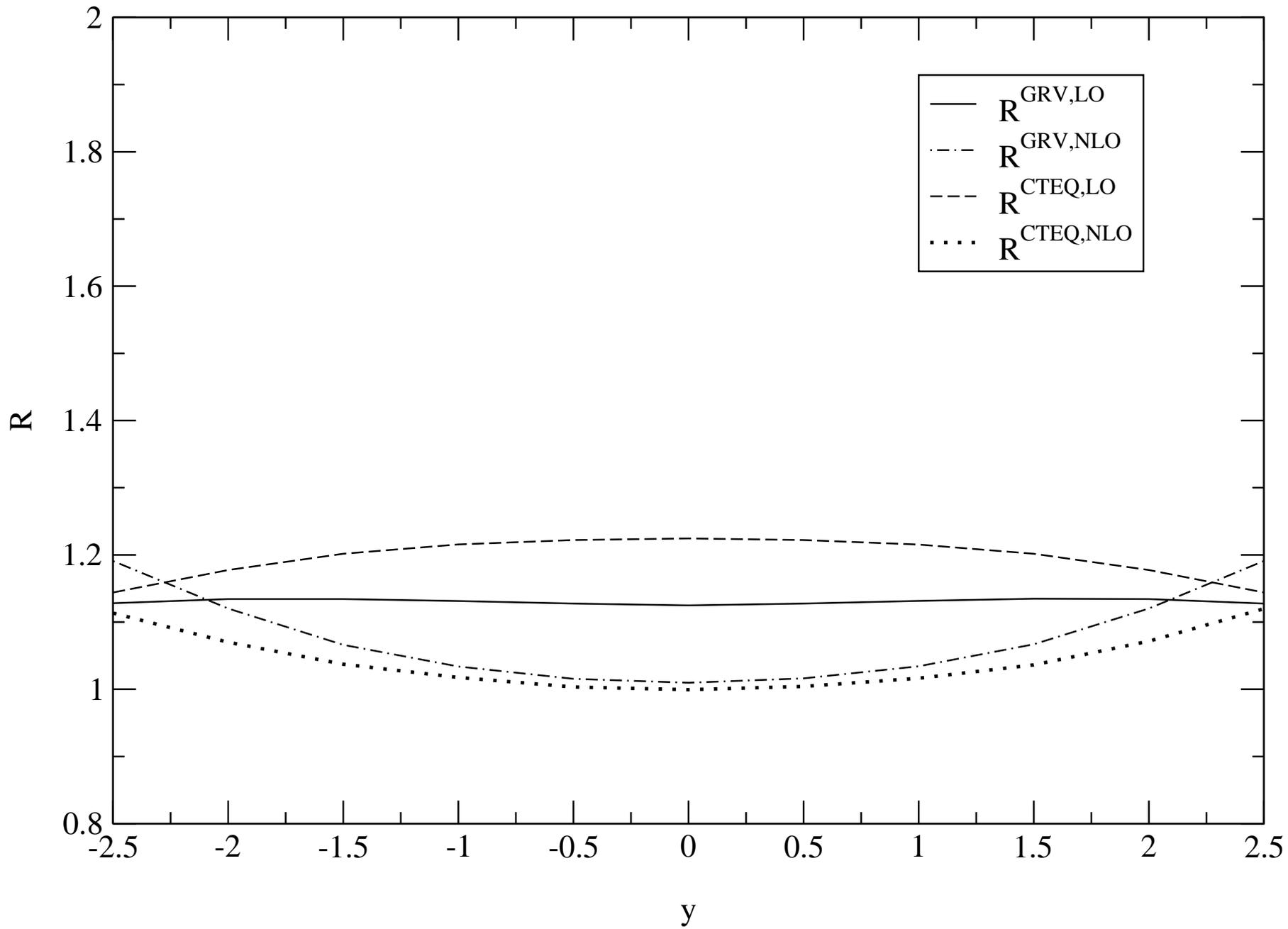
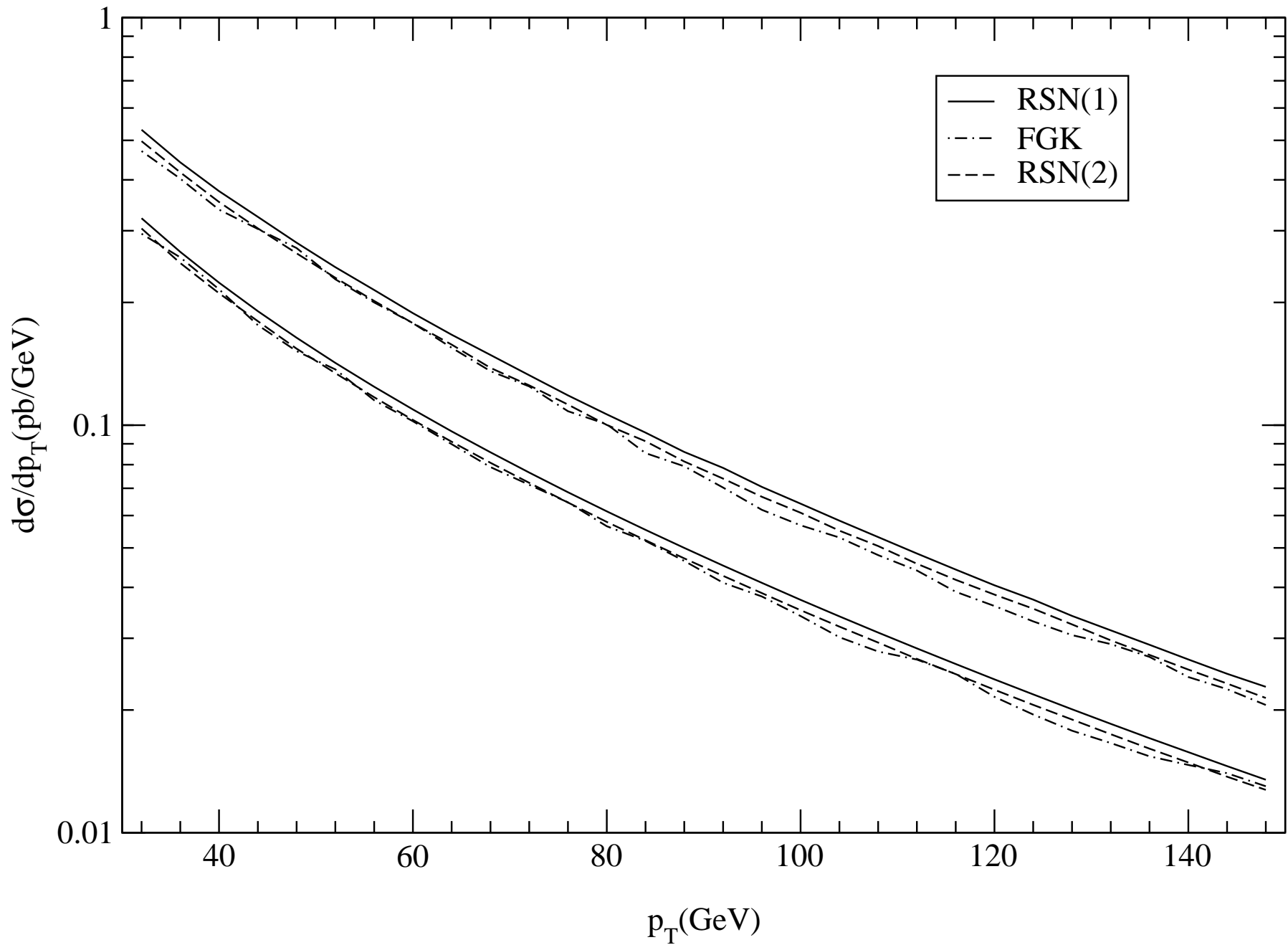
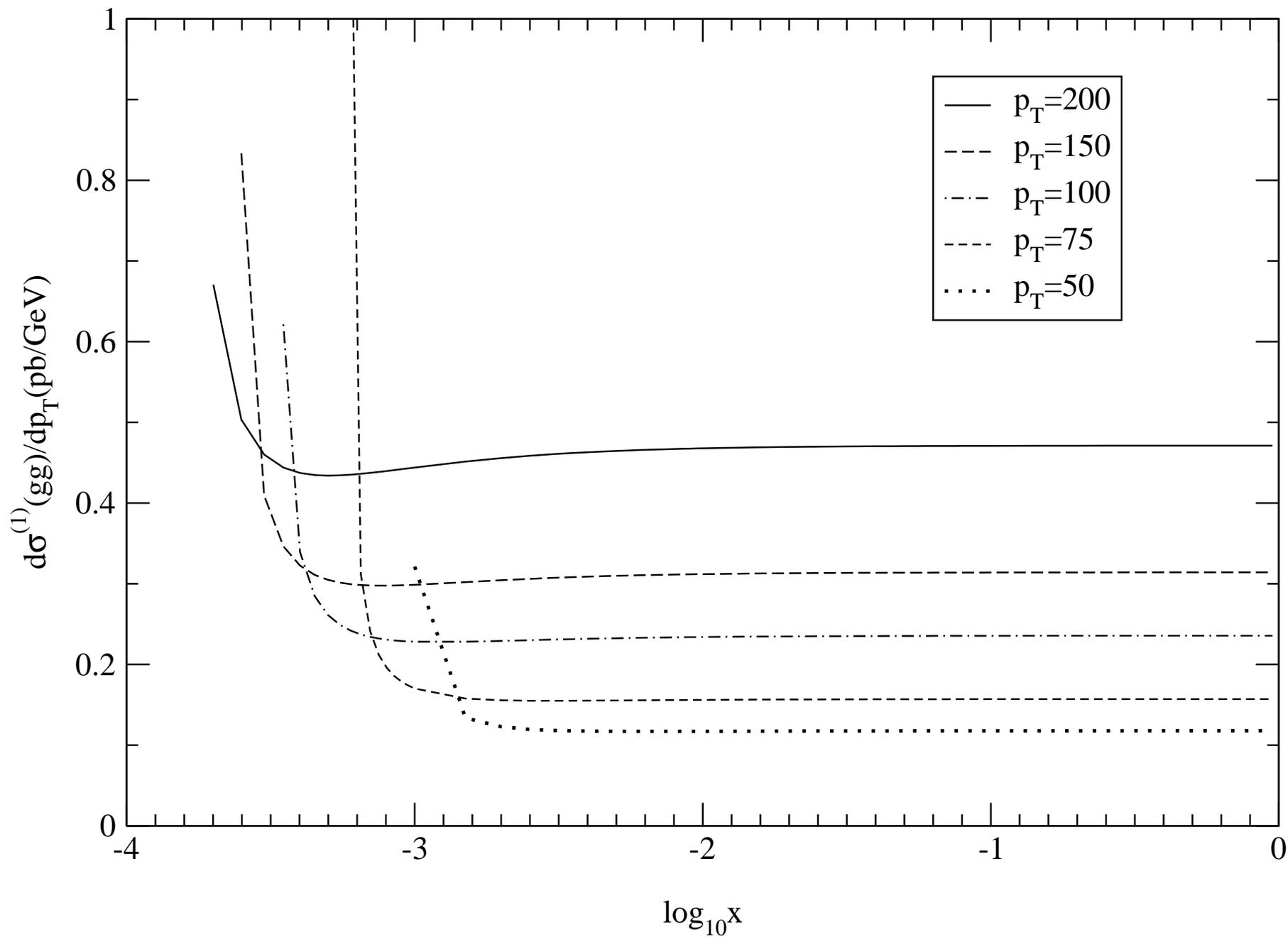
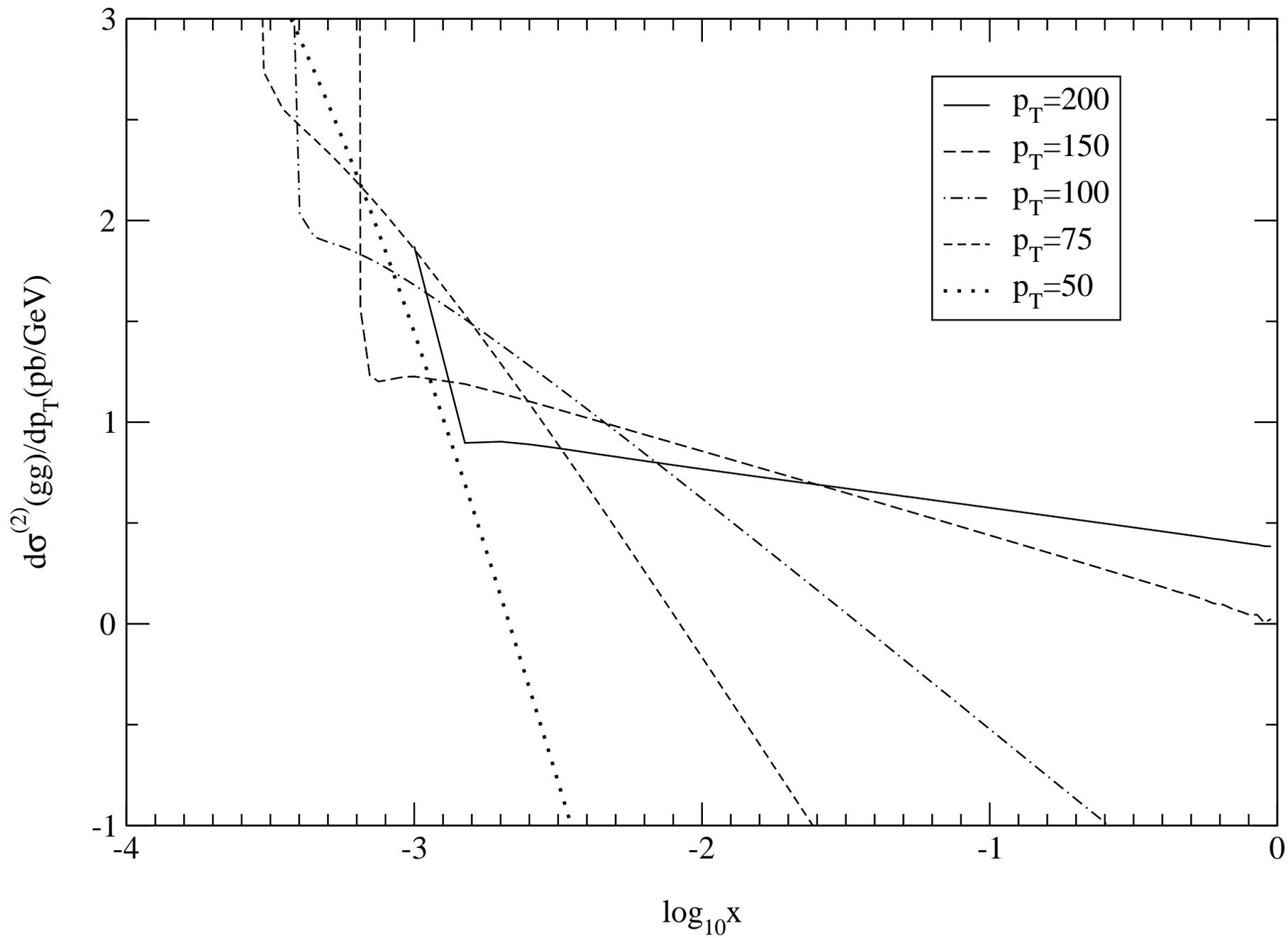
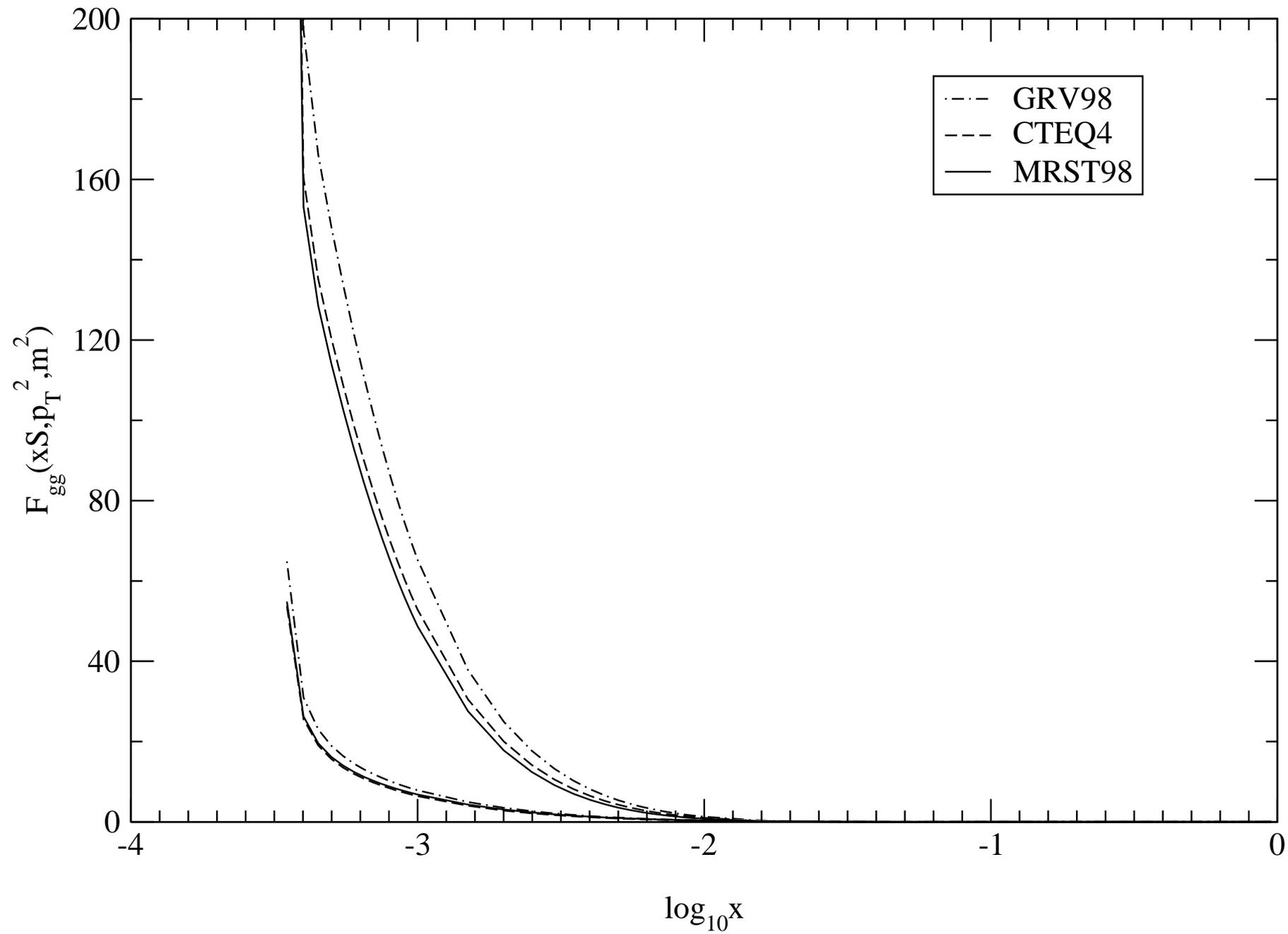
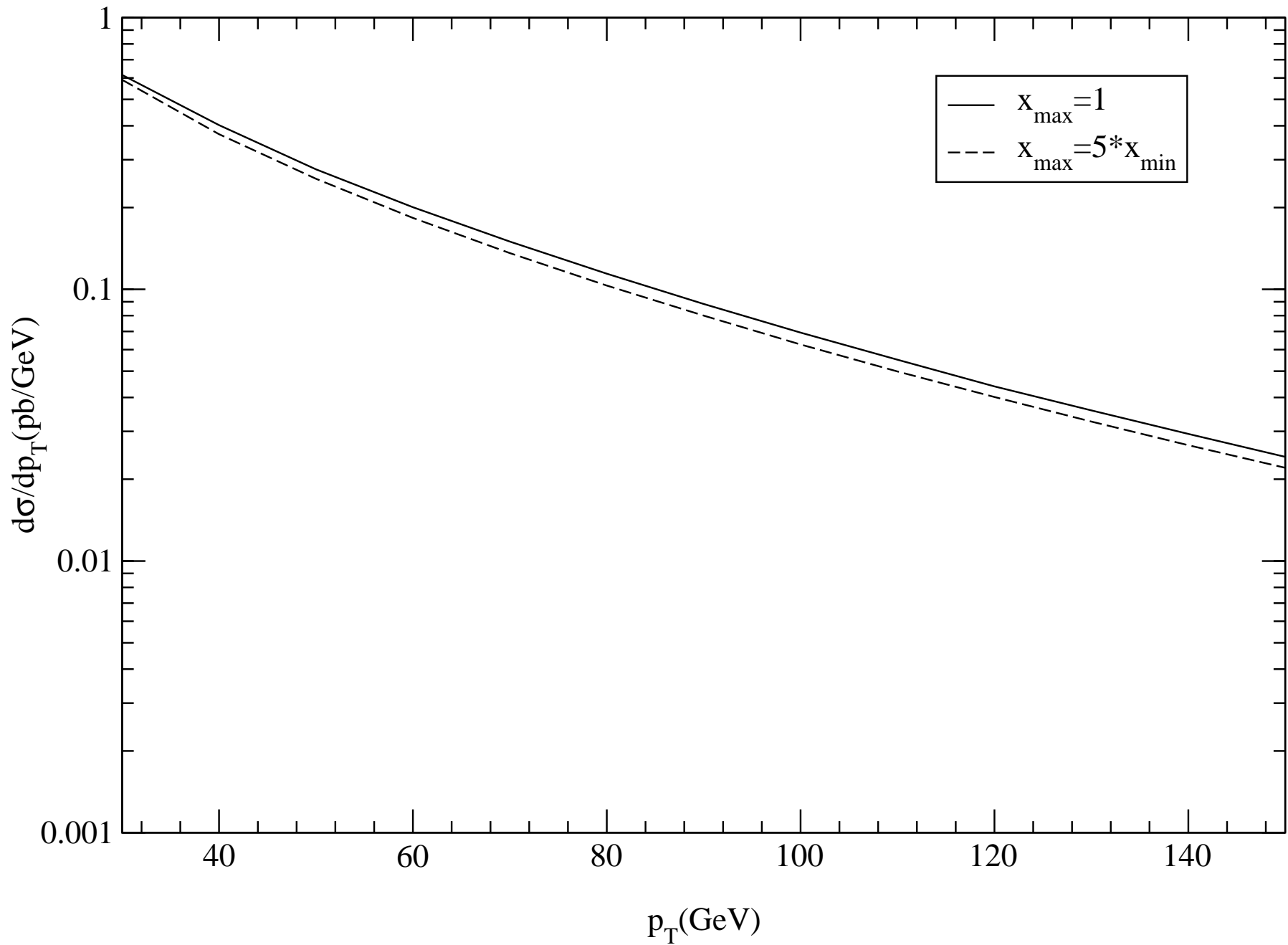
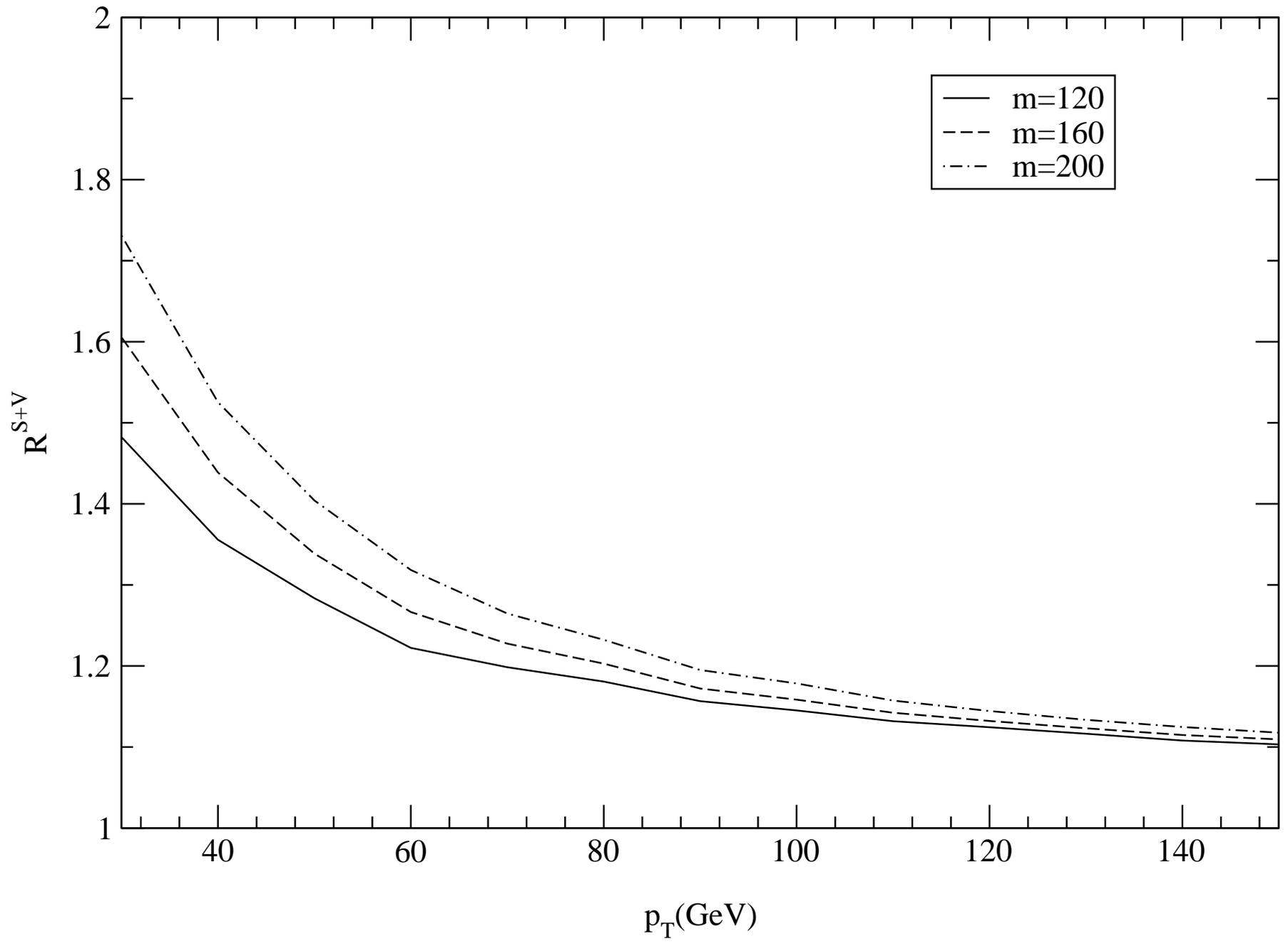
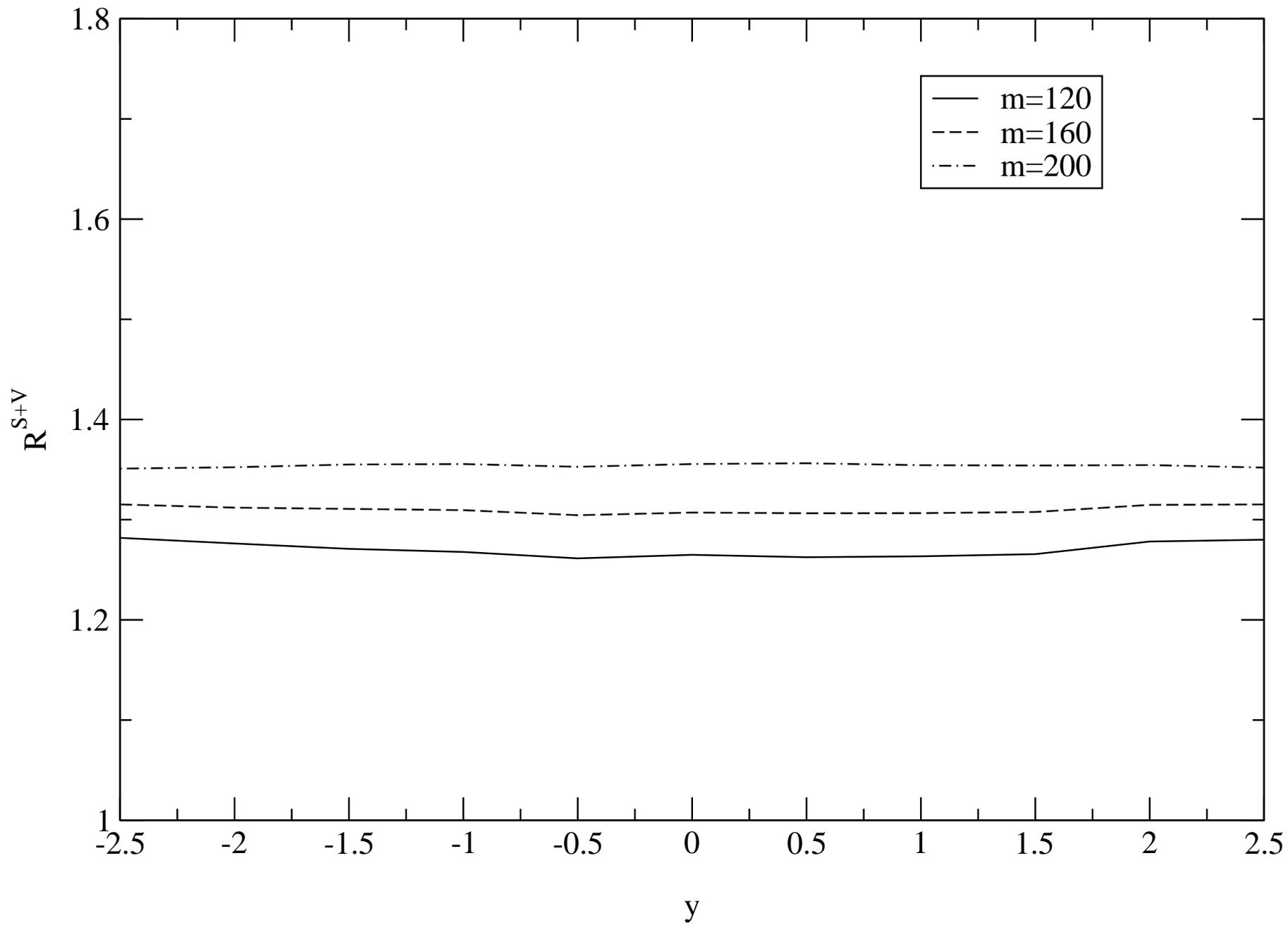
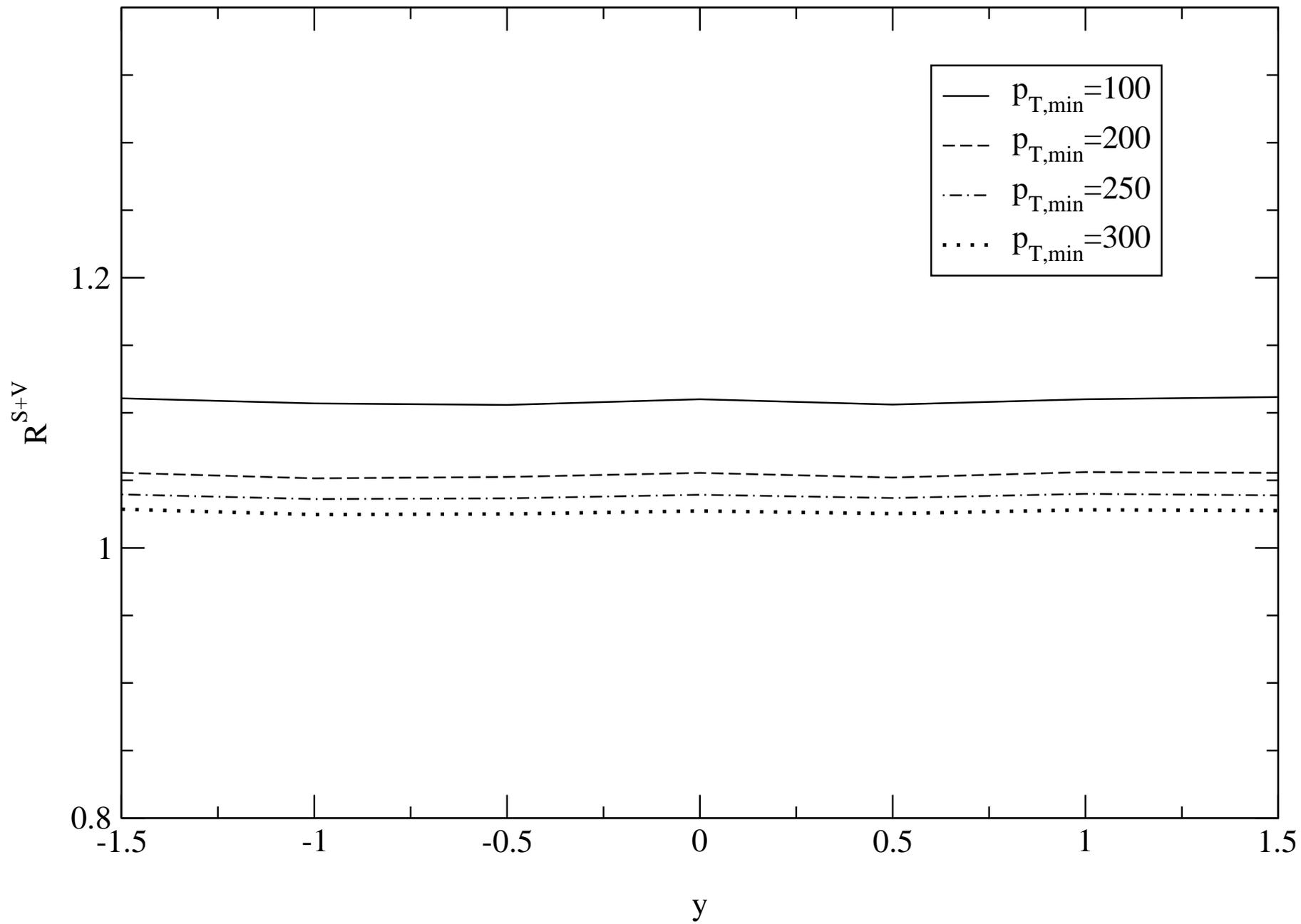
\end{document}